\documentclass[aps,prd,nofootinbib,showpacs,superscriptaddress,twocolumn,floatfix]{revtex4-1}
\usepackage{amsmath,amssymb,bm,color,fancyhdr,hyperref,lastpage,units}
\usepackage{graphicx}
\graphicspath{{figs/}}

\usepackage[latin1,utf8]{inputenc}
\usepackage[english]{babel}

\def \unitmatrix{1\!\!\!1}

\newcommand{\psib}{\bar{\psi}}
\newcommand{\chib}{\bar{\chi}}
\newcommand{\cM}{\mathcal{M}}

\newcommand{\Ns}{N_\sigma}
\newcommand{\Nt}{N_\tau}
\newcommand{\x}{\mathbf{x}}

\begin{document}
\title{Meson Screening Masses in (2+1)-Flavor QCD}

\author{A.\ Bazavov}
\affiliation{Department of Computational Mathematics, Science and Engineering and
Department of Physics and Astronomy,\\ Michigan State University, East Lansing, MI
48824, USA}

\author{S.\ Dentinger}
\affiliation{Fakult\"at f\"ur Physik, Universit\"at Bielefeld, D-33615 Bielefeld, Germany}

\author{H.-T.\ Ding}
\affiliation{Key Laboratory of Quark \& Lepton Physics (MOE) and Institute of
Particle Physics,\\ Central China Normal University, Wuhan 430079, China}

\author{P.\ Hegde}
\affiliation{Center for High Energy Physics, Indian Institute of Science, Bangalore 560012, India}

\author{O.\ Kaczmarek}
\affiliation{Fakult\"at f\"ur Physik, Universit\"at Bielefeld, D-33615 Bielefeld,
Germany}
\affiliation{Key Laboratory of Quark \& Lepton Physics (MOE) and Institute of
Particle Physics,\\ Central China Normal University, Wuhan 430079, China}

\author{\\ F.\ Karsch}
\affiliation{Fakult\"at f\"ur Physik, Universit\"at Bielefeld, D-33615 Bielefeld, Germany}

\author{E.\ Laermann}
\thanks{Deceased.}
\affiliation{Fakult\"at f\"ur Physik, Universit\"at Bielefeld, D-33615 Bielefeld, Germany}

\author{Anirban\ Lahiri}
\affiliation{Fakult\"at f\"ur Physik, Universit\"at Bielefeld, D-33615 Bielefeld, Germany}

\author{Swagato\ Mukherjee}
\affiliation{Physics Department, Brookhaven National Laboratory, Upton, NY 11973, USA}

\author{H.\ Ohno}
\affiliation{Center for Computational Sciences, University of Tsukuba, Tsukuba, Ibaraki 305-8577, Japan}

\author{\\ P.\ Petreczky}
\affiliation{Physics Department, Brookhaven National Laboratory, Upton, NY 11973, USA}

\author{R.\ Thakkar}
\affiliation{Center for High Energy Physics, Indian Institute of Science, Bangalore 560012, India}

\author{H.\ Sandmeyer}
\affiliation{Fakult\"at f\"ur Physik, Universit\"at Bielefeld, D-33615 Bielefeld, Germany}

\author{C.\ Schmidt}
\affiliation{Fakult\"at f\"ur Physik, Universit\"at Bielefeld, D-33615 Bielefeld, Germany}

\author{S.\ Sharma}
\affiliation{Department of Theoretical Physics, The Institute of Mathematical Sciences, Chennai 600113, India}

\author{P.\ Steinbrecher}
\affiliation{Physics Department, Brookhaven National Laboratory, Upton, NY 11973, USA}

\collaboration{HotQCD collaboration}

\begin{abstract}
We present lattice QCD results for mesonic screening masses in the temperature range 140 MeV $\lesssim T \lesssim 2500$~MeV.
Our calculations were carried out using (2+1)-flavors of the Highly Improved Staggered Quark (HISQ) action, with 
a physical value for the strange quark mass and two values of the light quark mass corresponding to pion masses
of 160 MeV and 140 MeV. Continuum-extrapolated results were obtained using calculations with
a variety of lattice spacings corresponding to temporal lattice extents $\Nt = 6$~--~16. We discuss the implications
of these results for the effective restoration of various symmetries in the high temperature phase of QCD, as well as
the approach toward the perturbative limit.
\end{abstract}

\date{\today}

\maketitle
\section{Introduction}
\label{sec:introduction}
At high temperatures
the properties of strong-interaction matter change from being controlled by
hadronic degrees of freedom to deconfined quarks and gluons.
While the thermodynamics in the low temperature phase of QCD resembles many 
features of a hadron resonance gas, with hadrons keeping their 
vacuum masses, this quickly changes at temperatures close to and above the 
crossover transition to the high temperature phase. In fact, the zero 
temperature hadronic degrees of freedom seem to provide a quite satisfactory 
description of thermal conditions close to the transition to the high 
temperature phase \cite{Andronic:2017pug}, although there is evidence of
thermal modification of the spectrum \cite{Aarts:2018glk}.
At high temperature, however, quarks and gluons deconfine, which also is 
reflected in properties of hadron correlation functions and the thermal
masses extracted from them (see e.g. \cite{Karsch:2003jg}). Resonance
peaks in spectral functions, which enter the integral representations
of thermal hadron correlation functions, broaden and shift with
temperature \cite{Aarts:2017rrl}.
In spatial correlation functions \cite{DeTar:1987ar} the finite temporal
extents, $0\le \tau\le 1/T$, of the Euclidean lattice acts on spatial
quark and anti-quark propagators like a finite volume effect, which
influences the long-distance behavior of these correlation functions.
Their exponential decay at large distances defines
screening masses, which differ substantially from the pole masses at
zero temperature, and approach multiples of $\pi T$ at high temperature,
which is characteristic for the propagation of free quark quasi-particles
in a thermal medium.

The chiral cross-over separating the low and high temperature regimes for
non-vanishing quark masses is characterized by a smooth but rapid change
of the chiral condensate around $T_{pc}$. The pseudo-critical temperature
$T_{pc}$, for the physical value of the ratio of light and strange
quark masses, has recently been determined from fluctuations
of various chiral observables :
$T_{pc} = (156.5\pm 1.5)~\textrm{MeV}$ \cite{Bazavov:2018mes}.

Despite a small explicit breaking of the chiral symmetry by the residual light 
quark masses, the chiral symmetry, which is spontaneously broken in the 
hadronic phase, gets effectively restored above $T_{pc}$. 
The deconfinement of the light quark and gluon degrees of freedom is believed 
to be strongly related to the drop of the chiral condensate and the resultant
effective restoration of the chiral symmetry.
If chiral symmetry is restored then the excitations of the plasma are
also expected to carry those informations in spatial hadron correlators.
In fact, the analysis of spatial hadron correlation functions and
their asymptotic large distance behavior \cite{DeTar:1987ar} is found 
to be a sensitive tool for studies of different patterns of chiral
symmetry restoration at high temperature. Generally it is found in
calculations at physical values of the quark masses that the temperature
dependence of screening masses, differs significantly in quantum number 
channels sensitive to the restoration of the $SU_L(2)\times SU_R(2)$ 
chiral flavor symmetry and the anomalous axial $U_A(1)$ symmetry, respectively.
While the former will be restored completely at chiral transition temperature
in the chiral limit, the latter remains broken also at high temperature by the
Adler-Bell-Jackiw anomaly~\cite{Adler:1969gk, Bell:1969ts, Adler:1969er}.
However, with the thermal suppression of non-perturbative breaking effects,
which at zero temperature arise, for instance, from the presence of
topologically non-trivial gauge field configurations \cite{Gross:1980br},
the anomalous axial symmetry may be \textquotedblleft effectively restored\textquotedblright.
It has been argued that the question whether or not the chiral symmetry and anomalous
axial symmetry get effectively restored at the same temperature may have
significant qualitative consequences for the structure of the QCD
phase diagram in the chiral limit \cite{Pisarski:1983ms}.

Calculations with staggered fermions \cite{Cheng:2010fe, Ohno:2012br} show evidence for
$U_A(1)$ symmetry breaking also above $T_{pc}$ and provide evidence
for the close relation between axial symmetry breaking and the density
of near-zero eigenmodes \cite{Dick:2015twa}. However, to
what extent the flavor singlet anomalous axial $U_A(1)$ symmetry gets effectively
restored at the chiral phase transition temperature,
$T_c^0= 132^{+3}_{-6} \textrm{MeV}$ \cite{Ding:2019prx}, which defines
the onset of a true phase transition in the chiral limit, is still an
open question~\cite{Shuryak:1993ee,Birse:1996dx,Lee:1996zy,Evans:1996wf}.

Several recent lattice QCD calculations performed in 2 and (2+1)-flavor
QCD with physical quark mass values utilizing overlap and M\"obius domain 
wall \cite{Buchoff:2013nra,Bazavov:2012qja,Suzuki:2018vbe,Tomiya:2016jwr,Chiu:2013wwa, Sharma:2016cmz}
 as well as Wilson \cite{Brandt:2016daq} fermions observe an effective restoration of the
$U_A(1)$ symmetry at temperatures above the pseudo-critical        
temperature $T_{pc}$, {\it i.e.}\ at about $(1.2-1.3) T_{pc}$. This is
in accordance with earlier findings in calculations of screening
masses with staggered fermions, where effective $U_A(1)$ restoration
has been observed through the degeneracy of scalar and pseudo-scalar
correlation functions and screening masses at temperatures
$T\gtrsim 1.3 T_{pc}$ \cite{Cheng:2010fe}.

One of the motivations of this study is to also determine the extent to which 
$U_A(1)$ is effectively restored at the chiral crossover temperature through
screening masses for which we have performed continuum extrapolation not
yet performed in earlier studies. At the level of screening correlators,
$U_A(1)$ restoration will lead to a degeneracy between the scalar ($S$)
and pseudoscalar ($PS$) correlators, while chiral symmetry restoration
yields a degeneracy between the vector ($V$) and axial vector ($AV$) correlators.
We calculate mesonic correlation functions numerically using 
(2+1)-flavor lattice QCD for all the possible flavor combinations including 
light and strange quarks, namely, light-light ($\bar u d$), light-strange
($\bar u s$) and strange-strange ($\bar ss$). Within each flavor combination, 
we determine scalar, pseudoscalar, vector and axial vector ground sate screening masses.
The temperature dependence of this set of meson correlation functions has been analyzed 
before~\cite{Cheng:2010fe}, including also charmonia~\cite{Karsch:2012na}, on coarse lattices using the
p4 discretization scheme for staggered fermions. With this calculation we
substantially improve over earlier work by using the Highly Improved
Staggered Quark (HISQ) action with physical values for the light and strange
quark masses and by performing calculations in a wide range of lattice spacings,
0.017 fm $\le a \le$ 0.234 fm that allows us to perform controlled
extrapolations to the continuum limit in the temperature range
140 MeV $\le T\le$ 974 MeV. Albeit not continuum extrapolated, we
extend the calculation of screening masses to temperatures as
large as $2.5$~GeV. Results for screening masses for charmonia, open
strange-charm as well as for $\bar s s$ channels, with the HISQ action
but for only a single lattice spacing corresponding to $\Nt = 12$,
have been reported before~\cite{Bazavov:2014cta}.

This paper is organized as follows: In the next section, we briefly review 
properties of spatial meson correlation functions and their evaluation using
the staggered fermion discretization scheme. We describe the staggered fermion
set-up for our calculations in Sec.~\ref{sec:calc_setup}. We then present our
results in Sec.~\ref{sec:results} where we start with updating our scale setting
in Sec.~\ref{ssec:scale_lop} and present some zero-temperature meson masses.
Staggered fermion specific cut-off effects, so-called taste splittings, for
$T=0$ are shown in Sec.~\ref{ssec:taste_splitting_hisq_action}.
We present results for temperatures around the chiral crossover regime in 
Sec.~\ref{ssec:low_temperature_results} where we also discuss
effective $U_A(1)$ restoration. In Sec.~\ref{ssec:high_temperature_results},
we present our results for the screening masses at high temperatures compared
to chiral crossover temperature and compare these with predictions from
resummed thermal perturbation theory. Finally we state our conclusions in
Sec.~\ref{sec:conclusions}. For completeness we have appendices where we start with
an update of the parametrization for scale setting in Appendix~\ref{app:fKparametrization} and
then in Appendix~\ref{app:statistics} and Appendix~\ref{app:screen_mass_values}, we summarize
our statistics and tabulate the continuum-extrapolated values of the screening masses, respectively.

\section{Spatial correlators and screening masses}
\label{sec:spatial_corr_screening_mass}

Properties of the hadron spectrum at zero and non-zero temperature are commonly
determined from an analysis of two-point correlation functions
$\langle\cM_\Gamma({\bf x}) \overline{\cM}_\Gamma({\bf y})\rangle$, where the operators 
$\cM_\Gamma$ project on to a specific set of quantum numbers and $\bf x$,
$\bf y$ are Euclidean space-time coordinates. At zero temperature
the lowest excitation (mass) in a given quantum number channel is
conveniently extracted from the asymptotic large Euclidean time behavior of the
correlation function. At finite temperature, the calculation of
correlators separated in Euclidean time is limited by the limited extent of this 
direction that determines the inverse temperature of the system, $\beta = 1/T$. 
In contrast there are no such restrictions for spatially separated correlators, 
also known as screening correlators.

In QCD, the finite temperature meson screening correlators, projected
onto zero transverse momentum (${\bf p}_\perp\equiv (p_x,p_y)=0$) and lowest Matsubara
frequency of a bosonic state ($p_0\equiv \omega_0=0$), are defined by
\begin{equation}
G_\Gamma(z,T) = \int_0^\beta d\tau \int dxdy\,\Big\langle \cM_\Gamma(x,y,z,\tau) \overline{\cM_\Gamma}(0,0,0,0) \Big\rangle,
\label{eq:definition}
\end{equation}
where $\cM_\Gamma \equiv \psib \Gamma \psi$ is a meson operator
that projects onto a quantum number channel $\Gamma$ selected by
$\Gamma = \Gamma_D \otimes t^a$ with Dirac matrices $\Gamma_D$ and 
a flavor matrix $t^a$. The angular brackets $\langle\cdots\rangle$,
denote the expectation value over the gauge field ensemble.
The correlators decay exponentially for large $z$,
\begin{equation}
	G_\Gamma(z,T)\hspace*{0.4cm} 
	{\raise0.6ex\hbox{$\sim$\kern-0.75em\raise-1.4ex\hbox{\hspace*{-9pt}$z \to \infty$}}}
	\hspace*{0.2cm}{\rm e}^{-m_\Gamma (T) z}\;\; , 
	\label{asymptotic}
\end{equation}
which defines the corresponding screening mass $m_\Gamma(T)$.
As already mentioned, for $T \to 0$, the screening masses tend to
the mass of the $T=0$ meson with the same quantum numbers.
For $T \to \infty$, they approach the common value $m_\Gamma = 2 \pi T$
irrespective of the spin and flavor \cite{DeTar:1987ar}, which indicates that the
dominant excitations consist of two almost free fermionic excitations (quarks)
which each have a lowest Matsubara frequency (energy) $\omega_0=\pi T$.
For non-zero $T$, the relation between screening mass and pole mass could be highly non-trivial
due to the emergence of non-analytic structures in the spectral function \cite{Hashimoto:1992np}.

\begin{table}[!t]
\begin{center}
\begin{tabular}{|c c|c c|c c|c c|} \hline
           & $\phi(\x)$& \multicolumn{2}{c|}{$\Gamma_D$}&
  \multicolumn{2}{c|}{$J^{PC}$}& \multicolumn{2}{c|}{states} \\
  && NO& O& NO& O& NO& O \\ \hline

  $\cM1$& $(-1)^{x+y+\tau}$& $\gamma_3\gamma_5$& $\unitmatrix$& $0^{-+}$& $0^{++}$&
  $\pi_2$& $a_0$ \\

  $\cM2$& $1$& $\gamma_5$& $\gamma_3$& $0^{-+}$& $0^{+-}$& $\pi$& -- \\

  $\cM3$& $(-1)^{y+\tau}$& $\gamma_1\gamma_3$& $\gamma_1\gamma_5$& $1^{--}$& $1^{++}$&
  $\rho^\mathcal{T}_2$& $a_1^\mathcal{T}$ \\
  
  $\cM4$& $(-1)^{x+\tau}$& $\gamma_2\gamma_3$& $\gamma_2\gamma_5$& $1^{--}$& $1^{++}$&
  $\rho^\mathcal{T}_2$& $a_1^\mathcal{T}$ \\
  
  $\cM5$& $(-1)^{x+y}$& $\gamma_4\gamma_3$& $\gamma_4\gamma_5$& $1^{--}$& $1^{++}$&
  $\rho^\mathcal{L}_2$& $a_1^\mathcal{L}$ \\

  $\cM6$& $(-1)^x$& $\gamma_1$& $\gamma_2\gamma_4$& $1^{--}$& $1^{+-}$&
  $\rho^\mathcal{T}_1$& $b_1^\mathcal{T}$ \\

  $\cM7$& $(-1)^y$& $\gamma_2$& $\gamma_1\gamma_4$& $1^{--}$& $1^{+-}$&
  $\rho^\mathcal{T}_1$& $b_1^\mathcal{T}$ \\

  $\cM8$& $(-1)^\tau$& $\gamma_4$& $\gamma_1\gamma_2$& $1^{--}$& $1^{+-}$&
  $\rho^\mathcal{L}_1$& $b_1^\mathcal{L}$ \\ \hline 
\end{tabular}
\end{center}
\caption{The list of local meson operators studied in this work. States associated with the
non-oscillating and the oscillating part of the screening correlators are designated
by the identifiers NO and O, respectively. Particle assignments of the corresponding
states are given only for the $\bar u d$ flavor combination. The superscripts $\mathcal{T}$ and
$\mathcal{L}$ stand for transverse and longitudinal, respectively.
The operators listed here are the same as in Ref.~\cite{Cheng:2010fe}.\label{tab:phase_factors}}
\end{table}

On the lattice, the continuum Dirac action must be replaced by a suitable discrete variant.
Staggered fermions, which we use in this work, are described by one-component spinors rather than the usual
four-component spinors. Because of this, they are relatively inexpensive to simulate. However the price to be paid
is that the relation to the continuum theory is subtle. The continuum limit of the theory is the Dirac theory of
four fermions rather than one. As a result, each meson too comes in sixteen degenerate 
copies which are known as tastes and the corresponding operators are of the form 
$\psib({\bf x})~(\Gamma_D \otimes \Gamma_T^*)~\psi({\bf x})$, where $\psi({\bf x})$
is the 16-component hypercubic spinor and $\Gamma_D$ and $\Gamma_T$ are Dirac
matrices in spin and taste space respectively. Although different tastes are degenerate in the continuum, on the lattice this degeneracy
is broken by gluonic interactions. The masses of the taste partners can be determined from the decay of correlation
functions of staggered meson operators $\cM({\bf x}) = \sum_{{\bf n, n'}}\phi({\bf n,n'})\chib({\bf x+n})\chi({\bf x+n'})$,
where $\bf x$ is the hypercube co-ordinate and $\bf n$ and $\bf n'$ point to the various vertices of the unit hypercube and
$\phi$ is a site-dependent phase factor whose form depends on the spin and taste quantum numbers of the
meson~\cite{Golterman:1985dz,Kilcup:1986dg,Altmeyer:1992dd}.

In this work, we will only consider \emph{local}
operators, \emph{i.e.}, operators with ${\bf n} = {\bf n'}$. 
In Table~\ref{tab:phase_factors} we list the eight local staggered meson 
operators that were studied in this work and their mapping to the familiar 
mesons of QCD. We note that the operators $\cM3$, $\cM4$ and $\cM5$ 
(respectively $\cM6$, $\cM7$ and $\cM8$) refer to the $x$, $y$ and $\tau$ 
components of the same axial vector (respectively vector) meson. In the
spatial correlation functions the meson operators were
separated along the $z$ direction. One thus may average over the $\cM3$ and 
$\cM4$ (respectively $\cM6$ and $\cM7$) 
components in order to improve the signal. Note however, that unlike at 
$T=0$, at finite temperature one cannot average over all three transverse
directions due to absence of Lorentz invariance in the definition of the
correlators~\cite{Gupta:1999hp}. In the vector and axial vector channels
we thus deal with two distinct correlation functions and resulting
screening masses, denoted as transverse and longitudinal.

A typical staggered meson correlator, for a fixed separation (in lattice unit) between source and sink,
is an oscillating correlator that simultaneously couples to two sets of mesons with the same spin but with 
opposite parities:
\begin{equation}
\begin{split}
	G_{\phi}(n_\sigma) &= \sum_i \left[ A_i^{(-)} \cosh
	\left(am_{\phi,i}^{(-)} \left(n_\sigma-\frac{N_\sigma}{2}\right) \right) \right.\\
 &- \left. (-1)^{n_\sigma} A_i^{(+)} 
 \cosh \left(am_{\phi,i}^{(+)} \left(n_\sigma-\frac{N_\sigma}{2}\right) \right) \right]\, 
\label{eq:meson}
\end{split}
\end{equation}
where $n_\sigma =z/a$ denotes the spatial separation of the source and sink
operators $\cM_\phi$. For large enough distances the correlator of Eq.~\ref{eq:meson}
may be constrained to a single term, {\it i.e.}\ $i=0$. In Eq.~\ref{eq:meson} we also
replaced the large distance exponential fall-off given in
Eq.~\ref{asymptotic} by a hyperbolic cosine which arises due to the periodic 
nature of correlators on lattices with finite spatial extent $N_\sigma$.

\section{Calculational setup}
\label{sec:calc_setup}
\subsection{Data sets}
We calculated the six distinct mesonic correlators, constructed from
local staggered fermion operators introduced in the previous
subsection, numerically using (2+1)-flavor gauge field ensembles 
generated with the HISQ action and a Symanzik improved gauge action. 
The HISQ action~\cite{Lepage:1998vj, Follana:2006rc, Bazavov:2009jc} 
is known to have the least amount of taste-splitting \cite{Bazavov:2010ru},
due to which it has been used in several precision studies both at $T = 0$
as well as at finite temperature~\cite{Bazavov:2010ru, Bazavov:2010pg, Follana:2006rc, Follana:2007uv, Bazavov:2011nk}.
The gauge ensembles for $\beta\le 7.825$, have been generated by HotQCD
collaboration and previously had been used to study the QCD equation of
state of strongly interacting matter~\cite{Bazavov:2014pvz, Bazavov:2017dus}. 
For $\beta > 7.825$, we have used the gauge ensembles from TUMQCD
collaboration, generated for the study of the expectation values of the
Polyakov loop and its correlators \cite{Bazavov:2016uvm, Bazavov:2018wmo}.
Gauge configurations have been generated on lattices of size $\Ns^3\times\Nt$,
where $\Nt = 6$, 8, 10, 12 and 16, and $\Ns=4\Nt$.
Most of the data for these five different values of the temporal 
lattice size, corresponding to five different values of the lattice spacing
$a$ at fixed value of the temperature $T=1/(\Nt a)$, have been collected
in a temperature range $140~{\rm MeV} \le T\le 172~{\rm MeV}$ using
physical values of the light ($m_l$) and strange ($m_s$) quark masses, {\it i.e.} a
quark mass ratio $1/27$. On lattices
with temporal extent $\Nt=8,~10$ and $12$ we also used data sets 
obtained with a slightly larger quark mass ratio, $1/20$. These data sets cover
a larger temperature range up to about $2.5$~GeV. The Goldstone pion masses for these two quark mass ratios
are 140 MeV for $m_l/m_s = 1/27$ and 160 MeV for $m_l/m_s = 1/20$.

All the above-mentioned gauge configurations used in this analysis have been generated with a strange quark 
mass tuned to its physical value by tuning the mass of the $\eta_{\bar ss}$ meson,
$M_{\eta_{\bar ss}} = 686$~MeV. This value is based on leading order chiral perturbation theory relation,
$M_{\eta_{\bar ss}}=\sqrt{2m_K^2-m_\pi^2}$, between the $\eta_{\bar ss}$, $\pi$ and $K$ masses.
Once the strange quark mass had been determined, the light quark mass was set to either $m_l = m_s/27$ or
$m_l = m_s/20$, as already discussed. The former choice of quark mass were used for temperatures below and 
near the chiral cross-over temperature, $T_{pc}$, while the higher quark mass was used at higher temperatures
($T \gtrsim 172$~MeV) where quark mass effects are negligible. The tuning of the strange quark masse, which
leads to our line of constant physics, is also discussed in detail in Ref.~\cite{Bazavov:2014pvz}.
All our simulation parameters and the number of gauge field configurations analyzed are summarized in 
Appendix~\ref{app:statistics}.

The conversion of hadron masses, calculated in lattice units, into physical 
units as well as the determination of our temperature scale requires the 
calculation of one physical observable that is used for the scale setting.
For this purpose we use the kaon decay constant, $f_K=156.1/\sqrt{2}$~MeV,
also used in other thermodynamics studies with the HISQ action.
We give the updated parametrizations of $f_Ka(\beta)$
in Appendix~\ref{app:fKparametrization}.

The purpose of the new calibration of the parametrization of $f_Ka(\beta)$
in Appendix~\ref{app:fKparametrization} is to improve on the scale at the larger $\beta$-values in this study.
Note that when compared to the previous scale \cite{Bazavov:2011nk,Bazavov:2014pvz}, this leads 
to a small $\sim$ 1\% decrease of the lattice spacing at the largest $\beta$-values
compared to the previous scale determination~\cite{Bazavov:2011nk,Bazavov:2014pvz}, while the
differences are negligible for $\beta \lesssim 7.0$.

\subsection{Hadron correlation functions}
A general meson correlator 
$\langle \cM({\bf x}) \overline{\cM}({\bf y}) \rangle$ 
consists of quark line connected and disconnected parts.
In this work we only focus on flavor non-singlet mesonic
correlators which do not have disconnected contribution.
The analysis of chiral symmetry restoration, including the $U_A(1)$ restoration,
can be performed using flavor non-singlet correlators alone 
\cite{Bazavov:2012qja,Hegde:2011np}.
The (fictitious) $\eta_{\bar ss}$ meson, whose mass was used to fix
the bare quark masses, also does not receive any contributions from
disconnected diagrams \cite{Bazavov:2014cta}.

\begin{figure}[!tb]
    \includegraphics[width=0.40\textwidth]{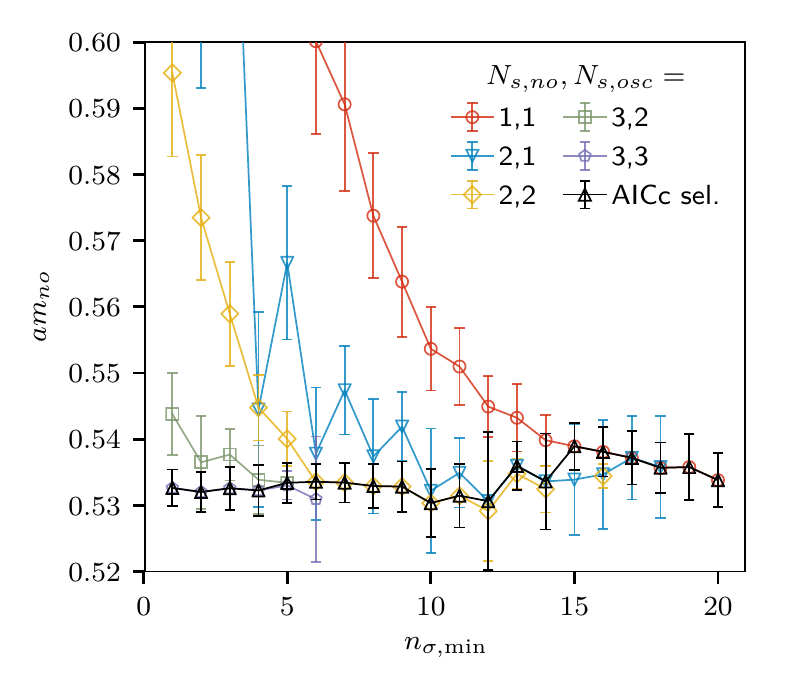}
    \caption{ Screening masses for in the vector channel with different number
              of states varying the fit interval for
              $48^3\times 12$ lattice for $\beta = 8.710$ which corresponds to
              $T=866\unit{MeV}$. Fit results selected by the AICc criterion
              (shown in black) for different values of minimum distance for
              the fits, seem to fall on a nice plateau.}
    \label{fig:AICc_selection}
\end{figure}

We generally had to retain up to 2-3 states in
Eq.\ \ref{eq:meson}. Such multi-state fits present a challenge as a 
straightforward fit is often highly unstable. For this purpose we
developed a routine to guess the initial parameters directly from
the data \cite{Sandmeyer:2019} for different terms of the sum in
Eq.\ \ref{eq:meson}.
We also developed \cite{Sandmeyer:2019}
a fit parameter estimation routine that works directly on the
oscillating correlators. This method relies on the fact that the
mass of the oscillating and non-oscillating part are usually roughly
of similar size and thus assumes their equality in the first step:

\begin{enumerate}
    \item At a small fit interval $[n_{\sigma,min}:n_{\sigma,max}=N_{\sigma}/2]$, perform one state fits on all even points
          of the correlator and we call the resulting fit parameters say $A_{\phi,0}^{even}$ and $m_{\phi,0}^{even}$.
          Repeat the same for the odd points ($A_{\phi,0}^{odd}$, $m_{\phi,0}^{odd}$).
    \item \label{item:even_odd_fit} Assuming similar size of the non-oscillating and oscillating mass, the fit parameters
          for the combined fit may be estimated with $A^{-}_{\phi,0} = (A_{\phi,0}^{even} + A_{\phi,0}^{odd}) / 2$,
          $A^{+}_{\phi,0} = (A_{\phi,0}^{even} - A_{\phi,0}^{odd}) / 2$ and
          $m^{-}_{\phi,0} = m^{+}_{\phi,0} = (m_{\phi,0}^{even} + m_{\phi,0}^{odd}) / 2$.
    \item \label{item:first_osc_fit} Using the parameters from step \ref{item:even_odd_fit} as initial
        guess, perform a full one state fit with oscillating and non-oscillating part.
    \item \label{item:guess_mass} Increase the fit interval. Guess the mass of the next excited state
        of either the even or the odd part (we used $m^{-/+}_{\phi,1} = 5/4m^{-/+}_{\phi,0}$).
        Adjust the corresponding amplitude ($A^{-}_{\phi,1}$ or $A^{+}_{\phi,1}$) such that the
        first point of the correlator in the fit interval is reproduced.
    \item \label{item:next_fit} Perform a full fit with higher states. Use the parameters
        from steps \ref{item:first_osc_fit} and \ref{item:guess_mass} as initial guess.
    \item Repeat steps \ref{item:guess_mass} to \ref{item:next_fit} until the desired number of states is reached.

\end{enumerate}

Having developed a method to perform automated multiple state fits, we still have to find which
set of fit parameters is the most reasonable one for a given fit interval. For that purpose we
have used the corrected Akaike information criterion (AICc)\cite{AIC,AICc}: For each fit interval we
have performed different multiple state fits (maximum 3 states for oscillating correlators and maximum
4 states for non-oscillating correlators) and selected the one which has the smallest AICc.
In Fig.~\ref{fig:AICc_selection} a comparison between the different multiple state fits and
the result that is selected by the AICc is shown. In contrast to the one state fit, this results
in an early onset of a  stable plateau. After the fits have been performed
the plateaus were manually selected for each correlator. The final value for the screening mass
and its uncertainty are determined by Gaussian bootstrapping. 
More technical details about the automated fitting procedure can be found in Ref.\ \cite{Sandmeyer:2019}.

\begin{figure}[!tb] 
\includegraphics[width=0.38\textwidth]{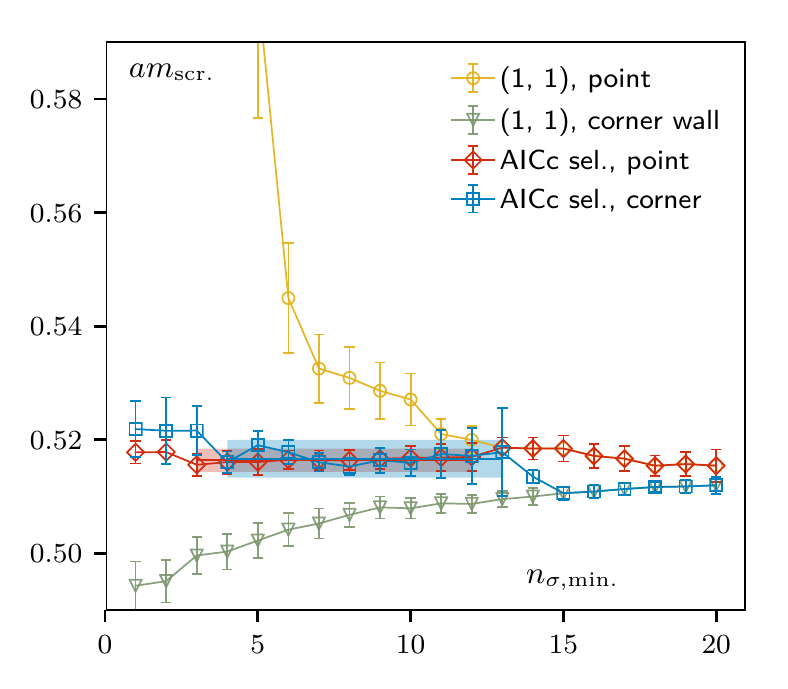}
\includegraphics[width=0.38\textwidth]{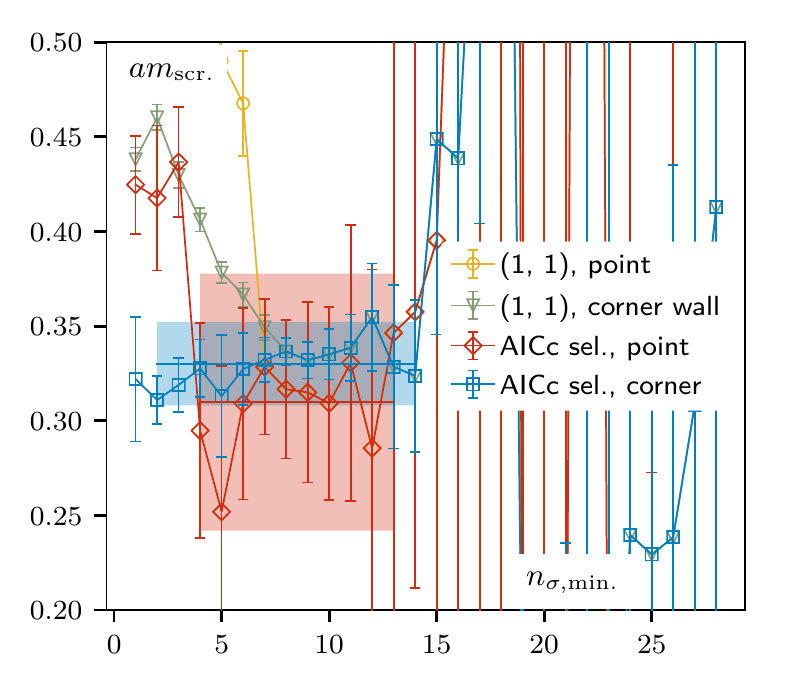}
\caption{Comparison of point versus corner wall sources for (top) the
scalar ($\cM1$) channel using a $48^3\times12$ lattice at $T = 769$~MeV and
(bottom) the vector (average of $\cM6$ and $\cM7$) channel using a $64^3\times16$ lattice at $T = 146$~MeV.
Numbers appearing in parenthesis corresponds to number of states taken in the fit for non-oscillating and
oscillating part of the correlator. The method of using AICc selection criterion to find a plateau
among various fits has been described in the main text.}
\label{fig:comp_point_cornerwall}
\end{figure}

We calculated screening correlation functions using point as well as corner wall sources.
The point source is the simplest type of source that one can use to calculate mesonic screening
functions and we have used one source for each color. However it does not suppress the excited
states, therefore isolating the ground state can be difficult unless the states are well-separated
or the lattice extent is large. The use of extended (smeared) sources can often help to suppress
excited state contributions, allowing to extract the ground state mass and amplitude even on smaller
lattices. Here we have used a corner wall source which means putting an unit source at the
origin of each $2^3$ cube on a chosen (in our case) $z$-slice \cite{Bernard:1991ah, Bernard:1993an, Bernard:2001av}.
In Fig.~\ref{fig:comp_point_cornerwall}, we show a comparison of a
mass calculation using point and corner wall sources at two different
temperatures. As discussed earlier, in both
cases we found that it is necessary to take into account contributions
from higher excited states to obtain reliable results for the ground
state screening masses. In Fig.~\ref{fig:comp_point_cornerwall} we have
only shown the fit results where we have taken one state for both oscillating
and non-oscillating part of the correlator (denoted by
self-explanatory notation \textquoteleft (1,1)\textquoteright)
and the AICc selected plateaus for the corresponding fit interval.
We found that the use of a corner wall source
provided advantages only for the noisy correlators, which in
particular are the vector and axial vector channels at low temperatures.
In the bottom panel of Fig.~\ref{fig:comp_point_cornerwall}, we provide an
example where a corner wall source yielded a better signal as compared
to a point source and one gets a longer plateau with smaller uncertainty
when the minimum distance for the fit, $n_{\sigma,min}$, is varied.
Therefore, we used the corner wall source only where
it was necessary, \emph{i.e.} for the vector and axial vector channels
below $T = 300$ MeV. For all the other cases however, we found that
higher state fits for the point source worked just as well and that
their results agreed with the corner wall results. We also found that
in the case of a corner wall source, the excited state often has a
negative amplitude and therefore, the influence of the higher
states is to shift the result for the screening mass downward
rather than upward as can be seen from the top panel of
Fig~\ref{fig:comp_point_cornerwall}.

\section{Results}
\label{sec:results}
\subsection{Scale setting and line of constant physics}
\label{ssec:scale_lop}
\begin{figure*}[tb]
\hspace{-0.03\textwidth}%
\includegraphics[width=0.33\textwidth]{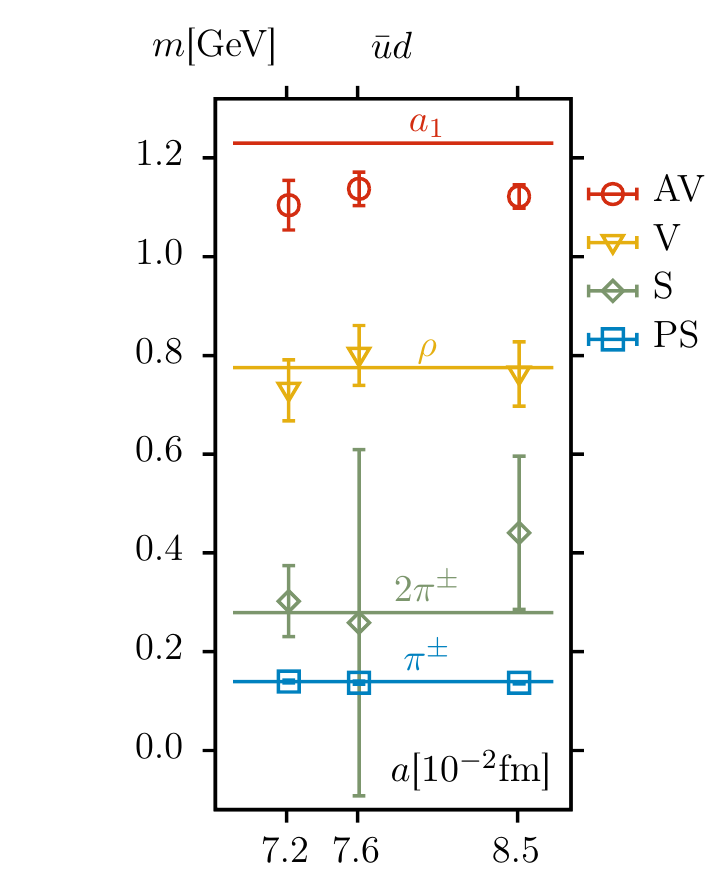}
\includegraphics[width=0.33\textwidth]{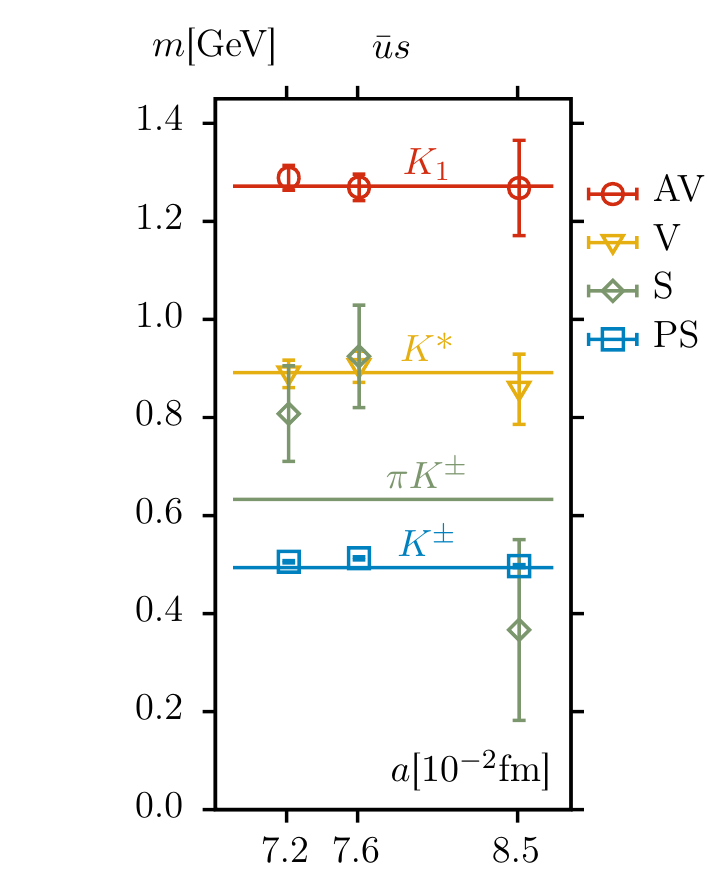}
\includegraphics[width=0.33\textwidth]{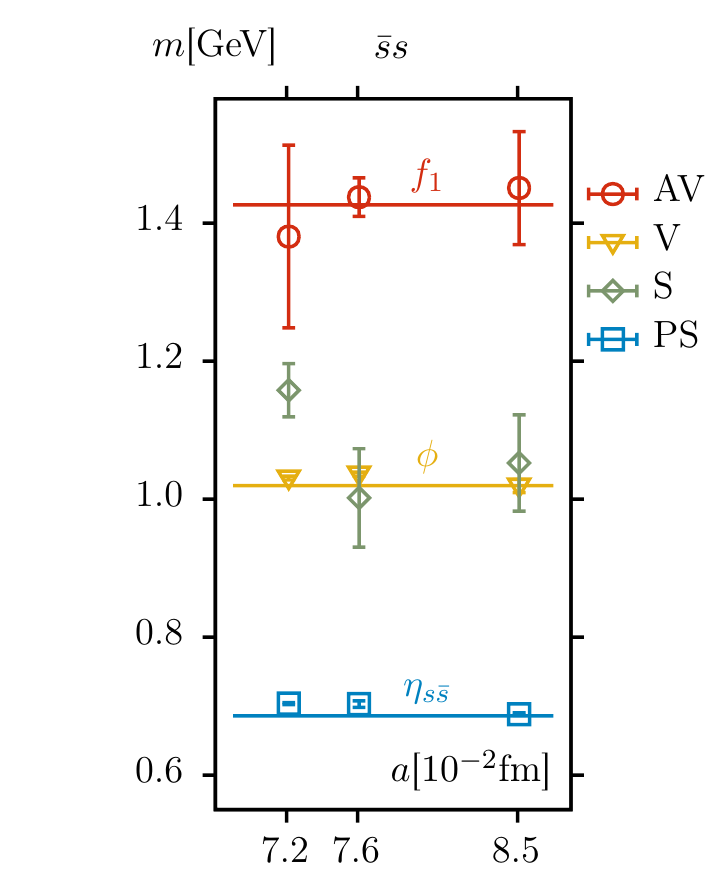}
\caption{$T=0$ masses of the four kinds of mesons studied in this work for the
$\bar u d$, $\bar u s$ and $\bar ss$ flavor channels, respectively. Horizontal lines correspond
to the physical values of the masses~\cite{Tanabashi:2018oca}. The scalar meson
mass is $2m_\pi$ instead of $m_{a0}$ (or $m_{\pi}+m_{\eta}$) due to a staggered artifact at finite lattice
spacing. This discrepancy vanishes when the continuum limit of the correlator
would be taken before calculating the screening mass
\cite{Prelovsek:2005rf, Prelovsek:2004jp}
(see Sec.~\ref{ssec:low_temperature_results}).}
\label{fig:zero_temp_masses}
\end{figure*}

As the scale setting calculations as well as the determination of the
line of constant physics had been performed prior to our current screening
mass analysis we tried to re-confirm the scales used in our calculation
through additional zero temperature calculations performed on lattices of
size $64^4$. We performed calculations at three values of the gauge coupling,
$\beta =7.01,\ 7.13$ and $7.188$. Using the parametrization of $f_Ka(\beta)$ 
given in Appendix~\ref{app:fKparametrization} this corresponds to lattice spacings
$a=0.085~{\rm fm},\ 0.076~{\rm fm}$ and $0.072~{\rm fm}$, respectively.
The strange quark mass has been fixed using $m_sa(\beta)$ from
Ref.\ \cite{Bazavov:2014pvz} and the light quark mass was taken
to be $m_l=m_s/27$. The resulting zero-temperature meson spectrum
is shown in Fig.~\ref{fig:zero_temp_masses}. 
The solid horizontal lines in the figures correspond to the 
experimentally determined values of the respective masses~\cite{Tanabashi:2018oca}.
The slight mismatch for $M_{\eta_{\bar ss}}$ ($m_K$),
arising from the slight mistuning of the strange quark mass, is
visible in right (middle) panel of Fig.~\ref{fig:zero_temp_masses}.
We note that results for most of the $PS$, $V$ and $AV$ mesons agree well
with the physical zero temperature spectrum within errors. The scalar meson,
in the ${\bar ud}$ sector however, seems to have twice the pseudoscalar mass
rather than the true scalar mass for ${\bar ud}$ sector. This is a well-known
staggered artifact \cite{Prelovsek:2005rf,Prelovsek:2004jp,Bernard:2007qf}
and we will also discuss its effect for non-zero temperatures in
Sec.~\ref{ssec:low_temperature_results}. However, such definite trend
is absent in heavier ${\bar us}$ and ${\bar ss}$ sectors.
A slight mismatch can also be observed for the $AV$ masses in ${\bar ud}$ sector
with no definite trend with decreasing lattice spacing.

\subsection{Taste splittings at $\bold{T=0}$}
\label{ssec:taste_splitting_hisq_action}
\begin{figure*}[!tbh]
\hspace{-0.03\textwidth}%
\includegraphics[width=0.33\textwidth]{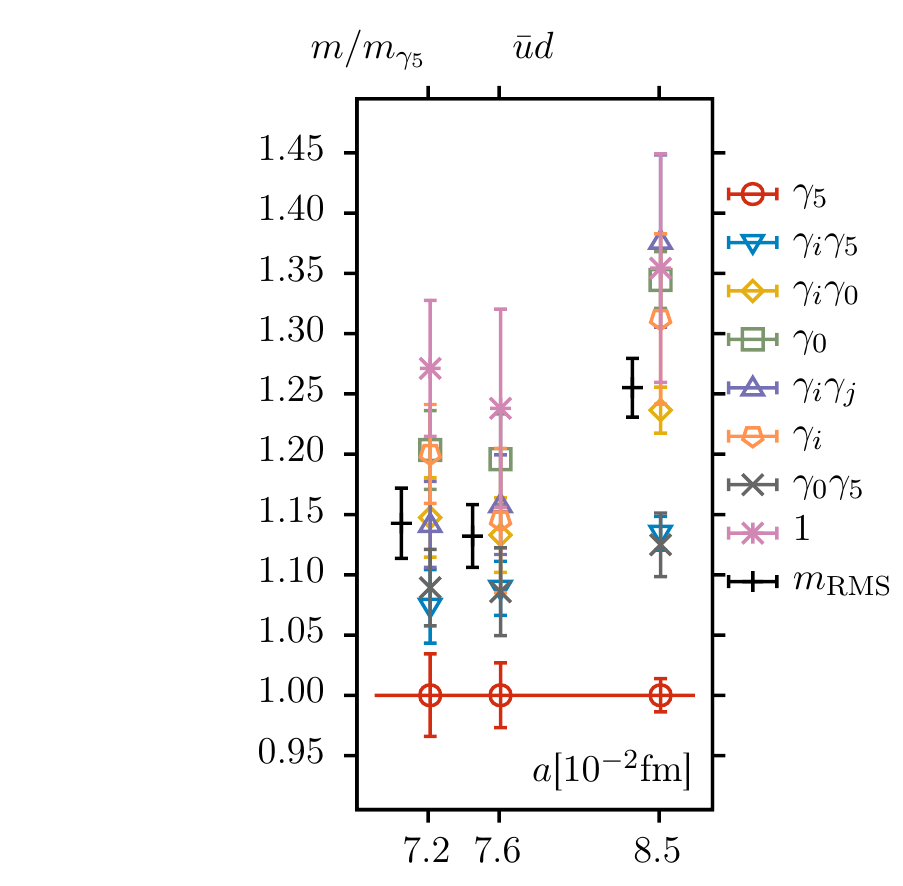}
\includegraphics[width=0.33\textwidth]{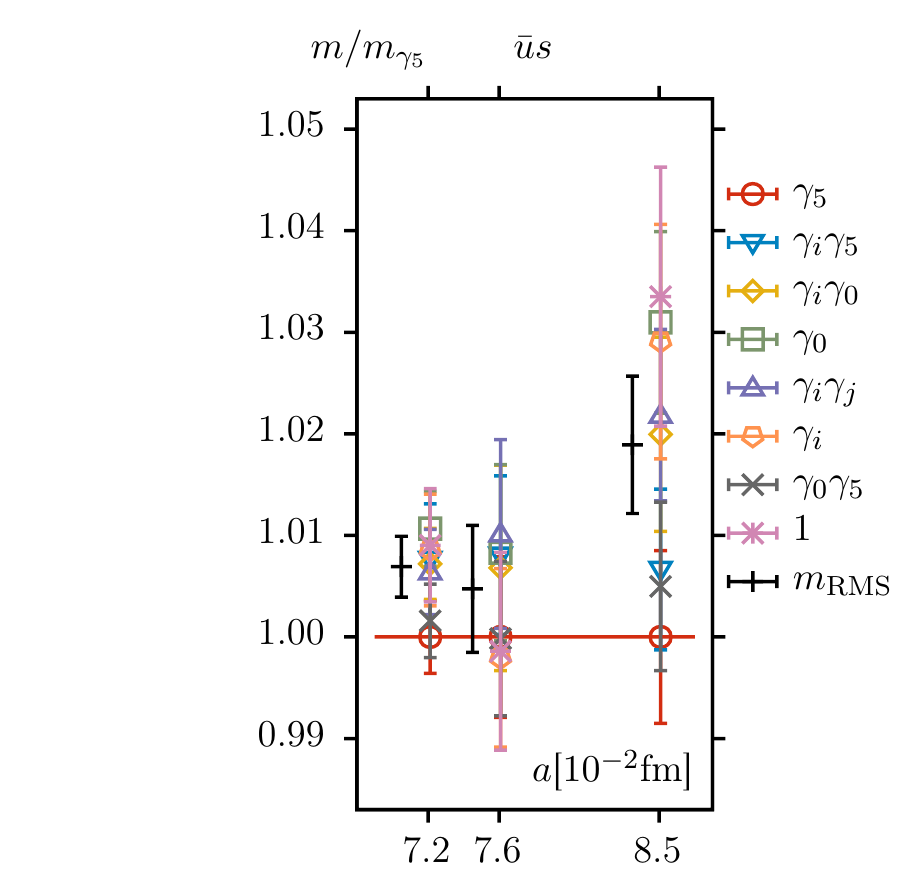}
\includegraphics[width=0.33\textwidth]{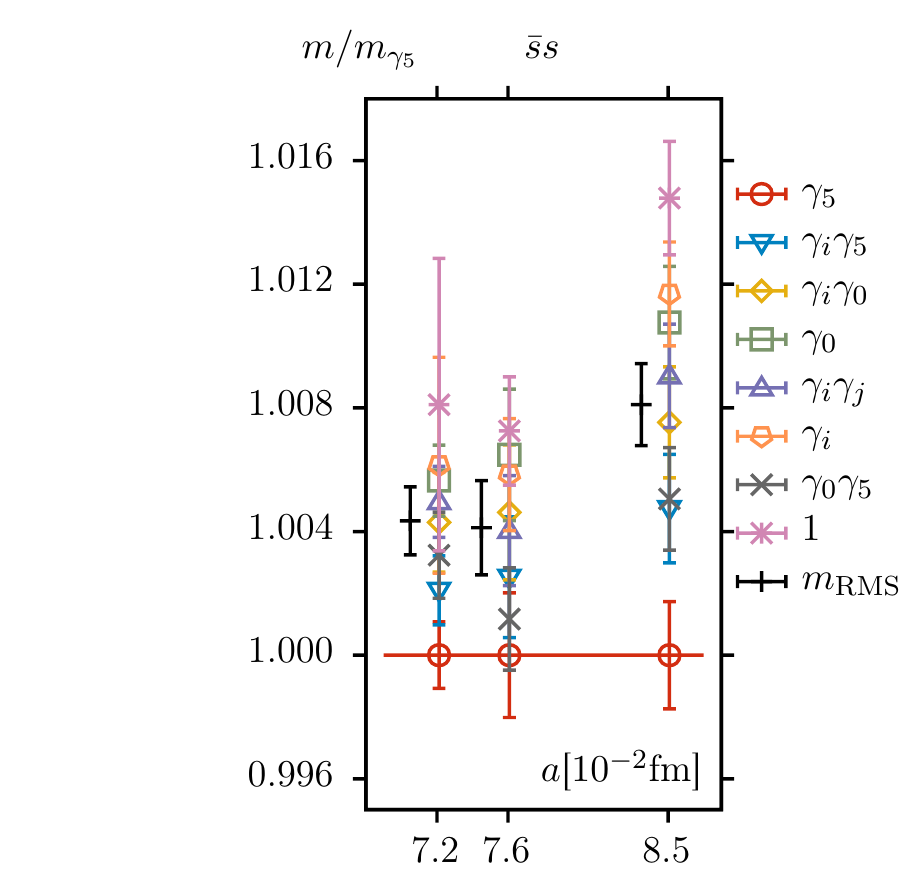}
\caption{Masses of the different taste partners of the pseudoscalar mesons, labeled by different $\Gamma_T$,
for light-light($\bar u d$), light-strange($\bar u s$), strange-strange($\bar s s$) sectors, normalized to
the corresponding Goldstone $\pi$, $K$ and $\eta_{\bar ss}$ masses, respectively. The lattice spacings considered
here range from approximately 0.085--0.072~fm. Also plotted are the RMS masses defined in Eq.~\eqref{eq:mPiRMS}.}
\label{fig:taste_splittings}
\end{figure*}

Although use of staggered quarks will lead to taste-splitting in every hadronic channels,
its effects are particularly severe in the pseudoscalar sector
($\pi$, $K$ and $\eta_{\bar ss}$), since these are the lightest states in the theory.
In Fig.~\ref{fig:taste_splittings}, we plot the masses of the sixteen different tastes 
of each of the three pseudoscalar mesons, {\it i.e.}, $\pi$, $K$ and $\eta_{\bar ss}$, for three
different values of the lattice spacing. The correlators for the different taste partners
are constructed using non-local operators \cite{Altmeyer:1992dd} with $\Gamma_D = \gamma_5$
and various $\Gamma_T$, as shown in Fig.~\ref{fig:taste_splittings}. In each case, the lightest
meson is the meson with the quantum numbers $\Gamma_T = \Gamma_D = \gamma_5$.
This meson is the only Goldstone boson that is massless in the chiral limit at finite lattice
spacing and the masses of the other fifteen mesons approach its mass in the continuum limit.
The masses of the other partners have been normalized to the mass of the corresponding
Goldstone boson for that particular lattice spacing. Our results extend the previous HISQ
results for taste-splitting to smaller lattice spacings. A more detailed discussion on
the taste splitting effects, also in comparison to other staggered discretization
schemes can be found in~\cite{Bazavov:2011nk,Bazavov:2010pg}.

One can define the root mean square (RMS) pion mass $m_{RMS}^{PS}$, as a measure of the taste splitting~\cite{Lee:1999zxa}:
\begin{widetext}
\begin{equation}
  m^{PS}_{RMS}= \sqrt{
  \frac{1}{16}\left(m_{\gamma_5}^2+m_{\gamma_0\gamma_5}^2
  +3m_{\gamma_i\gamma_5}^2+3m_{\gamma_i\gamma_j}^2
  +3m_{\gamma_i\gamma_0}^2+3m_{\gamma_i}^2
  +m_{\gamma_0}^2+m_{1}^2\right)}
\label{eq:mPiRMS}
\end{equation}
\end{widetext}
The $\gamma$-matrix suffixes in Eq.~\ref{eq:mPiRMS} refer to the taste structure of the mesons. The RMS mass
approaches the Goldstone mass in the continuum limit; hence its deviation from the Goldstone mass at a given
lattice spacing is a way of quantifying the taste-breaking effects. The sixteen tastes group into different 
multiplets, in a way understood from staggered chiral perturbation theory~\cite{Lee:1999zxa}. This is the 
reason for the factors of 3 in Eq.~\eqref{eq:mPiRMS}. We find that the RMS taste splitting is of the
order of 15-25~\% for the light-light($\bar u d$) sector but decreases to about 4-8~\% for the 
strange-strange($\bar{s}s$) sector. We also see that this splitting decreases as the lattice spacing
decreases, as expected. Lastly we note that the masses plotted here are consistent with the trend observed
in Fig.\ 2 of Ref.~\cite{Bazavov:2011nk} where the taste-splitting was calculated, with the same action but
for coarser lattices and a slightly heavier quark mass.

\subsection{Screening masses around the cross-over region}
\label{ssec:low_temperature_results}
\begin{figure*}[!tbh]
\hspace{-0.03\textwidth}%
\includegraphics[width=0.33\textwidth]{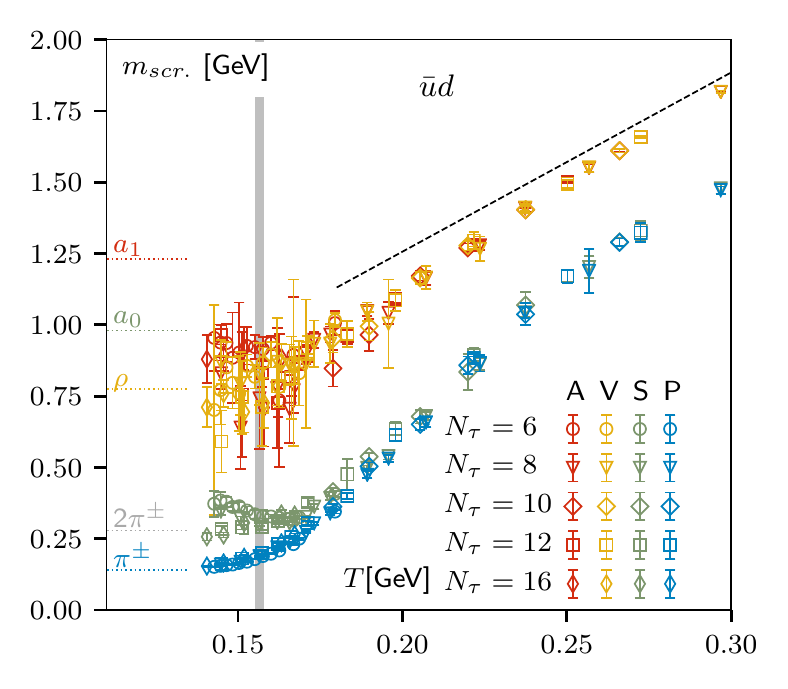}
\includegraphics[width=0.33\textwidth]{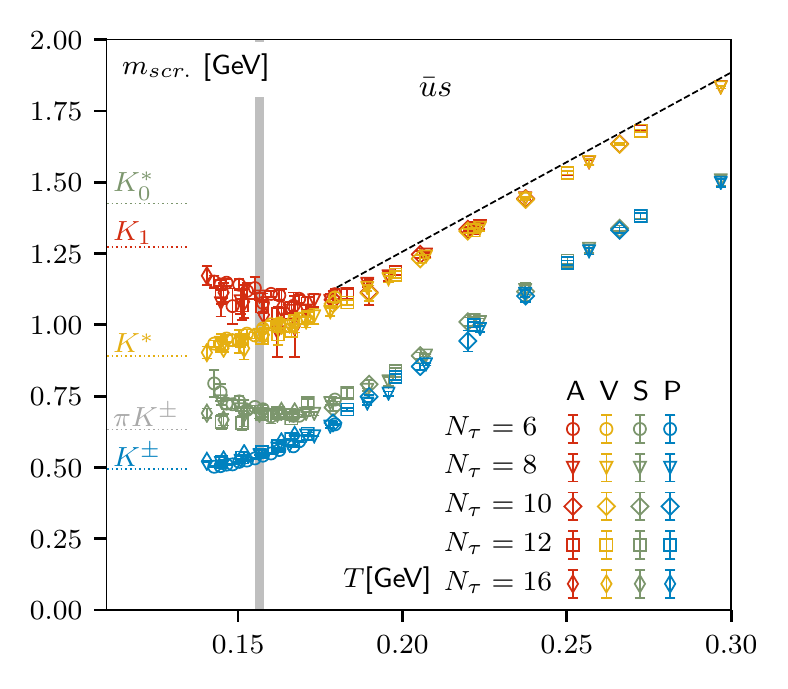}
\includegraphics[width=0.33\textwidth]{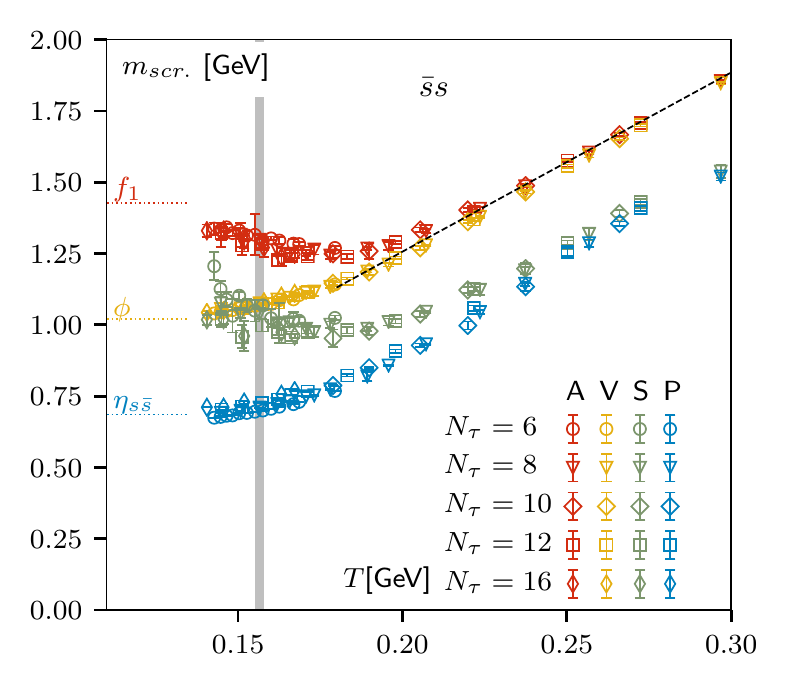}
\caption{(Left to right) Results for all four screening masses for the $\bar u d$, $\bar u s$ and $\bar{s}s$ flavor combinations.
The gray vertical band in all the figures represents the pseudo-critical temperature, $T_{pc} = 156.5(1.5)$~MeV~\cite{Bazavov:2018mes}.
The dashed lines corresponds to the free theory limit of $m = 2\pi T$.
\label{fig:all_results}}
\end{figure*}

We now present our results for screening masses calculated in a range of temperatures going from just below 
the chiral cross-over temperature, $T_{pc} = 156.5(1.5)$~MeV, to about $2T_{pc}$, namely $140~\text{MeV} \le T \le 300~\text{MeV}$.
This temperature range is important both from the phenomenological point of view as well as regarding
the restoration of chiral $SU_A(2)$ and axial $U_A(1)$ symmetries. As already mentioned earlier, our screening
masses were calculated at two values of the light quark mass viz.\ $m_l = m_s/27$ for $T \lesssim 172$~MeV and
$m_l = m_s/20$ for all higher temperatures.
It is worth to mention here that we have also calculated screening masses with $m_l = m_s/20$
for $T \lesssim 172$~MeV but we do not show them here because we have fewer statistics compared
to that for $m_l = m_s/27$.
For higher temperatures, the quark mass dependence is negligible and the heavier quark mass can be used
without affecting any of the conclusions.

\begin{figure}[!tbh]
    \hspace{-0.03\textwidth}%
    \includegraphics[width=0.35\textwidth]{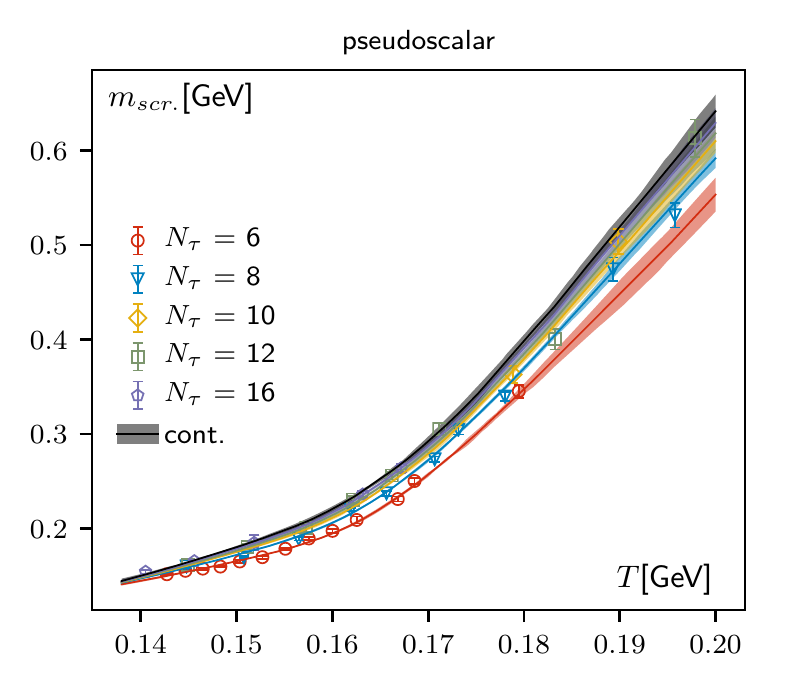}
    \hspace{0.10\textwidth}
    \includegraphics[width=0.35\textwidth]{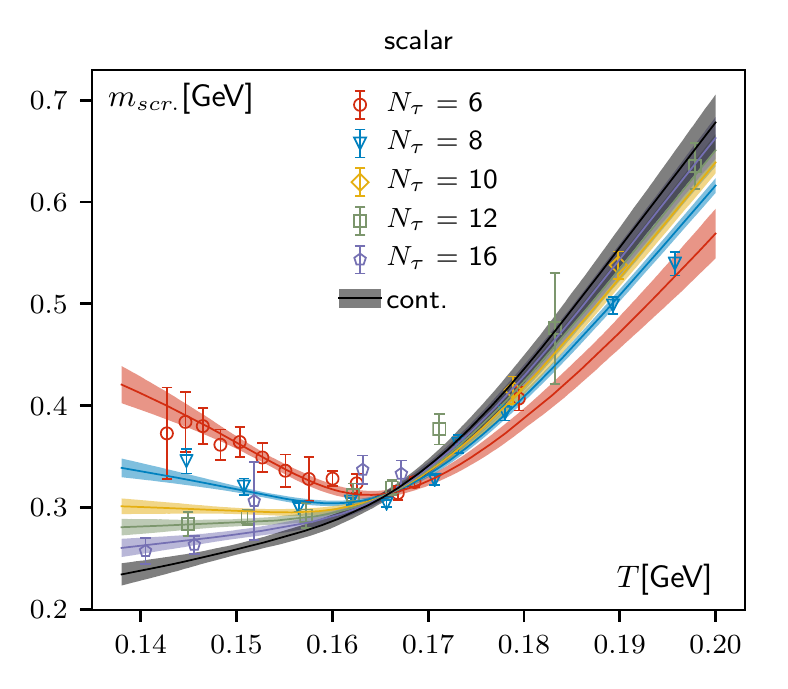}
    \caption{Examples for the continuum extrapolations for the pseudoscalar (top), scalar (bottom) screening masses
             in a reduced temperature range. The data for different $\Nt$ were fitted to an $\Nt$-dependent fit function.
             Also shown in each figure are the bands for each $\Nt$, obtained using the same fit function.}
    \label{fig:continuum_fits}
\end{figure}

Using the fitting procedure described in Sec.\ \ref{sec:calc_setup},
we calculated screening masses for five different values of the
lattice spacings corresponding to $N_{\tau}=6$, 8, 10, 12 and 16,
which allow for a continuum extrapolation.
As the temperatures do not agree among the different lattices,
the screening masses have to be interpolated between the different temperature values.
In our extrapolation method, the interpolation and the extrapolation are performed in one single fit:
For the interpolation we use simple splines. Then, the extrapolation is performed by replacing the
spline coefficients by a function linear in $1/N_\tau^2$ and performing a joint fit, that includes
all the data. The spline's knot positions are placed according to the density of data points.
The knots are positioned in such a way that the same number of data points lies between each pair of subsequent knots. 
This means in particular that more knots are used at the low temperature region, where the interpolation is more
curvy. To stabilize the spline, we use some of its coefficients, to constrain the spline's
derivative with respect to $T$ at some points. These constraints are placed far outside of the actual region where the
extrapolation is performed \cite{Sandmeyer:2019}.
The error band are computed using Gaussian bootstrapping and by performing the extrapolation on each sample.
Final values and errors are calculated using median and 68\%-percentiles of the bootstrap distribution.
In Fig.\ \ref{fig:continuum_fits} we show two examples of continuum extrapolations following the
above-mentioned procedure in the $PS$ and $S$ sector for a limited temperature range.
More technical details of the continuum extrapolations can be found in Ref.\ \cite{Sandmeyer:2019}.
Continuum extrapolated masses of all four channels for all three flavor combinations have
been tabulated in Appendix.\ \ref{app:screen_mass_values}.

\begin{figure*}[!tbh]

\hspace{-0.03\textwidth}%
\includegraphics[width=0.33\textwidth]{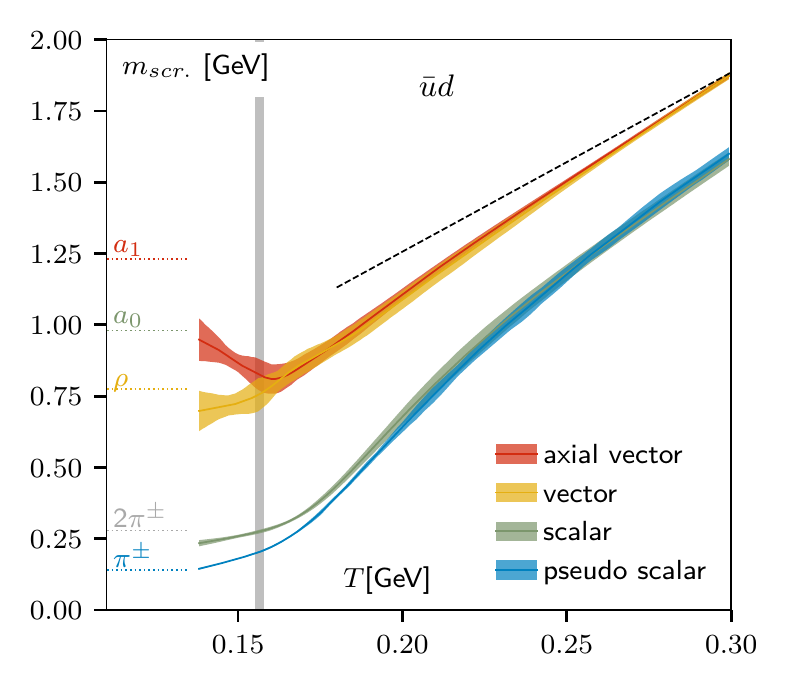}
\includegraphics[width=0.33\textwidth]{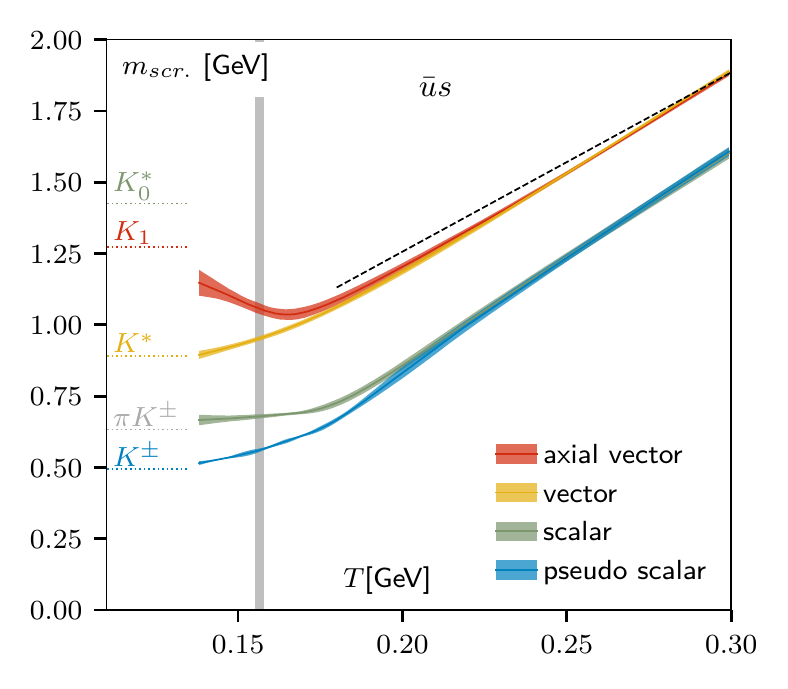}
\includegraphics[width=0.33\textwidth]{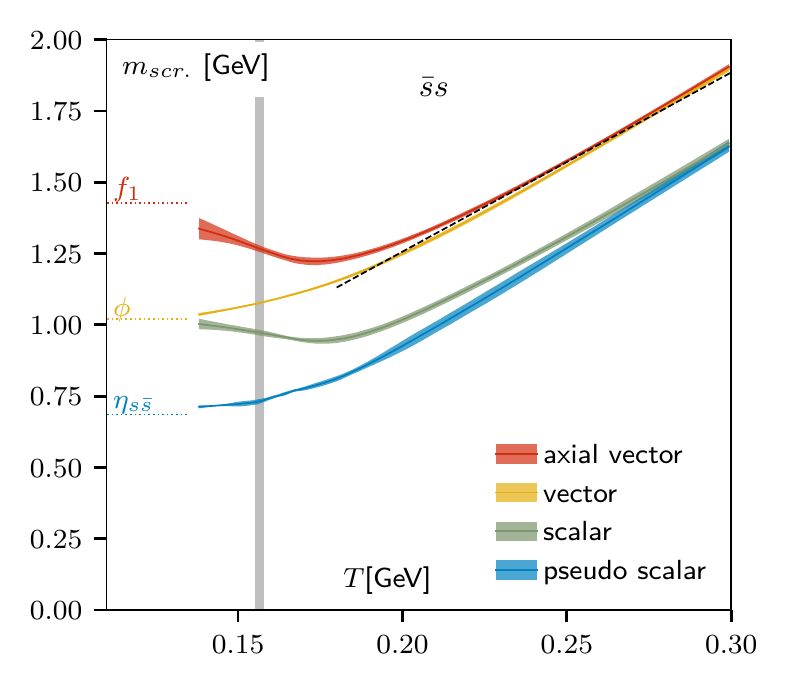}
\caption{Continuum bands for screening masses of all four types of mesons for $\bar u d$, $\bar u s$ and $\bar{s}s$ (left to right).
\label{fig:continuum_bands}}
\end{figure*}

We plot the screening masses for $140~\text{MeV} \le T \le 300~\text{MeV}$, for the different flavor sectors and for all lattice
spacings, in Fig.~\ref{fig:all_results}. The mesons with angular momentum $J=0$ ($S$ and $PS$) were easier to determine, especially for lower
temperatures, as compared to the $J=1$ mesons ($V$ and $AV$). We find some cut-off dependence in the scalar sector,
especially for smaller $\Nt$. For the other sectors, the cut-off dependence was
indistinguishable within the statistical error. We perform the continuum limit for all
the sectors, using data from five different values of the cut-off corresponding to our five different
values of the temporal lattice extent, mentioned earlier. The resulting continuum extrapolated bands
are plotted in Fig.~\ref{fig:continuum_bands}.
In Fig.~\ref{fig:all_results} and Fig.~\ref{fig:continuum_bands} we also show the
pseudo-critical temperature region as a gray vertical band. The massless infinite temperature limit 
$m^{\rm free}_{scr} = 2 \pi T$ is shown as a dashed line in each of the plots.

For $T \ll T_{pc}$ the screening masses are expected to approach the mass of the lightest zero
temperature meson with the same quantum numbers, e.g., the $\bar u d$ pseudoscalar screening mass
should approach the pion mass $m_\pi$. We see that this behavior is readily realized for the $PS$ and $V$ sectors.
Already for $T\lesssim 0.9 T_{pc}$ the corresponding zero temperature masses are approached in the
$\bar u d$, $\bar u s$ and $\bar{s}s$ sectors to better than 10\%.
Although the zero temperature limits are not yet reached in the $AV$ channel at this temperature, we see
clear indications for a rapid approach to the corresponding zero temperature masses for all combinations
of heavy and light quarks. These values are in all cases approached from below, i.e., at the
pseudo-critical temperature the $AV$ screening masses are smaller than the corresponding zero 
temperature masses. In the $\bar s s$ sector the screening mass of the $f_1$-meson is about 15\%
lower than the $f_1$-mass around $T_{pc}$ and reduces to about 7\% already at $T\lesssim 0.9 T_{pc}$.
The situation is similar in the $\bar u s$ sector. However, thermal effects are
substantially larger in the $\bar u d$ sector. Here we find that the screening mass
of the $a_1$ mesons at $T_{pc}$ differs by about 35\% from the corresponding zero temperature
mass and the two masses still differ by about 20\%  at $T\lesssim 0.9 T_{pc}$.
Note that also from our calculations for $m_l = m_s/20$, where we have results
at even lower temperatures, we found that the screening masses go towards
corresponding zero-temperature masses steeply. Similar behavior was also found in
calculations with staggered fermions utilizing the p4 discretization scheme \cite{Cheng:2010fe}.

The situation is far more complicated in the $S$ sector for finite lattice spacings. In nature, the lightest
flavored scalar meson is either the $a_0(980)$ or the $a_0(1450)$. Rather than either of these values, as can
be seen from the left panel of Fig.~\ref{fig:all_results} and Fig.~\ref{fig:continuum_bands}, the scalar
screening mass approaches the value $2m_\pi$ instead. The reason for this is that for staggered fermions,
the scalar can decay into two pions at finite lattice spacing~\cite{Prelovsek:2005rf}. This decay is forbidden
in nature due to parity, isospin and $G$-parity ($I^G$) conservation. The unphysical behavior in the staggered
discretization comes from the contribution of the different tastes in the intermediate states of loop diagrams.
If one takes the continuum limit for the correlator before calculating the screening mass, then the contribution
from different tastes cancels out and the physical behavior is recovered~\cite{Prelovsek:2005rf,Prelovsek:2004jp,Bernard:2007qf}.
Since we, however, calculate the screening masses first and then take the continuum limit, we obtain the unphysical
$\pi\pi$ state rather than the true scalar ground state or the physically allowed $\pi\eta$ decay. The unphysical
decay only occurs for mesons with isospin $I=1$. For the $\bar u s$ case ($I=1/2$), the decay to $K\pi$ actually 
occurs in nature. In Figs.~\ref{fig:all_results} and~\ref{fig:continuum_bands}, we see that the scalar screening
mass indeed tends to $m_\pi + m_K$ as $T \to 0$.

As the cross-over temperature is approached, the vector and axial vector screening masses should become equal
due to effective restoration of chiral symmetry. At $T = 0$, the axial vector meson $a_1$ is about 500 MeV
heavier than the vector meson $\rho$. As the temperature is increased, the $AV$ screening mass decreases
while the $V$ mass increases slightly until the two masses become degenerate right at the pseudo-critical
temperature (left panel of Fig.~\ref{fig:continuum_bands}). In contrast, in the $\bar u s$ and $\bar s s$
sectors, $AV$ and $V$ masses become equal at higher temperatures compared to $T_{pc}$. Moreover, the
relative change of $AV$ masses w.r.t.\ $V$ masses from low temperature towards degeneracy temperature progressively
decreases when one goes from $\bar u d$ to $\bar s s$. It must be noted that the approach is nevertheless
smoother compared to previous results that were obtained using the p4 discretization scheme for 
staggered fermions~\cite{Cheng:2010fe}.
cross-over temperature, as noted from Fig.~\ref{fig:continuum_bands}, is quite
similar to what has been seen in the calculation of nucleon masses, where the mass
of one particular parity (the one with higher zero-temperature mass) of nucleon changes
a lot and comes close to its parity partner, which on the contrary, hardly changes from
low temperature towards chiral cross-over temperature~\cite{Aarts:2017iai,Aarts:2018glk,Datta:2012fz}.

In Fig.~\ref{fig:continuum_bands}, we also see that the scalar and pseudoscalar screening masses in the $\bar u d$ sector
become degenerate around $T \sim 200$~MeV. Unfortunately, one cannot immediately draw any conclusions about an effective
$U_A(1)$ restoration from this due to the pathology of the $\bar u d$ scalar correlator that we have discussed above.
However, as we have already mentioned, the unphysical contribution cancels out if one would take the continuum limit for
the correlator first. Moreover, as the pion screening mass increases around the cross-over region while the
continuum scalar screening mass is expected to decrease around $T_{pc}$ before rising again at higher temperatures,
this unphysical decay channel might be closed around $T_{pc}$ due to lack of phase space.
Therefore the degeneracy of the screening masses in the $S$ and $PS$ channel around $T \sim 200$~MeV
is an indication towards an effective restoration of the $U_A(1)$.

Despite the above argument, we may nevertheless try and estimate the effective $U_A(1)$ restoration temperature directly 
from the correlators. Although it is difficult to calculate the continuum limit of staggered correlators due to
their oscillating behavior, one may instead consider the corresponding susceptibility, which is given
by the integrated correlator, and calculate its continuum limit instead. The staggered $\pi$ and $\delta$
susceptibilities are defined as
\begin{align} && 
\chi_\pi = \sum_{n_\sigma=0}^{\Ns-1} G_{\cM2}(n_\sigma), && 
\chi_{a_0} = - \sum_{n_\sigma=0}^{\Ns-1} (-1)^{n_\sigma} G_{\cM1}(n_\sigma). && 
\label{eq:chip_chid_def}
\end{align}

\begin{figure}[!tb]
\includegraphics[width=0.37\textwidth]{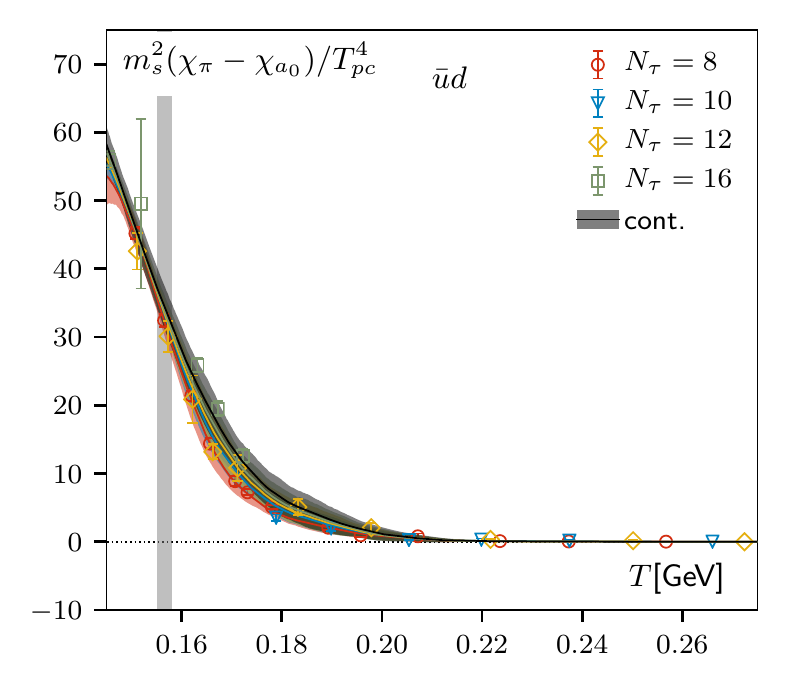}
\caption{Difference between the pseudoscalar and scalar susceptibilities as a function of the temperature.
The difference is multiplied by $m_s^2$ to renormalize and normalized to $1/T_{pc}^4$.
The continuum extrapolation is also shown in the Figure as a superimposed band.}
\label{fig:susc_diff_conex}
\end{figure}

We plot our results, along with the continuum extrapolations, for the difference of the scalar and pseudoscalar
susceptibilities for the $\bar u d$ sector in Fig.~\ref{fig:susc_diff_conex}. In order to be able to take
the continuum limit, we have renormalized the quantity with $m_s^2$. We have also normalized these numbers
to $T_{pc}^4$. For reference, we also show the pseudo-critical temperature region by a gray band in the figure. 
For the $\bar u d$ sector we see that the difference is non-zero around the pseudo-critical temperature and 
only goes to zero for $T \sim 200$~MeV. There are some theoretical arguments in favor of effective $U_A(1)$ 
restoration  at the chiral phase transition~\cite{Aoki:2012yj} in the chiral limit. 
On the other hand lattice calculations, performed away from the chiral limit, have found evidence
in favor of  this scenario~\cite{Tomiya:2016jwr,Suzuki:2017ifu,Buchoff:2013nra,Dick:2015twa}.

Before moving on, we note that the behavior of the screening masses and susceptibilities in the 
$\bar u s$ and $\bar{s}s$ sectors is qualitatively the same although the degeneracies discussed
above occur at progressively higher temperatures \cite{Sandmeyer:2019}. This mass ordering of degeneracy
temperatures is in complete accordance with what has been observed for even heavier sectors~\cite{Bazavov:2014cta},
although one has to keep in mind that the mass effects in the susceptibilities for heavier sectors are expected
to be much larger than the $U_A(1)$ breaking effects due to quantum fluctuations.

\subsection{Screening masses at high temperatures}
\label{ssec:high_temperature_results}
\begin{figure*}[!tbh]
\hspace{-0.03\textwidth}%
\includegraphics[width=0.33\textwidth]{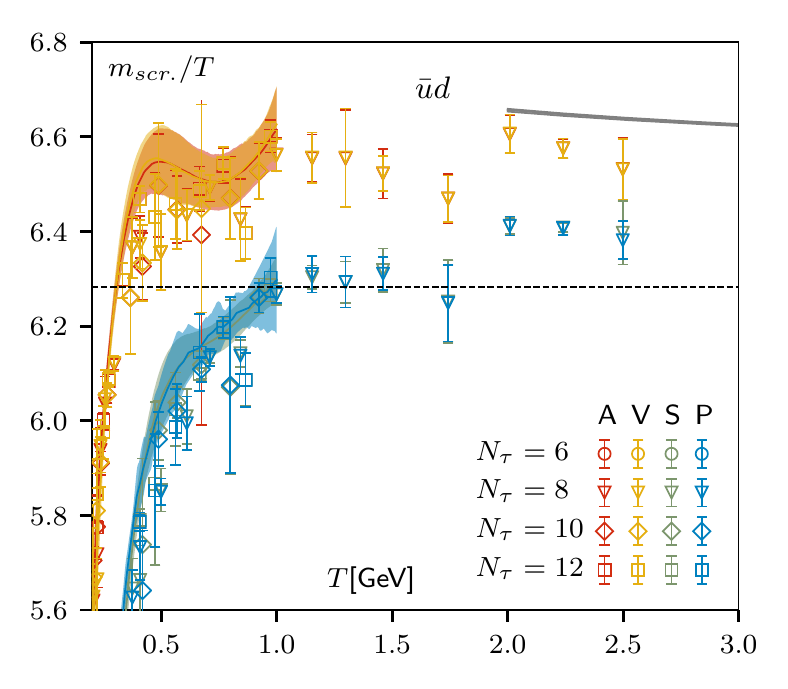}
\includegraphics[width=0.33\textwidth]{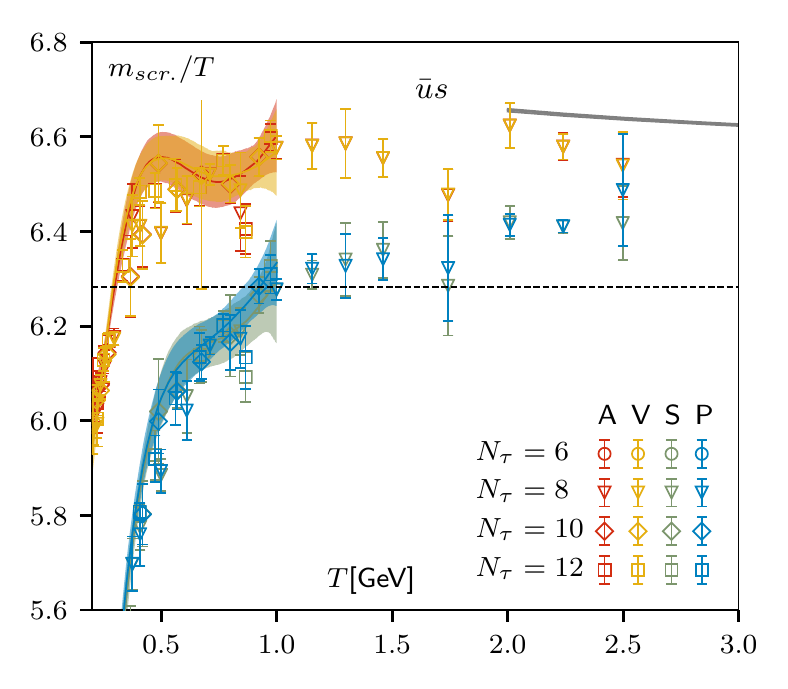}
\includegraphics[width=0.33\textwidth]{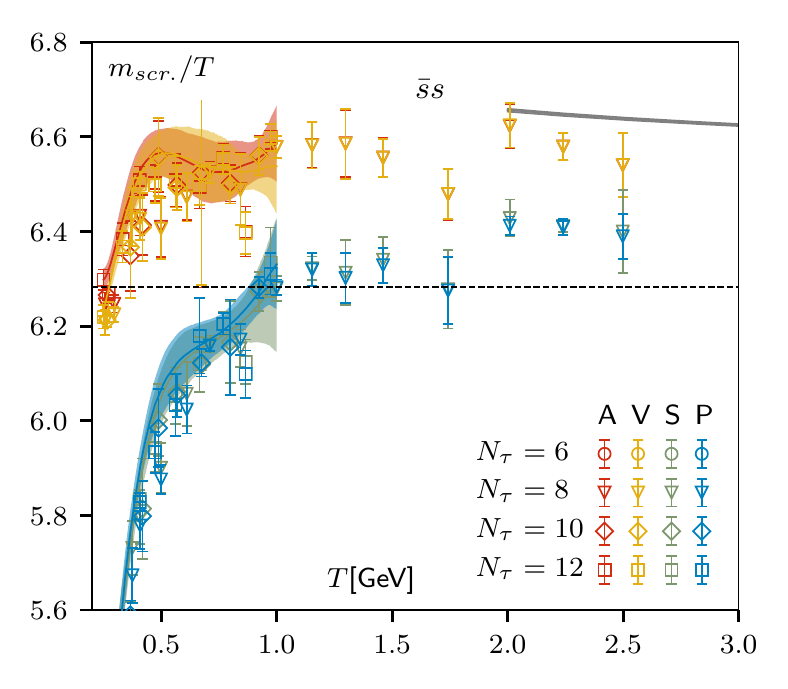}
\caption{Screening masses divided by the temperature, for temperatures $T \gtrsim 200$~MeV and for $\Nt = 6$, 8, 
10 and 12 for all four channels in different flavor sectors. The curves at the top right part of each figure depict 
the resummed perturbation theory predictions. Beyond $T \gtrsim 1$~GeV, only $\Nt = 8$
data exist as a result of which, a continuum extrapolation is not
possible. The lattice results are obtained with corner wall source for $V$ and $AV$ channels for $T \lesssim 300$~MeV
and with point source elsewhere.\label{fig:high_T}}
\end{figure*}

In the previous subsection we have seen that the temperature dependence of the screening masses at $T>250$ MeV
qualitatively follows the free theory expectations, namely the screening masses are proportional to the temperature,
with proportionality constant not very different from $2 \pi$. Furthermore, the AV and V screening masses are close
to the free theory expectations, while the $PS$ and $S$ screening masses are 10-20\% smaller. In this subsection
we will study the screening masses at higher temperature with the aim to see how the degeneracy of $PS(S)$ and
$AV(V)$ screening masses expected in the infinite temperature limit sets it.
We would like to see if contacts to the weak coupling calculations can be made at high temperatures.

Although attempts have been made
\cite{DeTar:1987ar, DeTar:1987xb, Born:1991zz, Banerjee:2011yd, Laermann:2012sr, Datta:2012fz, Gupta:2013vha,Brandt:2014uda}
to compare screening masses from lattice QCD to those from weak coupling calculations, it is not clear in
which temperature range weak coupling results can be reliable. For this reason it is important to perform
lattice calculations at as high temperatures as possible. 
Therefore, we extended the calculations of the meson screening masses to $T=1$ GeV using four lattice spacings
corresponding to $N_{\tau}=6,~8,~10$ and $12$, and performed the continuum extrapolations. The results are shown
in Fig.~\ref{fig:high_T}. We find that the lattice spacing dependence is very small
for $T>300$ MeV, and within errors the $N_{\tau}=8$ results agree with the continuum extrapolated values.
Therefore, for $1~ {\rm GeV} < T < 2.5$ GeV we calculated the screening masses using only $N_{\tau}=8$ lattices.
The results of these calculations are also shown in \ref{fig:high_T}. We clearly see from
the figure that the AV and V screening masses overshoot the free theory value around $T=400$ MeV and are almost
constant in temperature units. The $PS$ and $S$ screening masses overshoot the free theory expectation only at
temperature larger than $1$ GeV and remain smaller than the $AV$ and $V$ screening masses up to the highest
temperature considered. 

The behavior of the screening masses in the weak coupling picture beyond the free theory limit can be understood in 
terms of dimensionally reduced effective field theory, called electro-static QCD (EQCD) \cite{Braaten:1995jr}.
This approach turned out to be useful for understanding the lattice on the quark number susceptibilities
\cite{Bazavov:2013uja,Ding:2015fca},
the expectation value of Polyakov loop \cite{Bazavov:2016uvm} and the Polyakov loop correlators \cite{Bazavov:2018wmo}.
It is interesting to see if deviation of the screening masses at high temperature from $2 \pi T$ can be understood
within this framework. 

In EQCD the correction to the free theory value for the screening masses is obtained by solving the Schr\"odinger
equation in two spatial dimensions with appropriately defined potential \cite{Koch:1992nx,Shuryak:1993kg,Laine:2003bd}. 
At leading order the potential is proportional to the coupling constant of EQCD, $g_E^2$ \cite{Laine:2003bd},
which in turn can be expressed in terms of the QCD coupling constant $g^2=4 \pi \alpha_s$. At leading
order $g_E^2=g^2 T$, and $g_E^2$ has been calculated to 2-loops \cite{Laine:2005ai}. 
Moreover, at leading order the potential and the correction
to the free theory value are independent of the spin, i.e. the $PS(S)$ and $AV(V)$ screening masses receive the same
correction, that has been calculated in Ref. \cite{Laine:2003bd}. This correction is positive in
qualitative agreement with our lattice results.
In Fig. \ref{fig:high_T} we show the corresponding weak coupling result from EQCD. 
We used the 2-loop result for $g_E^2$ and the optimal choice for the renormalization scale
$\mu/T = 9.08$ \cite{Laine:2005ai}. We varied the scale $\mu$ by factor of two around this optimal value to
estimate the perturbative uncertainty, which turned out to be very small (the uncertainty corresponds
to the width of the weak coupling curve in Fig. \ref{fig:high_T}. We see that the
weak coupling results from EQCD are slightly larger than the $AV(V)$ screening masses and significantly
larger the the lattice results for $PS(S)$ screening masses. This is not completely
surprising because the EQCD coupling constant $g_E^2$ is not small except for very
high temperatures and thus higher order corrections may be important. Beyond $\mathcal{O}(g_E^2)$
the correction will be spin dependent \cite{Koch:1992nx,Shuryak:1993kg}.  
Since the coupling constant decreases logarithmically the screening masses will approach
$2 \pi T$ only for temperatures many orders of magnitude larger than those considered here,
when the $AV(V)$ and $PS(S)$ screening masses become degenerate. It would be interesting
to calculate the $\mathcal{O}(g_E^4)$ correction to meson screening masses and see whether EQCD predictions 
work quantitatively.

\section{Conclusions}
\label{sec:conclusions}
We have performed an in-depth analysis of mesonic screening masses
in (2+1)-flavor QCD with physical (degenerate) light and strange
quark masses. In the vicinity of the pseudo-critical temperature
for chiral symmetry restoration, $T_{pc}$ and up to about 1 GeV
we could perform controlled continuum extrapolations, using input
from five different values of the lattice cut-off. Comparing 
screening masses for chiral partners, related through the chiral
$SU_L(2)\times SU_R(2)$ and the axial $U_A(1)$ transformations,
respectively, we find in the case of light-light mesons evidence
for the degeneracy of screening masses related through the chiral
$SU_L(2)\times SU_R(2)$ at or very close to $T_{pc}$ while screening
masses related through an axial $U_A(1)$ transformation start becoming
degenerate only at about $1.3 T_{pc}$. In particular, the $V$ and $AV$ mesons ($J=1$), which are related
by chiral $SU_L(2)\times SU_R(2)$ transformations, become degenerate at $T\simeq T_{pc}$, while the $S$ and the
$PS$ ($J=0$) mesons, which are related by axial $U_A(1)$ transformations, only become degenerate around $1.3T_{pc}$.
The onset of these degeneracies also occurs in the light-strange and strange-strange meson sectors, 
but at higher temperatures.

At high temperatures the screening masses overshoot the free theory expectations
in qualitative agreement with the weak coupling calculations at $\mathcal{O}(g^2_E)$.
While mesonic screening masses in given angular momentum ($J$) channels
become degenerate, screening masses in
channels with different $J$, {\it e.g.}\ $J=0$ and $J=1$, stay well separated
even up to the highest temperature, $T=2.5$~GeV, that was analyzed by us.
We argued that it is necessary to go beyond
$\mathcal{O}(g^2_E)$ calculations in order to understand this feature within the EQCD framework.
This non-degeneracy has also been observed in Ref.~\cite{Rohrhofer:2019qwq}, 
where it was also shown that these two sets of mesons only become degenerate at asymptotically high temperatures.
This conclusion is in agreement with the results that we have presented in this paper in 
Sec.~\ref{ssec:high_temperature_results} (Fig.~\ref{fig:high_T}).

\acknowledgements
This material is based upon work supported by the U.S. Department of Energy, Office
of Science, Office of Nuclear Physics:
(i) Through the Contract No. DE-SC0012704;
(ii) Within the framework of the Beam Energy Scan Theory (BEST) Topical Collaboration;
(iii) Through the Scientific Discovery through Advanced Computing
(SciDAC) award Computing the Properties of Matter with Leadership Computing
Resources.

This research also was funded by--- 
(i) the Deutsche Forschungsgemeinschaft
(DFG, German Research Foundation) through the CRC-TR 211 \textquoteleft Strong-interaction
matter under extreme conditions\textquoteright\ – project number 315477589 – TRR 211;
(ii) Grant 05P18PBCA1 of the German Bundesministerium f\"ur Bildung und Forschung;
(iii) The grant 283286 of the European Union;
(iv) The U.S. National Science Foundation under award PHY-1812332,
(v) The Early Career Research Award of the Science and Engineering Research Board of the Government of India;
(vi) Ramanujan Fellowship of the Department of Science and Technology, Government of India,
(vii) The National Natural Science Foundation of China under grant numbers 11775096 and 11535012.

This research used awards of computer time: 
(i) Provided by the INCITE and ALCC programs Oak Ridge Leadership Computing Facility,
a DOE Office of Science User Facility operated under Contract No. DE-AC05-00OR22725;
(ii) Provided by the ALCC program at National Energy Research Scientific Computing Center, 
a U.S. Department of Energy Office of Science User Facility operated under Contract No. DE-AC02-05CH11231;
(iii) Provided by the INCITE program at Argonne Leadership Computing Facility, 
a U.S. Department of Energy Office of Science User Facility operated under Contract No. DE-AC02-06CH11357.
(iv) Provided by the USQCD consortium at its Jefferson Laboratory and Fermilab computing facilities.
This research also used computing resources made available through: 
(i) The GPU supercomputing cluster of Bielefeld University,
(ii) Piz Daint at CSCS, Switzerland, and Marconi at CINECA, Italy through PRACE grant,
(iii) JUWELS at NIC-J\"ulich, Germany.

\appendix

\section{Parametrization of \boldmath{$f_Ka(\beta)$} for scale setting}
\label{app:fKparametrization}

\begin{figure}[!bth]
 \hspace{-0.03\textwidth}%
 \includegraphics[width=0.45\textwidth]{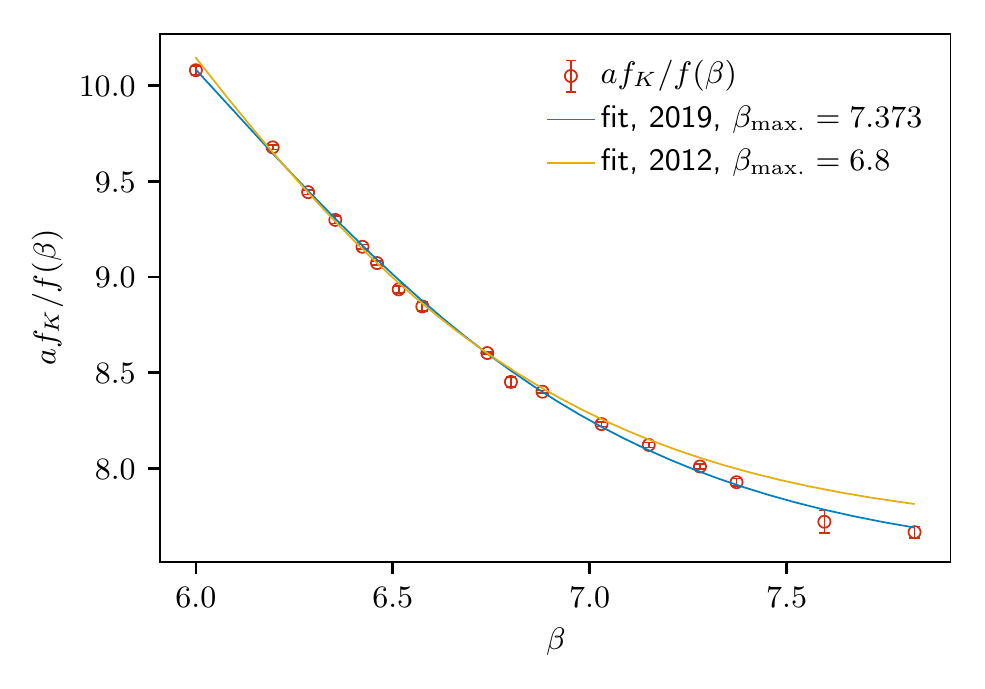}
 \caption{Comparison of updated $f_Ka(\beta)$ parametrization and the older one
 from Ref.~\cite{Bazavov:2011nk}.}
 \label{fig:fKparametrization}
\end{figure}

For the scale setting in this project we used the Kaon decay constant, {\it i.e.}\ $f_Ka(\beta)$.
Including the measurements up to $\beta=7.373$, listed in Ref.~\cite{Bazavov:2014pvz},
we have updated the parametrization used in Ref.~\cite{Bazavov:2011nk} :
\begin{equation}
 f_Ka(\beta)=\frac{c_0f(\beta)+c_2(10/\beta)f^3(\beta)}{1+d_2(10/\beta)f^2(\beta)}
 \label{eq:fKparametrization}
\end{equation}
where
\begin{equation*}
 f(\beta)=\left( \frac{10 b_0}{\beta} \right)^{-b_1/(2 b_0^2)} \exp(-\beta/(20 b_0))
\end{equation*}
with $b_0$ and $b_1$ being the coefficients of the two-loop beta function.
For the three-flavor case : $b_0=9/(16 \pi^2)$, $b_1=1/(4 \pi^4)$.
The updated fit renders the following parameters for the form described in
Eq.~(\ref{eq:fKparametrization}) :
$c_0 = 7.49415$, $c_2 = 46049(1248)$ and $d_2 = 3671(137)$.
We have not included the $f_Ka(\beta)$ measurements for two highest
$\beta$ values, shown in Fig.~\ref{fig:fKparametrization} because
of possible large finite volume effects.

In Fig.~\ref{fig:fKparametrization} we have compared the fit described
with Eq.~(\ref{eq:fKparametrization}) to the same from Ref.~\cite{Bazavov:2011nk}.
It can be seen from the plot that one overestimates $f_Ka(\beta)$ with the old
parametrization for $\beta \gtrsim 6.9$ by $\sim$ 1\%.
One can look in Ref.~\cite{Bazavov:2011nk,Bazavov:2014pvz}
for more details on this kind of parametrization.

\section{Summary of statistics for $\bold{m_l = m_s/20}$ and $\bold{m_l = m_s/27}$}
\label{app:statistics}
Here we summarize our data sets and the number of configurations on which point and wall source
correlators have been calculated are given in the last two columns of the tables which are labled
\textquoteleft point\textquoteright and \textquoteleft wall\textquoteright, respectively.

\begin{table}[!h]
\centering
\setlength\tabcolsep{4pt}
\begin{tabular}{|c|c|c|c|c|c|}
\hline
$\beta$ & $T$[MeV]& $m_l$ & $m_s$ & point & wall \\
\hline
5.850 & 119.19 & 0.00712 & 0.1424 & 1166 & 1166 \\
5.900 & 125.45 & 0.00660 & 0.1320 & 1000 & 1000 \\
5.950 & 132.07 & 0.00615 & 0.1230 & 1000 & 1000 \\
6.000 & 139.08 & 0.00569 & 0.1138 & 3073 & 3073 \\
6.025 & 142.73 & 0.00550 & 0.1100 & 1000 & 1000 \\
6.050 & 146.48 & 0.00532 & 0.1064 & 1000 & 1000 \\
6.062 & 148.32 & 0.005235 & 0.1047 & 1000 & 1000 \\
6.075 & 150.33 & 0.00518 & 0.1036 & 1000 & 1000 \\
6.090 & 152.70 & 0.00504 & 0.1008 & 1001 & 1001 \\
6.100 & 154.29 & 0.00499 & 0.0998 & 3363 & 3363 \\
6.120 & 157.54 & 0.004845 & 0.0969 & 1001 & 1001 \\
6.125 & 158.36 & 0.00483 & 0.0966 & 1003 & 1003 \\
6.150 & 162.54 & 0.00468 & 0.0936 & 1000 & 1000 \\
6.165 & 165.10 & 0.00457 & 0.0914 & 1000 & 1000 \\
6.185 & 168.58 & 0.004455 & 0.0891 & 1000 & 1000 \\
6.195 & 170.35 & 0.00440 & 0.0880 & 1000 & 1000 \\
6.245 & 179.46 & 0.00415 & 0.0830 & 1000 & 1000 \\
\hline
\end{tabular}

\caption{Summary of statistics for $m_l = m_s/20$, $24^3\times6$ lattices.\label{tab:mso20nt6}}
\end{table}

\begin{table}[!h]
\centering
\setlength\tabcolsep{4pt}
\begin{tabular}{|c|c|c|c|c|c|}
\hline
$\beta$ & $T$[MeV]& $m_l$ & $m_s$ & point & wall \\
\hline
6.050 & 109.86 & 0.00532 & 0.1064 & 2108 & 2108 \\
6.125 & 118.77 & 0.00483 & 0.0966 & 2241 & 2241 \\
6.195 & 127.76 & 0.00440 & 0.0880 & 1690 & 1690 \\
6.245 & 134.60 & 0.00415 & 0.0830 & 2710 & 2710 \\
6.285 & 140.32 & 0.00395 & 0.0790 & 2000 & 2000 \\
6.341 & 148.74 & 0.00370 & 0.0740 & 1713 & 1713 \\
6.354 & 150.76 & 0.00364 & 0.0728 & 1249 & 1249 \\
6.390 & 156.50 & 0.00347 & 0.0694 & 2604 & 2604 \\
6.423 & 161.93 & 0.00335 & 0.0670 & 2031 & 2031 \\
6.460 & 168.24 & 0.00320 & 0.0640 & 1644 & 1644 \\
6.488 & 173.16 & 0.00310 & 0.0620 & 1790 & 1790 \\
6.515 & 178.03 & 0.00302 & 0.0604 & 3067 & 3067 \\
6.575 & 189.29 & 0.00282 & 0.0564 & 3206 & 3206 \\
6.608 & 195.75 & 0.00271 & 0.0542 & 2379 & 2379 \\
6.664 & 207.17 & 0.00257 & 0.0514 & 2001 & 2001 \\
6.740 & 223.58 & 0.00238 & 0.0476 & 831 & 831 \\
6.800 & 237.32 & 0.00224 & 0.0448 & 500 & 500 \\
6.880 & 256.75 & 0.00206 & 0.0412 & 500 & 500 \\
7.030 & 296.81 & 0.00178 & 0.0356 & 500 & 500 \\
7.280 & 375.26 & 0.00142 & 0.0284 & 500 & 500 \\
7.373 & 408.63 & 0.00125 & 0.0250 & 500 & 500 \\
7.596 & 499.30 & 0.00101 & 0.0202 & 500 & 500 \\
7.825 & 610.60 & 0.00082 & 0.0164 & 500 & 500 \\
8.000 & 710.45 & 0.00070 & 0.0140 & 500 & 500 \\
8.200 & 843.20 & 0.0005835 & 0.0116 & 250 & 250 \\
8.400 & 999.39 & 0.0004875 & 0.00975 & 250 & 250 \\
8.570 & 1153.83 & 0.0004188 & 0.008376 & 200 & 200 \\
8.710 & 1298.31 & 0.0003697 & 0.007394 & 200 & 200 \\
8.850 & 1460.54 & 0.0003264 & 0.006528 & 200 & 200 \\
9.060 & 1742.17 & 0.0002417 & 0.004834 & 200 & 0 \\
9.230 & 2009.14 & 0.0002074 & 0.004148 & 200 & 200 \\
9.360 & 2240.48 & 0.00018455 & 0.003691 & 200 & 200 \\
9.490 & 2498.41 & 0.00016425 & 0.003285 & 200 & 200 \\
9.670 & 2905.28 & 0.00013990 & 0.002798 & 0 & 200 \\
\hline
\end{tabular}

\caption{Summary of statistics for $m_l = m_s/20$, $32^3\times8$ lattices.\label{tab:mso20nt8}}
\end{table}

\begin{table}[!h]
\centering
\setlength\tabcolsep{4pt}
\begin{tabular}{|c|c|c|c|c|c|}
\hline
$\beta$ & $T$[MeV]& $m_l$ & $m_s$ & point & wall \\
\hline
6.488 & 138.53 & 0.00310 & 0.0620 & 9534 & 9534 \\
6.515 & 142.42 & 0.00302 & 0.0604 & 2525 & 2525 \\
6.575 & 151.43 & 0.00282 & 0.0564 & 2512 & 2512 \\
6.608 & 156.60 & 0.00271 & 0.0542 & 2685 & 2685 \\
6.664 & 165.73 & 0.00257 & 0.0514 & 1071 & 1071 \\
6.740 & 178.86 & 0.00238 & 0.0476 & 1021 & 1021 \\
6.800 & 189.85 & 0.00224 & 0.0448 & 800 & 800 \\
6.880 & 205.40 & 0.00206 & 0.0412 & 650 & 650 \\
6.950 & 219.87 & 0.00193 & 0.0386 & 500 & 500 \\
7.030 & 237.45 & 0.00178 & 0.0356 & 600 & 600 \\
7.150 & 266.03 & 0.00160 & 0.0320 & 500 & 500 \\
7.500 & 366.65 & 0.00111 & 0.0222 & 450 & 450 \\
7.650 & 419.00 & 0.00096 & 0.0192 & 250 & 250 \\
7.825 & 488.48 & 0.00082 & 0.016 & 250 & 250 \\
8.000 & 568.36 & 0.00070 & 0.0140 & 500 & 500 \\
8.200 & 674.56 & 0.0005835 & 0.0116 & 551 & 551 \\
8.400 & 799.51 & 0.0004875 & 0.00975 & 300 & 300 \\
8.570 & 923.07 & 0.0004188 & 0.008376 & 250 & 250 \\
\hline
\end{tabular}

\caption{Summary of statistics for $m_l = m_s/20$, $40^3\times10$ lattices.\label{tab:mso20nt10}}
\end{table}

\begin{table}[!h]
\centering
\setlength\tabcolsep{4pt}
\begin{tabular}{|c|c|c|c|c|c|}
\hline
$\beta$ & $T$[MeV]& $m_l$ & $m_s$ & point & wall \\
\hline
6.664 & 138.11 & 0.00257 & 0.0514 & 372 & 372 \\
6.700 & 143.20 & 0.00248 & 0.0496 & 649 & 649 \\
6.740 & 149.05 & 0.00238 & 0.0476 & 2214 & 2214 \\
6.800 & 158.21 & 0.00224 & 0.0448 & 2008 & 2008 \\
6.880 & 171.17 & 0.00206 & 0.0412 & 2001 & 2001 \\
6.950 & 183.22 & 0.00193 & 0.0386 & 1300 & 1300 \\
7.030 & 197.87 & 0.00178 & 0.0356 & 1000 & 1000 \\
7.150 & 221.69 & 0.00160 & 0.0320 & 730 & 730 \\
7.280 & 250.18 & 0.00142 & 0.0284 & 800 & 800 \\
7.373 & 272.42 & 0.00125 & 0.0250 & 800 & 800 \\
7.596 & 332.87 & 0.00101 & 0.0202 & 800 & 800 \\
7.825 & 407.06 & 0.00082 & 0.0164 & 900 & 900 \\
8.000 & 473.63 & 0.00070 & 0.0140 & 310 & 310 \\
8.200 & 562.13 & 0.0005835 & 0.0116 & 500 & 500 \\
8.400 & 666.26 & 0.0004875 & 0.00975 & 500 & 500 \\
8.570 & 769.22 & 0.0004188 & 0.008376 & 250 & 250 \\
8.710 & 865.54 & 0.0003697 & 0.007394 & 250 & 250 \\
8.850 & 973.70 & 0.0003264 & 0.006528 & 250 & 250 \\
\hline
\end{tabular}

\caption{Summary of statistics for $m_l = m_s/20$, $48^3\times12$ lattices.\label{tab:mso20nt12}}
\end{table}

\begin{table}[!h]
\centering
\setlength\tabcolsep{4pt}
\begin{tabular}{|c|c|c|c|c|c|}
\hline
$\beta$ & $T$[MeV]& $m_l$ & $m_s$ & point & wall \\
\hline
6.025 & 142.73 & 0.004074 & 0.1100 & 990 & 990 \\
6.038 & 144.66 & 0.004 & 0.1082 & 1581 & 1581 \\
6.050 & 146.48 & 0.003941 & 0.1064 & 1649 & 1649 \\
6.062 & 148.32 & 0.003878 & 0.1047 & 1650 & 1650 \\
6.075 & 150.33 & 0.003837 & 0.1036 & 1393 & 1749 \\
6.090 & 152.70 & 0.003733 & 0.1008 & 1386 & 1386 \\
6.105 & 155.10 & 0.003659 & 0.0988 & 1749 & 1749 \\
6.120 & 157.54 & 0.003589 & 0.0969 & 1649 & 1649 \\
6.135 & 160.02 & 0.003519 & 0.0950 & 1749 & 1749 \\
6.150 & 162.54 & 0.003467 & 0.0936 & 990 & 990 \\
6.175 & 166.83 & 0.003356 & 0.0906 & 1472 & 1472 \\
6.185 & 168.58 & 0.0033 & 0.0891 & 1475 & 1550 \\
\hline
\end{tabular}

\caption{Summary of statistics for $m_l = m_s/27$, $24^3\times6$ lattices.\label{tab:mso27nt6}}
\end{table}

\begin{table}[!h]
\centering
\setlength\tabcolsep{4pt}
\begin{tabular}{|c|c|c|c|c|c|}
\hline
$\beta$ & $T$[MeV]& $m_l$ & $m_s$ & point & wall \\
\hline
6.315 & 144.77 & 0.00281 & 0.0759 & 1115 & 1115 \\
6.354 & 150.76 & 0.00270 & 0.0728 & 3731 & 3731 \\
6.390 & 156.50 & 0.00257 & 0.0694 & 3514 & 3514 \\
6.423 & 161.93 & 0.00248 & 0.0670 & 3250 & 3250 \\
6.445 & 165.66 & 0.00241 & 0.0652 & 1912 & 2373 \\
6.474 & 170.68 & 0.00234 & 0.0632 & 1937 & 2425 \\
\hline
\end{tabular}

\caption{Summary of statistics for $m_l = m_s/27$, $32^3\times8$ lattices.\label{tab:mso27nt8}}
\end{table}

\begin{table}[!h]
\centering
\setlength\tabcolsep{4pt}
\begin{tabular}{|c|c|c|c|c|c|}
\hline
$\beta$ & $T$[MeV]& $m_l$ & $m_s$ & point & wall \\
\hline
6.712 & 144.94 & 0.00181 & 0.0490 & 1955 & 1955 \\
6.754 & 151.15 & 0.00173 & 0.0468 & 1484 & 1484 \\
6.794 & 157.28 & 0.00167 & 0.0450 & 1407 & 1407 \\
6.825 & 162.17 & 0.00161 & 0.0436 & 1946 & 1946 \\
6.850 & 166.21 & 0.00157 & 0.0424 & 2081 & 2081 \\
6.880 & 171.17 & 0.00153 & 0.0412 & 1960 & 1960 \\
\hline
\end{tabular}

\caption{Summary of statistics for $m_l = m_s/27$, $48^3\times12$ lattices.\label{tab:mso27nt12}}
\end{table}

\begin{table}[!h]
\centering
\setlength\tabcolsep{4pt}
\begin{tabular}{|c|c|c|c|c|c|}
\hline
$\beta$ & $T$[MeV]& $m_l$ & $m_s$ & point & wall \\
\hline
6.973 & 140.50 & 0.00139 & 0.0376 & 4817 & 2757 \\
7.010 & 145.59 & 0.00132 & 0.0357 & 5919 & 6168 \\
7.054 & 151.84 & 0.00129 & 0.0348 & 123 & 622 \\
7.095 & 157.87 & 0.00124 & 0.0334 & 0 & 308 \\
7.130 & 163.17 & 0.00119 & 0.0322 & 3697 & 3697 \\
7.156 & 167.20 & 0.00116 & 0.0314 & 5774 & 6107 \\
7.188 & 172.29 & 0.00113 & 0.0306 & 4451 & 4324 \\
\hline
\end{tabular}

\caption{Summary of statistics for $m_l = m_s/27$, $64^3\times16$ lattices.\label{tab:mso27nt16}}
\end{table}

\clearpage
\section{Continuum-extrapolated values of the screening masses}
\label{app:screen_mass_values}
Here we have tabulated the continuum extrapolated screening masses
of $PS$, $S$, $V$ and $AV$ channels and in each channel for all three
flavor combinations {\it i.e.}\ $\bar ud$, $\bar us$ and $\bar ss$.

\begin{table}[!ht]
\begin{tabular}{|c|c|c|c|c|}
\hline
$T$ [GeV] & $m_\mathrm{P}$ [GeV] & $m_\mathrm{V}$ [GeV] & $m_\mathrm{S}$ [GeV] & $m_\mathrm{A}$ [GeV] \\
\hline
0.132 &
$0.129(5)$ & $0.7(2)$ & $0.22(2)$ & $1.0(2)$ \\
0.136 &
$0.139(4)$ & $0.69(9)$ & $0.23(2)$ & $0.96(9)$ \\
0.140 &
$0.150(2)$ & $0.70(7)$ & $0.24(1)$ & $0.94(7)$ \\
0.144 &
$0.1615(9)$ & $0.71(5)$ & $0.245(8)$ & $0.91(5)$ \\
0.148 &
$0.174(2)$ & $0.72(4)$ & $0.254(6)$ & $0.88(4)$ \\
0.152 &
$0.187(2)$ & $0.73(5)$ & $0.263(6)$ & $0.85(4)$ \\
0.156 &
$0.202(3)$ & $0.75(6)$ & $0.274(7)$ & $0.83(6)$ \\
0.160 &
$0.221(3)$ & $0.78(5)$ & $0.286(7)$ & $0.81(6)$ \\
0.164 &
$0.245(2)$ & $0.82(4)$ & $0.303(6)$ & $0.82(5)$ \\
0.168 &
$0.275(4)$ & $0.85(5)$ & $0.326(6)$ & $0.84(4)$ \\
0.172 &
$0.310(7)$ & $0.88(4)$ & $0.356(9)$ & $0.87(4)$ \\
0.176 &
$0.352(8)$ & $0.90(4)$ & $0.39(2)$ & $0.90(4)$ \\
0.180 &
$0.399(7)$ & $0.93(4)$ & $0.44(2)$ & $0.94(4)$ \\
0.184 &
$0.445(9)$ & $0.96(4)$ & $0.48(2)$ & $0.97(3)$ \\
0.188 &
$0.50(1)$ & $0.99(4)$ & $0.53(2)$ & $1.00(3)$ \\
0.192 &
$0.54(1)$ & $1.02(4)$ & $0.58(3)$ & $1.04(3)$ \\
0.196 &
$0.59(2)$ & $1.05(4)$ & $0.63(3)$ & $1.07(3)$ \\
0.200 &
$0.64(2)$ & $1.09(4)$ & $0.68(3)$ & $1.11(3)$ \\
0.240 &
$1.08(4)$ & $1.41(2)$ & $1.10(4)$ & $1.43(1)$ \\
0.280 &
$1.45(3)$ & $1.73(1)$ & $1.43(3)$ & $1.729(8)$ \\
0.320 &
$1.76(2)$ & $2.03(2)$ & $1.74(3)$ & $2.03(2)$ \\
0.360 &
$2.06(2)$ & $2.32(2)$ & $2.04(2)$ & $2.32(2)$ \\
0.400 &
$2.34(3)$ & $2.61(3)$ & $2.33(2)$ & $2.60(2)$ \\
0.440 &
$2.61(3)$ & $2.88(3)$ & $2.61(3)$ & $2.87(3)$ \\
0.480 &
$2.88(3)$ & $3.15(4)$ & $2.89(4)$ & $3.14(4)$ \\
0.520 &
$3.15(4)$ & $3.41(4)$ & $3.16(4)$ & $3.40(4)$ \\
0.560 &
$3.42(5)$ & $3.66(5)$ & $3.42(4)$ & $3.66(5)$ \\
0.600 &
$3.68(4)$ & $3.92(5)$ & $3.68(4)$ & $3.92(5)$ \\
0.640 &
$3.94(4)$ & $4.17(4)$ & $3.93(3)$ & $4.17(5)$ \\
0.680 &
$4.19(4)$ & $4.43(4)$ & $4.19(3)$ & $4.43(5)$ \\
0.720 &
$4.45(4)$ & $4.68(4)$ & $4.44(3)$ & $4.68(5)$ \\
0.760 &
$4.71(4)$ & $4.94(4)$ & $4.70(3)$ & $4.94(5)$ \\
0.800 &
$4.97(4)$ & $5.21(5)$ & $4.96(3)$ & $5.21(5)$ \\
0.840 &
$5.23(4)$ & $5.48(5)$ & $5.22(4)$ & $5.48(6)$ \\
0.880 &
$5.49(4)$ & $5.76(5)$ & $5.49(3)$ & $5.75(5)$ \\
0.920 &
$5.76(6)$ & $6.04(5)$ & $5.75(4)$ & $6.03(6)$ \\
0.960 &
$6.02(9)$ & $6.33(6)$ & $6.03(4)$ & $6.32(6)$ \\
1.000 &
$6.3(2)$ & $6.63(9)$ & $6.30(5)$ & $6.62(9)$ \\
\hline
\end{tabular}

\caption{Continuum-extrapolated values of the light-light screening masses.\label{tab:masses_ll}}
\end{table}

\begin{table}[!ht]
\centering
\begin{tabular}{|c|c|c|c|c|}
\hline
$T$ [GeV] & $m_\mathrm{P}$ [GeV] & $m_\mathrm{V}$ [GeV] & $m_\mathrm{S}$ [GeV] & $m_\mathrm{A}$ [GeV] \\
\hline
0.132 &
$0.50(2)$ & $0.88(2)$ & $0.66(3)$ & $1.17(6)$ \\
0.136 &
$0.51(1)$ & $0.89(2)$ & $0.67(3)$ & $1.16(6)$ \\
0.140 &
$0.519(5)$ & $0.90(2)$ & $0.67(2)$ & $1.14(5)$ \\
0.144 &
$0.527(2)$ & $0.91(2)$ & $0.67(2)$ & $1.12(3)$ \\
0.148 &
$0.537(4)$ & $0.923(9)$ & $0.67(2)$ & $1.10(3)$ \\
0.152 &
$0.547(9)$ & $0.936(9)$ & $0.675(9)$ & $1.08(2)$ \\
0.156 &
$0.559(7)$ & $0.950(9)$ & $0.679(8)$ & $1.06(2)$ \\
0.160 &
$0.574(4)$ & $0.965(9)$ & $0.682(7)$ & $1.04(2)$ \\
0.164 &
$0.590(7)$ & $0.982(9)$ & $0.686(5)$ & $1.04(2)$ \\
0.168 &
$0.604(4)$ & $1.00(1)$ & $0.690(6)$ & $1.04(2)$ \\
0.172 &
$0.621(6)$ & $1.020(9)$ & $0.698(8)$ & $1.05(2)$ \\
0.176 &
$0.642(9)$ & $1.041(9)$ & $0.71(2)$ & $1.07(2)$ \\
0.180 &
$0.667(9)$ & $1.063(9)$ & $0.73(2)$ & $1.09(2)$ \\
0.184 &
$0.697(9)$ & $1.086(9)$ & $0.75(2)$ & $1.11(2)$ \\
0.188 &
$0.73(2)$ & $1.11(1)$ & $0.77(2)$ & $1.13(2)$ \\
0.192 &
$0.76(2)$ & $1.13(1)$ & $0.80(2)$ & $1.15(2)$ \\
0.196 &
$0.80(2)$ & $1.16(1)$ & $0.83(2)$ & $1.18(2)$ \\
0.200 &
$0.83(3)$ & $1.19(2)$ & $0.86(2)$ & $1.20(2)$ \\
0.240 &
$1.16(2)$ & $1.461(8)$ & $1.16(2)$ & $1.463(9)$ \\
0.280 &
$1.46(2)$ & $1.748(7)$ & $1.46(2)$ & $1.743(7)$ \\
0.320 &
$1.76(2)$ & $2.04(2)$ & $1.75(2)$ & $2.03(2)$ \\
0.360 &
$2.05(2)$ & $2.32(2)$ & $2.04(2)$ & $2.32(2)$ \\
0.400 &
$2.34(2)$ & $2.60(2)$ & $2.33(2)$ & $2.60(2)$ \\
0.440 &
$2.62(2)$ & $2.88(3)$ & $2.61(2)$ & $2.88(3)$ \\
0.480 &
$2.89(3)$ & $3.15(3)$ & $2.89(3)$ & $3.15(3)$ \\
0.520 &
$3.16(3)$ & $3.41(3)$ & $3.16(4)$ & $3.41(3)$ \\
0.560 &
$3.42(4)$ & $3.67(3)$ & $3.42(4)$ & $3.67(4)$ \\
0.600 &
$3.68(4)$ & $3.93(4)$ & $3.68(4)$ & $3.92(4)$ \\
0.640 &
$3.93(4)$ & $4.18(4)$ & $3.94(4)$ & $4.17(4)$ \\
0.680 &
$4.19(4)$ & $4.44(4)$ & $4.19(4)$ & $4.43(4)$ \\
0.720 &
$4.45(4)$ & $4.69(4)$ & $4.44(4)$ & $4.68(4)$ \\
0.760 &
$4.71(4)$ & $4.95(5)$ & $4.69(5)$ & $4.94(5)$ \\
0.800 &
$4.97(4)$ & $5.21(4)$ & $4.95(5)$ & $5.21(5)$ \\
0.840 &
$5.23(4)$ & $5.48(4)$ & $5.21(5)$ & $5.48(5)$ \\
0.880 &
$5.50(4)$ & $5.75(4)$ & $5.47(5)$ & $5.75(4)$ \\
0.920 &
$5.77(5)$ & $6.02(5)$ & $5.74(6)$ & $6.03(5)$ \\
0.960 &
$6.05(7)$ & $6.29(7)$ & $6.02(8)$ & $6.31(6)$ \\
1.000 &
$6.3(1)$ & $6.6(1)$ & $6.3(2)$ & $6.60(8)$ \\
\hline
\end{tabular}

\caption{Continuum-extrapolated values of the strange-light screening masses.\label{tab:masses_sl}}
\end{table}

\begin{table}[!ht]
\centering
\begin{tabular}{|c|c|c|c|c|}
\hline
$T$ [GeV] & $m_\mathrm{P}$ [GeV] & $m_\mathrm{V}$ [GeV] & $m_\mathrm{S}$ [GeV] & $m_\mathrm{A}$ [GeV] \\
\hline
0.132 &
$0.71(2)$ & $1.026(7)$ & $1.01(3)$ & $1.36(5)$ \\
0.136 &
$0.711(8)$ & $1.032(6)$ & $1.01(2)$ & $1.34(5)$ \\
0.140 &
$0.714(4)$ & $1.040(5)$ & $1.00(2)$ & $1.33(4)$ \\
0.144 &
$0.717(1)$ & $1.048(4)$ & $0.99(2)$ & $1.32(3)$ \\
0.148 &
$0.720(6)$ & $1.056(3)$ & $0.99(2)$ & $1.30(2)$ \\
0.152 &
$0.724(9)$ & $1.065(3)$ & $0.98(2)$ & $1.29(2)$ \\
0.156 &
$0.730(9)$ & $1.075(3)$ & $0.97(1)$ & $1.27(2)$ \\
0.160 &
$0.744(6)$ & $1.086(3)$ & $0.965(9)$ & $1.25(2)$ \\
0.164 &
$0.758(6)$ & $1.098(3)$ & $0.957(7)$ & $1.24(2)$ \\
0.168 &
$0.772(5)$ & $1.110(4)$ & $0.949(7)$ & $1.23(2)$ \\
0.172 &
$0.783(8)$ & $1.124(4)$ & $0.944(9)$ & $1.22(2)$ \\
0.176 &
$0.796(9)$ & $1.138(5)$ & $0.94(2)$ & $1.22(2)$ \\
0.180 &
$0.81(1)$ & $1.154(5)$ & $0.95(2)$ & $1.23(2)$ \\
0.184 &
$0.831(9)$ & $1.171(6)$ & $0.96(2)$ & $1.24(2)$ \\
0.188 &
$0.85(1)$ & $1.189(6)$ & $0.97(2)$ & $1.25(1)$ \\
0.192 &
$0.88(2)$ & $1.208(7)$ & $0.98(2)$ & $1.26(1)$ \\
0.196 &
$0.90(2)$ & $1.229(7)$ & $1.00(2)$ & $1.277(9)$ \\
0.200 &
$0.93(2)$ & $1.250(7)$ & $1.02(2)$ & $1.294(9)$ \\
0.240 &
$1.20(3)$ & $1.492(7)$ & $1.25(2)$ & $1.512(7)$ \\
0.280 &
$1.48(2)$ & $1.763(7)$ & $1.50(2)$ & $1.772(7)$ \\
0.320 &
$1.78(2)$ & $2.04(2)$ & $1.78(2)$ & $2.05(2)$ \\
0.360 &
$2.07(2)$ & $2.32(2)$ & $2.06(2)$ & $2.33(2)$ \\
0.400 &
$2.35(2)$ & $2.60(3)$ & $2.34(2)$ & $2.61(3)$ \\
0.440 &
$2.63(3)$ & $2.88(3)$ & $2.62(3)$ & $2.88(3)$ \\
0.480 &
$2.90(3)$ & $3.15(3)$ & $2.89(3)$ & $3.15(3)$ \\
0.520 &
$3.17(4)$ & $3.41(3)$ & $3.16(4)$ & $3.41(3)$ \\
0.560 &
$3.43(4)$ & $3.68(4)$ & $3.42(4)$ & $3.67(4)$ \\
0.600 &
$3.68(4)$ & $3.93(4)$ & $3.68(4)$ & $3.93(4)$ \\
0.640 &
$3.94(4)$ & $4.19(5)$ & $3.93(4)$ & $4.19(5)$ \\
0.680 &
$4.19(4)$ & $4.45(6)$ & $4.19(4)$ & $4.44(5)$ \\
0.720 &
$4.45(3)$ & $4.71(6)$ & $4.44(4)$ & $4.70(5)$ \\
0.760 &
$4.70(3)$ & $4.96(6)$ & $4.70(4)$ & $4.96(5)$ \\
0.800 &
$4.96(4)$ & $5.22(5)$ & $4.95(4)$ & $5.22(5)$ \\
0.840 &
$5.23(4)$ & $5.48(4)$ & $5.21(4)$ & $5.49(5)$ \\
0.880 &
$5.50(4)$ & $5.74(4)$ & $5.48(6)$ & $5.76(4)$ \\
0.920 &
$5.77(5)$ & $6.01(5)$ & $5.75(8)$ & $6.03(4)$ \\
0.960 &
$6.05(6)$ & $6.27(7)$ & $6.0(2)$ & $6.31(6)$ \\
1.000 &
$6.3(1)$ & $6.5(2)$ & $6.3(2)$ & $6.59(9)$ \\
\hline
\end{tabular}

\caption{Continuum-extrapolated values of the strange-strange screening masses.\label{tab:masses_ss}}
\end{table}

\bibliography{refs}

\begin{thebibliography}{74}%
\makeatletter
\providecommand \@ifxundefined [1]{%
 \@ifx{#1\undefined}
}%
\providecommand \@ifnum [1]{%
 \ifnum #1\expandafter \@firstoftwo
 \else \expandafter \@secondoftwo
 \fi
}%
\providecommand \@ifx [1]{%
 \ifx #1\expandafter \@firstoftwo
 \else \expandafter \@secondoftwo
 \fi
}%
\providecommand \natexlab [1]{#1}%
\providecommand \enquote  [1]{``#1''}%
\providecommand \bibnamefont  [1]{#1}%
\providecommand \bibfnamefont [1]{#1}%
\providecommand \citenamefont [1]{#1}%
\providecommand \href@noop [0]{\@secondoftwo}%
\providecommand \href [0]{\begingroup \@sanitize@url \@href}%
\providecommand \@href[1]{\@@startlink{#1}\@@href}%
\providecommand \@@href[1]{\endgroup#1\@@endlink}%
\providecommand \@sanitize@url [0]{\catcode `\\12\catcode `\$12\catcode
  `\&12\catcode `\#12\catcode `\^12\catcode `\_12\catcode `\%12\relax}%
\providecommand \@@startlink[1]{}%
\providecommand \@@endlink[0]{}%
\providecommand \url  [0]{\begingroup\@sanitize@url \@url }%
\providecommand \@url [1]{\endgroup\@href {#1}{\urlprefix }}%
\providecommand \urlprefix  [0]{URL }%
\providecommand \Eprint [0]{\href }%
\providecommand \doibase [0]{http://dx.doi.org/}%
\providecommand \selectlanguage [0]{\@gobble}%
\providecommand \bibinfo  [0]{\@secondoftwo}%
\providecommand \bibfield  [0]{\@secondoftwo}%
\providecommand \translation [1]{[#1]}%
\providecommand \BibitemOpen [0]{}%
\providecommand \bibitemStop [0]{}%
\providecommand \bibitemNoStop [0]{.\EOS\space}%
\providecommand \EOS [0]{\spacefactor3000\relax}%
\providecommand \BibitemShut  [1]{\csname bibitem#1\endcsname}%
\let\auto@bib@innerbib\@empty
\bibitem [{\citenamefont {Andronic}\ \emph {et~al.}(2018)\citenamefont
  {Andronic}, \citenamefont {Braun-Munzinger}, \citenamefont {Redlich},\ and\
  \citenamefont {Stachel}}]{Andronic:2017pug}%
  \BibitemOpen
  \bibfield  {author} {\bibinfo {author} {\bibfnamefont {A.}~\bibnamefont
  {Andronic}}, \bibinfo {author} {\bibfnamefont {P.}~\bibnamefont
  {Braun-Munzinger}}, \bibinfo {author} {\bibfnamefont {K.}~\bibnamefont
  {Redlich}}, \ and\ \bibinfo {author} {\bibfnamefont {J.}~\bibnamefont
  {Stachel}},\ }\href {\doibase 10.1038/s41586-018-0491-6} {\bibfield
  {journal} {\bibinfo  {journal} {Nature}\ }\textbf {\bibinfo {volume} {561}},\
  \bibinfo {pages} {321} (\bibinfo {year} {2018})},\ \Eprint
  {http://arxiv.org/abs/1710.09425} {arXiv:1710.09425} \BibitemShut {NoStop}%
\bibitem [{\citenamefont {Aarts}\ \emph {et~al.}(2019)\citenamefont {Aarts},
  \citenamefont {Allton}, \citenamefont {De~Boni},\ and\ \citenamefont
  {Jäger}}]{Aarts:2018glk}%
  \BibitemOpen
  \bibfield  {author} {\bibinfo {author} {\bibfnamefont {G.}~\bibnamefont
  {Aarts}}, \bibinfo {author} {\bibfnamefont {C.}~\bibnamefont {Allton}},
  \bibinfo {author} {\bibfnamefont {D.}~\bibnamefont {De~Boni}}, \ and\
  \bibinfo {author} {\bibfnamefont {B.}~\bibnamefont {Jäger}},\ }\href
  {\doibase 10.1103/PhysRevD.99.074503} {\bibfield  {journal} {\bibinfo
  {journal} {Phys. Rev.}\ }\textbf {\bibinfo {volume} {D99}},\ \bibinfo {pages}
  {074503} (\bibinfo {year} {2019})},\ \Eprint
  {http://arxiv.org/abs/1812.07393} {arXiv:1812.07393 [hep-lat]} \BibitemShut
  {NoStop}%
\bibitem [{\citenamefont {Karsch}\ and\ \citenamefont
  {Laermann}(2003)}]{Karsch:2003jg}%
  \BibitemOpen
  \bibfield  {author} {\bibinfo {author} {\bibfnamefont {F.}~\bibnamefont
  {Karsch}}\ and\ \bibinfo {author} {\bibfnamefont {E.}~\bibnamefont
  {Laermann}},\ }\href@noop {} {\ ,\ \bibinfo {pages} {In *Hwa, R.C. (ed.) et
  al.: Quark gluon plasma* 1} (\bibinfo {year} {2003})},\ \Eprint
  {http://arxiv.org/abs/hep-lat/0305025} {arXiv:hep-lat/0305025} \BibitemShut
  {NoStop}%
\bibitem [{\citenamefont {Aarts}\ \emph
  {et~al.}(2017{\natexlab{a}})\citenamefont {Aarts}, \citenamefont {Allton},
  \citenamefont {De~Boni}, \citenamefont {Hands}, \citenamefont {J{\"a}ger},
  \citenamefont {Praki},\ and\ \citenamefont {Skullerud}}]{Aarts:2017rrl}%
  \BibitemOpen
  \bibfield  {author} {\bibinfo {author} {\bibfnamefont {G.}~\bibnamefont
  {Aarts}}, \bibinfo {author} {\bibfnamefont {C.}~\bibnamefont {Allton}},
  \bibinfo {author} {\bibfnamefont {D.}~\bibnamefont {De~Boni}}, \bibinfo
  {author} {\bibfnamefont {S.}~\bibnamefont {Hands}}, \bibinfo {author}
  {\bibfnamefont {B.}~\bibnamefont {J{\"a}ger}}, \bibinfo {author}
  {\bibfnamefont {C.}~\bibnamefont {Praki}}, \ and\ \bibinfo {author}
  {\bibfnamefont {J.-I.}\ \bibnamefont {Skullerud}},\ }\href {\doibase
  10.1007/JHEP06(2017)034} {\bibfield  {journal} {\bibinfo  {journal} {JHEP}\
  }\textbf {\bibinfo {volume} {06}},\ \bibinfo {pages} {034} (\bibinfo {year}
  {2017}{\natexlab{a}})},\ \Eprint {http://arxiv.org/abs/1703.09246}
  {arXiv:1703.09246} \BibitemShut {NoStop}%
\bibitem [{\citenamefont {Detar}\ and\ \citenamefont
  {Kogut}(1987{\natexlab{a}})}]{DeTar:1987ar}%
  \BibitemOpen
  \bibfield  {author} {\bibinfo {author} {\bibfnamefont {C.~E.}\ \bibnamefont
  {Detar}}\ and\ \bibinfo {author} {\bibfnamefont {J.~B.}\ \bibnamefont
  {Kogut}},\ }\href {\doibase 10.1103/PhysRevLett.59.399} {\bibfield  {journal}
  {\bibinfo  {journal} {Phys. Rev. Lett.}\ }\textbf {\bibinfo {volume} {59}},\
  \bibinfo {pages} {399} (\bibinfo {year} {1987}{\natexlab{a}})}\BibitemShut
  {NoStop}%
\bibitem [{\citenamefont {Bazavov}\ \emph {et~al.}(2019)\citenamefont {Bazavov}
  \emph {et~al.}}]{Bazavov:2018mes}%
  \BibitemOpen
  \bibfield  {author} {\bibinfo {author} {\bibfnamefont {A.}~\bibnamefont
  {Bazavov}} \emph {et~al.} (\bibinfo {collaboration} {HotQCD}),\ }\href
  {\doibase 10.1016/j.physletb.2019.05.013} {\bibfield  {journal} {\bibinfo
  {journal} {Phys. Lett.}\ }\textbf {\bibinfo {volume} {B795}},\ \bibinfo
  {pages} {15} (\bibinfo {year} {2019})},\ \Eprint
  {http://arxiv.org/abs/1812.08235} {arXiv:1812.08235} \BibitemShut {NoStop}%
\bibitem [{\citenamefont {Adler}(1969)}]{Adler:1969gk}%
  \BibitemOpen
  \bibfield  {author} {\bibinfo {author} {\bibfnamefont {S.~L.}\ \bibnamefont
  {Adler}},\ }\href {\doibase 10.1103/PhysRev.177.2426} {\bibfield  {journal}
  {\bibinfo  {journal} {Phys. Rev.}\ }\textbf {\bibinfo {volume} {177}},\
  \bibinfo {pages} {2426} (\bibinfo {year} {1969})},\ \bibinfo {note}
  {[,241(1969)]}\BibitemShut {NoStop}%
\bibitem [{\citenamefont {Bell}\ and\ \citenamefont
  {Jackiw}(1969)}]{Bell:1969ts}%
  \BibitemOpen
  \bibfield  {author} {\bibinfo {author} {\bibfnamefont {J.~S.}\ \bibnamefont
  {Bell}}\ and\ \bibinfo {author} {\bibfnamefont {R.}~\bibnamefont {Jackiw}},\
  }\href {\doibase 10.1007/BF02823296} {\bibfield  {journal} {\bibinfo
  {journal} {Nuovo Cim.}\ }\textbf {\bibinfo {volume} {A60}},\ \bibinfo {pages}
  {47} (\bibinfo {year} {1969})}\BibitemShut {NoStop}%
\bibitem [{\citenamefont {Adler}\ and\ \citenamefont
  {Bardeen}(1969)}]{Adler:1969er}%
  \BibitemOpen
  \bibfield  {author} {\bibinfo {author} {\bibfnamefont {S.~L.}\ \bibnamefont
  {Adler}}\ and\ \bibinfo {author} {\bibfnamefont {W.~A.}\ \bibnamefont
  {Bardeen}},\ }\href {\doibase 10.1103/PhysRev.182.1517} {\bibfield  {journal}
  {\bibinfo  {journal} {Phys. Rev.}\ }\textbf {\bibinfo {volume} {182}},\
  \bibinfo {pages} {1517} (\bibinfo {year} {1969})},\ \bibinfo {note}
  {[,268(1969)]}\BibitemShut {NoStop}%
\bibitem [{\citenamefont {Gross}\ \emph {et~al.}(1981)\citenamefont {Gross},
  \citenamefont {Pisarski},\ and\ \citenamefont {Yaffe}}]{Gross:1980br}%
  \BibitemOpen
  \bibfield  {author} {\bibinfo {author} {\bibfnamefont {D.~J.}\ \bibnamefont
  {Gross}}, \bibinfo {author} {\bibfnamefont {R.~D.}\ \bibnamefont {Pisarski}},
  \ and\ \bibinfo {author} {\bibfnamefont {L.~G.}\ \bibnamefont {Yaffe}},\
  }\href {\doibase 10.1103/RevModPhys.53.43} {\bibfield  {journal} {\bibinfo
  {journal} {Rev. Mod. Phys.}\ }\textbf {\bibinfo {volume} {53}},\ \bibinfo
  {pages} {43} (\bibinfo {year} {1981})}\BibitemShut {NoStop}%
\bibitem [{\citenamefont {Pisarski}\ and\ \citenamefont
  {Wilczek}(1984)}]{Pisarski:1983ms}%
  \BibitemOpen
  \bibfield  {author} {\bibinfo {author} {\bibfnamefont {R.~D.}\ \bibnamefont
  {Pisarski}}\ and\ \bibinfo {author} {\bibfnamefont {F.}~\bibnamefont
  {Wilczek}},\ }\href {\doibase 10.1103/PhysRevD.29.338} {\bibfield  {journal}
  {\bibinfo  {journal} {Phys. Rev.}\ }\textbf {\bibinfo {volume} {D29}},\
  \bibinfo {pages} {338} (\bibinfo {year} {1984})}\BibitemShut {NoStop}%
\bibitem [{\citenamefont {Cheng}\ \emph {et~al.}(2011)\citenamefont {Cheng}
  \emph {et~al.}}]{Cheng:2010fe}%
  \BibitemOpen
  \bibfield  {author} {\bibinfo {author} {\bibfnamefont {M.}~\bibnamefont
  {Cheng}} \emph {et~al.},\ }\href {\doibase 10.1140/epjc/s10052-011-1564-y}
  {\bibfield  {journal} {\bibinfo  {journal} {Eur. Phys. J.}\ }\textbf
  {\bibinfo {volume} {C71}},\ \bibinfo {pages} {1564} (\bibinfo {year}
  {2011})},\ \Eprint {http://arxiv.org/abs/1010.1216} {arXiv:1010.1216}
  \BibitemShut {NoStop}%
\bibitem [{\citenamefont {Ohno}\ \emph {et~al.}(2012)\citenamefont {Ohno},
  \citenamefont {Heller}, \citenamefont {Karsch},\ and\ \citenamefont
  {Mukherjee}}]{Ohno:2012br}%
  \BibitemOpen
  \bibfield  {author} {\bibinfo {author} {\bibfnamefont {H.}~\bibnamefont
  {Ohno}}, \bibinfo {author} {\bibfnamefont {U.~M.}\ \bibnamefont {Heller}},
  \bibinfo {author} {\bibfnamefont {F.}~\bibnamefont {Karsch}}, \ and\ \bibinfo
  {author} {\bibfnamefont {S.}~\bibnamefont {Mukherjee}},\ }\bibfield
  {booktitle} {\emph {\bibinfo {booktitle} {{Proceedings, 30th International
  Symposium on Lattice Field Theory (Lattice 2012): Cairns, Australia, June
  24-29, 2012}}},\ }\href {\doibase 10.22323/1.164.0095} {\bibfield  {journal}
  {\bibinfo  {journal} {PoS}\ }\textbf {\bibinfo {volume} {LATTICE2012}},\
  \bibinfo {pages} {095} (\bibinfo {year} {2012})},\ \Eprint
  {http://arxiv.org/abs/1211.2591} {arXiv:1211.2591 [hep-lat]} \BibitemShut
  {NoStop}%
\bibitem [{\citenamefont {Dick}\ \emph {et~al.}(2015)\citenamefont {Dick},
  \citenamefont {Karsch}, \citenamefont {Laermann}, \citenamefont {Mukherjee},\
  and\ \citenamefont {Sharma}}]{Dick:2015twa}%
  \BibitemOpen
  \bibfield  {author} {\bibinfo {author} {\bibfnamefont {V.}~\bibnamefont
  {Dick}}, \bibinfo {author} {\bibfnamefont {F.}~\bibnamefont {Karsch}},
  \bibinfo {author} {\bibfnamefont {E.}~\bibnamefont {Laermann}}, \bibinfo
  {author} {\bibfnamefont {S.}~\bibnamefont {Mukherjee}}, \ and\ \bibinfo
  {author} {\bibfnamefont {S.}~\bibnamefont {Sharma}},\ }\href {\doibase
  10.1103/PhysRevD.91.094504} {\bibfield  {journal} {\bibinfo  {journal} {Phys.
  Rev.}\ }\textbf {\bibinfo {volume} {D91}},\ \bibinfo {pages} {094504}
  (\bibinfo {year} {2015})},\ \Eprint {http://arxiv.org/abs/1502.06190}
  {arXiv:1502.06190} \BibitemShut {NoStop}%
\bibitem [{\citenamefont {Ding}\ \emph {et~al.}(2019)\citenamefont {Ding} \emph
  {et~al.}}]{Ding:2019prx}%
  \BibitemOpen
  \bibfield  {author} {\bibinfo {author} {\bibfnamefont {H.~T.}\ \bibnamefont
  {Ding}} \emph {et~al.},\ }\href {\doibase 10.1103/PhysRevLett.123.062002}
  {\bibfield  {journal} {\bibinfo  {journal} {Phys. Rev. Lett.}\ }\textbf
  {\bibinfo {volume} {123}},\ \bibinfo {pages} {062002} (\bibinfo {year}
  {2019})},\ \Eprint {http://arxiv.org/abs/1903.04801} {arXiv:1903.04801}
  \BibitemShut {NoStop}%
\bibitem [{\citenamefont {Shuryak}(1994)}]{Shuryak:1993ee}%
  \BibitemOpen
  \bibfield  {author} {\bibinfo {author} {\bibfnamefont {E.~V.}\ \bibnamefont
  {Shuryak}},\ }\href@noop {} {\bibfield  {journal} {\bibinfo  {journal}
  {Comments Nucl. Part. Phys.}\ }\textbf {\bibinfo {volume} {21}},\ \bibinfo
  {pages} {235} (\bibinfo {year} {1994})},\ \Eprint
  {http://arxiv.org/abs/hep-ph/9310253} {arXiv:hep-ph/9310253} \BibitemShut
  {NoStop}%
\bibitem [{\citenamefont {Birse}\ \emph {et~al.}(1996)\citenamefont {Birse},
  \citenamefont {Cohen},\ and\ \citenamefont {McGovern}}]{Birse:1996dx}%
  \BibitemOpen
  \bibfield  {author} {\bibinfo {author} {\bibfnamefont {M.~C.}\ \bibnamefont
  {Birse}}, \bibinfo {author} {\bibfnamefont {T.~D.}\ \bibnamefont {Cohen}}, \
  and\ \bibinfo {author} {\bibfnamefont {J.~A.}\ \bibnamefont {McGovern}},\
  }\href {\doibase 10.1016/0370-2693(96)01151-3} {\bibfield  {journal}
  {\bibinfo  {journal} {Phys. Lett.}\ }\textbf {\bibinfo {volume} {B388}},\
  \bibinfo {pages} {137} (\bibinfo {year} {1996})},\ \Eprint
  {http://arxiv.org/abs/hep-ph/9608255} {arXiv:hep-ph/9608255} \BibitemShut
  {NoStop}%
\bibitem [{\citenamefont {Lee}\ and\ \citenamefont
  {Hatsuda}(1996)}]{Lee:1996zy}%
  \BibitemOpen
  \bibfield  {author} {\bibinfo {author} {\bibfnamefont {S.~H.}\ \bibnamefont
  {Lee}}\ and\ \bibinfo {author} {\bibfnamefont {T.}~\bibnamefont {Hatsuda}},\
  }\href {\doibase 10.1103/PhysRevD.54.R1871} {\bibfield  {journal} {\bibinfo
  {journal} {Phys. Rev.}\ }\textbf {\bibinfo {volume} {D54}},\ \bibinfo {pages}
  {R1871} (\bibinfo {year} {1996})},\ \Eprint
  {http://arxiv.org/abs/hep-ph/9601373} {arXiv:hep-ph/9601373} \BibitemShut
  {NoStop}%
\bibitem [{\citenamefont {Evans}\ \emph {et~al.}(1996)\citenamefont {Evans},
  \citenamefont {Hsu},\ and\ \citenamefont {Schwetz}}]{Evans:1996wf}%
  \BibitemOpen
  \bibfield  {author} {\bibinfo {author} {\bibfnamefont {N.~J.}\ \bibnamefont
  {Evans}}, \bibinfo {author} {\bibfnamefont {S.~D.~H.}\ \bibnamefont {Hsu}}, \
  and\ \bibinfo {author} {\bibfnamefont {M.}~\bibnamefont {Schwetz}},\ }\href
  {\doibase 10.1016/0370-2693(96)00280-8} {\bibfield  {journal} {\bibinfo
  {journal} {Phys. Lett.}\ }\textbf {\bibinfo {volume} {B375}},\ \bibinfo
  {pages} {262} (\bibinfo {year} {1996})},\ \Eprint
  {http://arxiv.org/abs/hep-ph/9601361} {arXiv:hep-ph/9601361} \BibitemShut
  {NoStop}%
\bibitem [{\citenamefont {Buchoff}\ \emph {et~al.}(2014)\citenamefont {Buchoff}
  \emph {et~al.}}]{Buchoff:2013nra}%
  \BibitemOpen
  \bibfield  {author} {\bibinfo {author} {\bibfnamefont {M.~I.}\ \bibnamefont
  {Buchoff}} \emph {et~al.},\ }\href {\doibase 10.1103/PhysRevD.89.054514}
  {\bibfield  {journal} {\bibinfo  {journal} {Phys. Rev.}\ }\textbf {\bibinfo
  {volume} {D89}},\ \bibinfo {pages} {054514} (\bibinfo {year} {2014})},\
  \Eprint {http://arxiv.org/abs/1309.4149} {arXiv:1309.4149} \BibitemShut
  {NoStop}%
\bibitem [{\citenamefont {Bazavov}\ \emph
  {et~al.}(2012{\natexlab{a}})\citenamefont {Bazavov} \emph
  {et~al.}}]{Bazavov:2012qja}%
  \BibitemOpen
  \bibfield  {author} {\bibinfo {author} {\bibfnamefont {A.}~\bibnamefont
  {Bazavov}} \emph {et~al.} (\bibinfo {collaboration} {HotQCD}),\ }\href
  {\doibase 10.1103/PhysRevD.86.094503} {\bibfield  {journal} {\bibinfo
  {journal} {Phys. Rev.}\ }\textbf {\bibinfo {volume} {D86}},\ \bibinfo {pages}
  {094503} (\bibinfo {year} {2012}{\natexlab{a}})},\ \Eprint
  {http://arxiv.org/abs/1205.3535} {arXiv:1205.3535} \BibitemShut {NoStop}%
\bibitem [{\citenamefont {Suzuki}\ \emph
  {et~al.}(2018{\natexlab{a}})\citenamefont {Suzuki}, \citenamefont {Aoki},
  \citenamefont {Aoki}, \citenamefont {Cossu}, \citenamefont {Fukaya},\ and\
  \citenamefont {Hashimoto}}]{Suzuki:2018vbe}%
  \BibitemOpen
  \bibfield  {author} {\bibinfo {author} {\bibfnamefont {K.}~\bibnamefont
  {Suzuki}}, \bibinfo {author} {\bibfnamefont {S.}~\bibnamefont {Aoki}},
  \bibinfo {author} {\bibfnamefont {Y.}~\bibnamefont {Aoki}}, \bibinfo {author}
  {\bibfnamefont {G.}~\bibnamefont {Cossu}}, \bibinfo {author} {\bibfnamefont
  {H.}~\bibnamefont {Fukaya}}, \ and\ \bibinfo {author} {\bibfnamefont
  {S.}~\bibnamefont {Hashimoto}} (\bibinfo {collaboration} {JLQCD})\ }(\bibinfo
  {year} {2018})\ \Eprint {http://arxiv.org/abs/1812.06621} {arXiv:1812.06621}
  \BibitemShut {NoStop}%
\bibitem [{\citenamefont {Tomiya}\ \emph {et~al.}(2017)\citenamefont {Tomiya},
  \citenamefont {Cossu}, \citenamefont {Aoki}, \citenamefont {Fukaya},
  \citenamefont {Hashimoto}, \citenamefont {Kaneko},\ and\ \citenamefont
  {Noaki}}]{Tomiya:2016jwr}%
  \BibitemOpen
  \bibfield  {author} {\bibinfo {author} {\bibfnamefont {A.}~\bibnamefont
  {Tomiya}}, \bibinfo {author} {\bibfnamefont {G.}~\bibnamefont {Cossu}},
  \bibinfo {author} {\bibfnamefont {S.}~\bibnamefont {Aoki}}, \bibinfo {author}
  {\bibfnamefont {H.}~\bibnamefont {Fukaya}}, \bibinfo {author} {\bibfnamefont
  {S.}~\bibnamefont {Hashimoto}}, \bibinfo {author} {\bibfnamefont
  {T.}~\bibnamefont {Kaneko}}, \ and\ \bibinfo {author} {\bibfnamefont
  {J.}~\bibnamefont {Noaki}},\ }\href {\doibase 10.1103/PhysRevD.96.034509,
  10.1103/PhysRevD.96.079902} {\bibfield  {journal} {\bibinfo  {journal} {Phys.
  Rev.}\ }\textbf {\bibinfo {volume} {D96}},\ \bibinfo {pages} {034509}
  (\bibinfo {year} {2017})},\ \bibinfo {note} {[Addendum: Phys.
  Rev.D96,no.7,079902(2017)]},\ \Eprint {http://arxiv.org/abs/1612.01908}
  {arXiv:1612.01908} \BibitemShut {NoStop}%
\bibitem [{\citenamefont {Chiu}\ \emph {et~al.}(2014)\citenamefont {Chiu},
  \citenamefont {Chen}, \citenamefont {Chen}, \citenamefont {Chou},\ and\
  \citenamefont {Hsieh}}]{Chiu:2013wwa}%
  \BibitemOpen
  \bibfield  {author} {\bibinfo {author} {\bibfnamefont {T.-W.}\ \bibnamefont
  {Chiu}}, \bibinfo {author} {\bibfnamefont {W.-P.}\ \bibnamefont {Chen}},
  \bibinfo {author} {\bibfnamefont {Y.-C.}\ \bibnamefont {Chen}}, \bibinfo
  {author} {\bibfnamefont {H.-Y.}\ \bibnamefont {Chou}}, \ and\ \bibinfo
  {author} {\bibfnamefont {T.-H.}\ \bibnamefont {Hsieh}} (\bibinfo
  {collaboration} {TWQCD}),\ }\href {\doibase 10.22323/1.187.0165} {\bibfield
  {journal} {\bibinfo  {journal} {PoS}\ }\textbf {\bibinfo {volume}
  {LATTICE2013}},\ \bibinfo {pages} {165} (\bibinfo {year} {2014})},\ \Eprint
  {http://arxiv.org/abs/1311.6220} {arXiv:1311.6220} \BibitemShut {NoStop}%
\bibitem [{\citenamefont {Sharma}\ \emph {et~al.}(2016)\citenamefont {Sharma},
  \citenamefont {Dick}, \citenamefont {Karsch}, \citenamefont {Laermann},\ and\
  \citenamefont {Mukherjee}}]{Sharma:2016cmz}%
  \BibitemOpen
  \bibfield  {author} {\bibinfo {author} {\bibfnamefont {S.}~\bibnamefont
  {Sharma}}, \bibinfo {author} {\bibfnamefont {V.}~\bibnamefont {Dick}},
  \bibinfo {author} {\bibfnamefont {F.}~\bibnamefont {Karsch}}, \bibinfo
  {author} {\bibfnamefont {E.}~\bibnamefont {Laermann}}, \ and\ \bibinfo
  {author} {\bibfnamefont {S.}~\bibnamefont {Mukherjee}},\ }\bibfield
  {booktitle} {\emph {\bibinfo {booktitle} {{Proceedings, 25th International
  Conference on Ultra-Relativistic Nucleus-Nucleus Collisions (Quark Matter
  2015): Kobe, Japan, September 27-October 3, 2015}}},\ }\href {\doibase
  10.1016/j.nuclphysa.2016.02.013} {\bibfield  {journal} {\bibinfo  {journal}
  {Nucl. Phys.}\ }\textbf {\bibinfo {volume} {A956}},\ \bibinfo {pages} {793}
  (\bibinfo {year} {2016})},\ \Eprint {http://arxiv.org/abs/1602.02197}
  {arXiv:1602.02197 [hep-lat]} \BibitemShut {NoStop}%
\bibitem [{\citenamefont {Brandt}\ \emph {et~al.}(2016)\citenamefont {Brandt},
  \citenamefont {Francis}, \citenamefont {Meyer}, \citenamefont {Philipsen},
  \citenamefont {Robaina},\ and\ \citenamefont {Wittig}}]{Brandt:2016daq}%
  \BibitemOpen
  \bibfield  {author} {\bibinfo {author} {\bibfnamefont {B.~B.}\ \bibnamefont
  {Brandt}}, \bibinfo {author} {\bibfnamefont {A.}~\bibnamefont {Francis}},
  \bibinfo {author} {\bibfnamefont {H.~B.}\ \bibnamefont {Meyer}}, \bibinfo
  {author} {\bibfnamefont {O.}~\bibnamefont {Philipsen}}, \bibinfo {author}
  {\bibfnamefont {D.}~\bibnamefont {Robaina}}, \ and\ \bibinfo {author}
  {\bibfnamefont {H.}~\bibnamefont {Wittig}},\ }\href {\doibase
  10.1007/JHEP12(2016)158} {\bibfield  {journal} {\bibinfo  {journal} {JHEP}\
  }\textbf {\bibinfo {volume} {12}},\ \bibinfo {pages} {158} (\bibinfo {year}
  {2016})},\ \Eprint {http://arxiv.org/abs/1608.06882} {arXiv:1608.06882}
  \BibitemShut {NoStop}%
\bibitem [{\citenamefont {Karsch}\ \emph {et~al.}(2012)\citenamefont {Karsch},
  \citenamefont {Laermann}, \citenamefont {Mukherjee},\ and\ \citenamefont
  {Petreczky}}]{Karsch:2012na}%
  \BibitemOpen
  \bibfield  {author} {\bibinfo {author} {\bibfnamefont {F.}~\bibnamefont
  {Karsch}}, \bibinfo {author} {\bibfnamefont {E.}~\bibnamefont {Laermann}},
  \bibinfo {author} {\bibfnamefont {S.}~\bibnamefont {Mukherjee}}, \ and\
  \bibinfo {author} {\bibfnamefont {P.}~\bibnamefont {Petreczky}},\ }\href
  {\doibase 10.1103/PhysRevD.85.114501} {\bibfield  {journal} {\bibinfo
  {journal} {Phys. Rev.}\ }\textbf {\bibinfo {volume} {D85}},\ \bibinfo {pages}
  {114501} (\bibinfo {year} {2012})},\ \Eprint {http://arxiv.org/abs/1203.3770}
  {arXiv:1203.3770} \BibitemShut {NoStop}%
\bibitem [{\citenamefont {Bazavov}\ \emph {et~al.}(2015)\citenamefont
  {Bazavov}, \citenamefont {Karsch}, \citenamefont {Maezawa}, \citenamefont
  {Mukherjee},\ and\ \citenamefont {Petreczky}}]{Bazavov:2014cta}%
  \BibitemOpen
  \bibfield  {author} {\bibinfo {author} {\bibfnamefont {A.}~\bibnamefont
  {Bazavov}}, \bibinfo {author} {\bibfnamefont {F.}~\bibnamefont {Karsch}},
  \bibinfo {author} {\bibfnamefont {Y.}~\bibnamefont {Maezawa}}, \bibinfo
  {author} {\bibfnamefont {S.}~\bibnamefont {Mukherjee}}, \ and\ \bibinfo
  {author} {\bibfnamefont {P.}~\bibnamefont {Petreczky}},\ }\href {\doibase
  10.1103/PhysRevD.91.054503} {\bibfield  {journal} {\bibinfo  {journal} {Phys.
  Rev.}\ }\textbf {\bibinfo {volume} {D91}},\ \bibinfo {pages} {054503}
  (\bibinfo {year} {2015})},\ \Eprint {http://arxiv.org/abs/1411.3018}
  {arXiv:1411.3018} \BibitemShut {NoStop}%
\bibitem [{\citenamefont {Hashimoto}\ \emph {et~al.}(1993)\citenamefont
  {Hashimoto}, \citenamefont {Nakamura},\ and\ \citenamefont
  {Stamatescu}}]{Hashimoto:1992np}%
  \BibitemOpen
  \bibfield  {author} {\bibinfo {author} {\bibfnamefont {T.}~\bibnamefont
  {Hashimoto}}, \bibinfo {author} {\bibfnamefont {A.}~\bibnamefont {Nakamura}},
  \ and\ \bibinfo {author} {\bibfnamefont {I.~O.}\ \bibnamefont {Stamatescu}},\
  }\href {\doibase 10.1016/0550-3213(93)90407-G} {\bibfield  {journal}
  {\bibinfo  {journal} {Nucl. Phys.}\ }\textbf {\bibinfo {volume} {B400}},\
  \bibinfo {pages} {267} (\bibinfo {year} {1993})}\BibitemShut {NoStop}%
\bibitem [{\citenamefont {Golterman}(1986)}]{Golterman:1985dz}%
  \BibitemOpen
  \bibfield  {author} {\bibinfo {author} {\bibfnamefont {M.~F.~L.}\
  \bibnamefont {Golterman}},\ }\href {\doibase 10.1016/0550-3213(86)90383-4}
  {\bibfield  {journal} {\bibinfo  {journal} {Nucl. Phys.}\ }\textbf {\bibinfo
  {volume} {B273}},\ \bibinfo {pages} {663} (\bibinfo {year}
  {1986})}\BibitemShut {NoStop}%
\bibitem [{\citenamefont {Kilcup}\ and\ \citenamefont
  {Sharpe}(1987)}]{Kilcup:1986dg}%
  \BibitemOpen
  \bibfield  {author} {\bibinfo {author} {\bibfnamefont {G.~W.}\ \bibnamefont
  {Kilcup}}\ and\ \bibinfo {author} {\bibfnamefont {S.~R.}\ \bibnamefont
  {Sharpe}},\ }\href {\doibase 10.1016/0550-3213(87)90285-9} {\bibfield
  {journal} {\bibinfo  {journal} {Nucl. Phys.}\ }\textbf {\bibinfo {volume}
  {B283}},\ \bibinfo {pages} {493} (\bibinfo {year} {1987})}\BibitemShut
  {NoStop}%
\bibitem [{\citenamefont {Altmeyer}\ \emph {et~al.}(1993)\citenamefont
  {Altmeyer}, \citenamefont {Born}, \citenamefont {Gockeler}, \citenamefont
  {Horsley}, \citenamefont {Laermann},\ and\ \citenamefont
  {Schierholz}}]{Altmeyer:1992dd}%
  \BibitemOpen
  \bibfield  {author} {\bibinfo {author} {\bibfnamefont {R.}~\bibnamefont
  {Altmeyer}}, \bibinfo {author} {\bibfnamefont {K.~D.}\ \bibnamefont {Born}},
  \bibinfo {author} {\bibfnamefont {M.}~\bibnamefont {Gockeler}}, \bibinfo
  {author} {\bibfnamefont {R.}~\bibnamefont {Horsley}}, \bibinfo {author}
  {\bibfnamefont {E.}~\bibnamefont {Laermann}}, \ and\ \bibinfo {author}
  {\bibfnamefont {G.}~\bibnamefont {Schierholz}} (\bibinfo {collaboration}
  {MT(c)}),\ }\href {\doibase 10.1016/0550-3213(93)90328-M} {\bibfield
  {journal} {\bibinfo  {journal} {Nucl. Phys.}\ }\textbf {\bibinfo {volume}
  {B389}},\ \bibinfo {pages} {445} (\bibinfo {year} {1993})}\BibitemShut
  {NoStop}%
\bibitem [{\citenamefont {Gupta}(1999)}]{Gupta:1999hp}%
  \BibitemOpen
  \bibfield  {author} {\bibinfo {author} {\bibfnamefont {S.}~\bibnamefont
  {Gupta}},\ }\href {\doibase 10.1103/PhysRevD.60.094505} {\bibfield  {journal}
  {\bibinfo  {journal} {Phys. Rev.}\ }\textbf {\bibinfo {volume} {D60}},\
  \bibinfo {pages} {094505} (\bibinfo {year} {1999})},\ \Eprint
  {http://arxiv.org/abs/hep-lat/9903019} {arXiv:hep-lat/9903019} \BibitemShut
  {NoStop}%
\bibitem [{\citenamefont {Lepage}(1999)}]{Lepage:1998vj}%
  \BibitemOpen
  \bibfield  {author} {\bibinfo {author} {\bibfnamefont {G.~P.}\ \bibnamefont
  {Lepage}},\ }\href {\doibase 10.1103/PhysRevD.59.074502} {\bibfield
  {journal} {\bibinfo  {journal} {Phys. Rev.}\ }\textbf {\bibinfo {volume}
  {D59}},\ \bibinfo {pages} {074502} (\bibinfo {year} {1999})},\ \Eprint
  {http://arxiv.org/abs/hep-lat/9809157} {arXiv:hep-lat/9809157} \BibitemShut
  {NoStop}%
\bibitem [{\citenamefont {Follana}\ \emph {et~al.}(2007)\citenamefont
  {Follana}, \citenamefont {Mason}, \citenamefont {Davies}, \citenamefont
  {Hornbostel}, \citenamefont {Lepage}, \citenamefont {Shigemitsu},
  \citenamefont {Trottier},\ and\ \citenamefont {Wong}}]{Follana:2006rc}%
  \BibitemOpen
  \bibfield  {author} {\bibinfo {author} {\bibfnamefont {E.}~\bibnamefont
  {Follana}}, \bibinfo {author} {\bibfnamefont {Q.}~\bibnamefont {Mason}},
  \bibinfo {author} {\bibfnamefont {C.}~\bibnamefont {Davies}}, \bibinfo
  {author} {\bibfnamefont {K.}~\bibnamefont {Hornbostel}}, \bibinfo {author}
  {\bibfnamefont {G.~P.}\ \bibnamefont {Lepage}}, \bibinfo {author}
  {\bibfnamefont {J.}~\bibnamefont {Shigemitsu}}, \bibinfo {author}
  {\bibfnamefont {H.}~\bibnamefont {Trottier}}, \ and\ \bibinfo {author}
  {\bibfnamefont {K.}~\bibnamefont {Wong}} (\bibinfo {collaboration} {HPQCD,
  UKQCD}),\ }\href {\doibase 10.1103/PhysRevD.75.054502} {\bibfield  {journal}
  {\bibinfo  {journal} {Phys. Rev.}\ }\textbf {\bibinfo {volume} {D75}},\
  \bibinfo {pages} {054502} (\bibinfo {year} {2007})},\ \Eprint
  {http://arxiv.org/abs/hep-lat/0610092} {arXiv:hep-lat/0610092} \BibitemShut
  {NoStop}%
\bibitem [{\citenamefont {Bazavov}\ \emph {et~al.}(2008)\citenamefont {Bazavov}
  \emph {et~al.}}]{Bazavov:2009jc}%
  \BibitemOpen
  \bibfield  {author} {\bibinfo {author} {\bibfnamefont {A.}~\bibnamefont
  {Bazavov}} \emph {et~al.} (\bibinfo {collaboration} {MILC}),\ }\href
  {\doibase 10.22323/1.066.0033} {\bibfield  {journal} {\bibinfo  {journal}
  {PoS}\ }\textbf {\bibinfo {volume} {LATTICE2008}},\ \bibinfo {pages} {033}
  (\bibinfo {year} {2008})},\ \Eprint {http://arxiv.org/abs/0903.0874}
  {arXiv:0903.0874} \BibitemShut {NoStop}%
\bibitem [{\citenamefont {Bazavov}\ \emph {et~al.}(2010)\citenamefont {Bazavov}
  \emph {et~al.}}]{Bazavov:2010ru}%
  \BibitemOpen
  \bibfield  {author} {\bibinfo {author} {\bibfnamefont {A.}~\bibnamefont
  {Bazavov}} \emph {et~al.} (\bibinfo {collaboration} {MILC}),\ }\href
  {\doibase 10.1103/PhysRevD.82.074501} {\bibfield  {journal} {\bibinfo
  {journal} {Phys. Rev.}\ }\textbf {\bibinfo {volume} {D82}},\ \bibinfo {pages}
  {074501} (\bibinfo {year} {2010})},\ \Eprint {http://arxiv.org/abs/1004.0342}
  {arXiv:1004.0342} \BibitemShut {NoStop}%
\bibitem [{\citenamefont {Bazavov}\ and\ \citenamefont
  {Petreczky}(2010)}]{Bazavov:2010pg}%
  \BibitemOpen
  \bibfield  {author} {\bibinfo {author} {\bibfnamefont {A.}~\bibnamefont
  {Bazavov}}\ and\ \bibinfo {author} {\bibfnamefont {P.}~\bibnamefont
  {Petreczky}} (\bibinfo {collaboration} {HotQCD}),\ }\href {\doibase
  10.22323/1.105.0169} {\bibfield  {journal} {\bibinfo  {journal} {PoS}\
  }\textbf {\bibinfo {volume} {LATTICE2010}},\ \bibinfo {pages} {169} (\bibinfo
  {year} {2010})},\ \Eprint {http://arxiv.org/abs/1012.1257} {arXiv:1012.1257}
  \BibitemShut {NoStop}%
\bibitem [{\citenamefont {Follana}\ \emph {et~al.}(2008)\citenamefont
  {Follana}, \citenamefont {Davies}, \citenamefont {Lepage},\ and\
  \citenamefont {Shigemitsu}}]{Follana:2007uv}%
  \BibitemOpen
  \bibfield  {author} {\bibinfo {author} {\bibfnamefont {E.}~\bibnamefont
  {Follana}}, \bibinfo {author} {\bibfnamefont {C.~T.~H.}\ \bibnamefont
  {Davies}}, \bibinfo {author} {\bibfnamefont {G.~P.}\ \bibnamefont {Lepage}},
  \ and\ \bibinfo {author} {\bibfnamefont {J.}~\bibnamefont {Shigemitsu}}
  (\bibinfo {collaboration} {HPQCD, UKQCD}),\ }\href {\doibase
  10.1103/PhysRevLett.100.062002} {\bibfield  {journal} {\bibinfo  {journal}
  {Phys. Rev. Lett.}\ }\textbf {\bibinfo {volume} {100}},\ \bibinfo {pages}
  {062002} (\bibinfo {year} {2008})},\ \Eprint {http://arxiv.org/abs/0706.1726}
  {arXiv:0706.1726} \BibitemShut {NoStop}%
\bibitem [{\citenamefont {Bazavov}\ \emph
  {et~al.}(2012{\natexlab{b}})\citenamefont {Bazavov} \emph
  {et~al.}}]{Bazavov:2011nk}%
  \BibitemOpen
  \bibfield  {author} {\bibinfo {author} {\bibfnamefont {A.}~\bibnamefont
  {Bazavov}} \emph {et~al.},\ }\href {\doibase 10.1103/PhysRevD.85.054503}
  {\bibfield  {journal} {\bibinfo  {journal} {Phys. Rev.}\ }\textbf {\bibinfo
  {volume} {D85}},\ \bibinfo {pages} {054503} (\bibinfo {year}
  {2012}{\natexlab{b}})},\ \Eprint {http://arxiv.org/abs/1111.1710}
  {arXiv:1111.1710} \BibitemShut {NoStop}%
\bibitem [{\citenamefont {Bazavov}\ \emph {et~al.}(2014)\citenamefont {Bazavov}
  \emph {et~al.}}]{Bazavov:2014pvz}%
  \BibitemOpen
  \bibfield  {author} {\bibinfo {author} {\bibfnamefont {A.}~\bibnamefont
  {Bazavov}} \emph {et~al.} (\bibinfo {collaboration} {HotQCD}),\ }\href
  {\doibase 10.1103/PhysRevD.90.094503} {\bibfield  {journal} {\bibinfo
  {journal} {Phys. Rev.}\ }\textbf {\bibinfo {volume} {D90}},\ \bibinfo {pages}
  {094503} (\bibinfo {year} {2014})},\ \Eprint {http://arxiv.org/abs/1407.6387}
  {arXiv:1407.6387} \BibitemShut {NoStop}%
\bibitem [{\citenamefont {Bazavov}\ \emph {et~al.}(2017)\citenamefont {Bazavov}
  \emph {et~al.}}]{Bazavov:2017dus}%
  \BibitemOpen
  \bibfield  {author} {\bibinfo {author} {\bibfnamefont {A.}~\bibnamefont
  {Bazavov}} \emph {et~al.},\ }\href {\doibase 10.1103/PhysRevD.95.054504}
  {\bibfield  {journal} {\bibinfo  {journal} {Phys. Rev.}\ }\textbf {\bibinfo
  {volume} {D95}},\ \bibinfo {pages} {054504} (\bibinfo {year} {2017})},\
  \Eprint {http://arxiv.org/abs/1701.04325} {arXiv:1701.04325} \BibitemShut
  {NoStop}%
\bibitem [{\citenamefont {Bazavov}\ \emph {et~al.}(2016)\citenamefont
  {Bazavov}, \citenamefont {Brambilla}, \citenamefont {Ding}, \citenamefont
  {Petreczky}, \citenamefont {Schadler}, \citenamefont {Vairo},\ and\
  \citenamefont {Weber}}]{Bazavov:2016uvm}%
  \BibitemOpen
  \bibfield  {author} {\bibinfo {author} {\bibfnamefont {A.}~\bibnamefont
  {Bazavov}}, \bibinfo {author} {\bibfnamefont {N.}~\bibnamefont {Brambilla}},
  \bibinfo {author} {\bibfnamefont {H.~T.}\ \bibnamefont {Ding}}, \bibinfo
  {author} {\bibfnamefont {P.}~\bibnamefont {Petreczky}}, \bibinfo {author}
  {\bibfnamefont {H.~P.}\ \bibnamefont {Schadler}}, \bibinfo {author}
  {\bibfnamefont {A.}~\bibnamefont {Vairo}}, \ and\ \bibinfo {author}
  {\bibfnamefont {J.~H.}\ \bibnamefont {Weber}},\ }\href {\doibase
  10.1103/PhysRevD.93.114502} {\bibfield  {journal} {\bibinfo  {journal} {Phys.
  Rev.}\ }\textbf {\bibinfo {volume} {D93}},\ \bibinfo {pages} {114502}
  (\bibinfo {year} {2016})},\ \Eprint {http://arxiv.org/abs/1603.06637}
  {arXiv:1603.06637} \BibitemShut {NoStop}%
\bibitem [{\citenamefont {Bazavov}\ \emph {et~al.}(2018)\citenamefont
  {Bazavov}, \citenamefont {Brambilla}, \citenamefont {Petreczky},
  \citenamefont {Vairo},\ and\ \citenamefont {Weber}}]{Bazavov:2018wmo}%
  \BibitemOpen
  \bibfield  {author} {\bibinfo {author} {\bibfnamefont {A.}~\bibnamefont
  {Bazavov}}, \bibinfo {author} {\bibfnamefont {N.}~\bibnamefont {Brambilla}},
  \bibinfo {author} {\bibfnamefont {P.}~\bibnamefont {Petreczky}}, \bibinfo
  {author} {\bibfnamefont {A.}~\bibnamefont {Vairo}}, \ and\ \bibinfo {author}
  {\bibfnamefont {J.~H.}\ \bibnamefont {Weber}} (\bibinfo {collaboration}
  {TUMQCD}),\ }\href {\doibase 10.1103/PhysRevD.98.054511} {\bibfield
  {journal} {\bibinfo  {journal} {Phys. Rev.}\ }\textbf {\bibinfo {volume}
  {D98}},\ \bibinfo {pages} {054511} (\bibinfo {year} {2018})},\ \Eprint
  {http://arxiv.org/abs/1804.10600} {arXiv:1804.10600} \BibitemShut {NoStop}%
\bibitem [{\citenamefont {Hegde}(2011)}]{Hegde:2011np}%
  \BibitemOpen
  \bibfield  {author} {\bibinfo {author} {\bibfnamefont {P.}~\bibnamefont
  {Hegde}},\ }\href {\doibase 10.22323/1.139.0014} {\bibfield  {journal}
  {\bibinfo  {journal} {PoS}\ }\textbf {\bibinfo {volume} {LATTICE2011}},\
  \bibinfo {pages} {014} (\bibinfo {year} {2011})},\ \Eprint
  {http://arxiv.org/abs/1112.0364} {arXiv:1112.0364} \BibitemShut {NoStop}%
\bibitem [{\citenamefont {Sandmeyer}(2019)}]{Sandmeyer:2019}%
  \BibitemOpen
  \bibfield  {author} {\bibinfo {author} {\bibfnamefont {H.}~\bibnamefont
  {Sandmeyer}},\ }\href {\doibase 10.4119/unibi/2936264} {\  (\bibinfo {year}
  {PhD thesis 2019}),\ 10.4119/unibi/2936264}\BibitemShut {NoStop}%
\bibitem [{\citenamefont {Akaike}(1974)}]{AIC}%
  \BibitemOpen
  \bibfield  {author} {\bibinfo {author} {\bibfnamefont {H.}~\bibnamefont
  {Akaike}},\ }\href {\doibase 10.1109/TAC.1974.1100705} {\bibfield  {journal}
  {\bibinfo  {journal} {IEEE Transactions on Automatic Control}\ }\textbf
  {\bibinfo {volume} {19}},\ \bibinfo {pages} {716} (\bibinfo {year}
  {1974})}\BibitemShut {NoStop}%
\bibitem [{\citenamefont {Cavanaugh}(1997)}]{AICc}%
  \BibitemOpen
  \bibfield  {author} {\bibinfo {author} {\bibfnamefont {J.~E.}\ \bibnamefont
  {Cavanaugh}},\ }\href {\doibase
  https://doi.org/10.1016/S0167-7152(96)00128-9} {\bibfield  {journal}
  {\bibinfo  {journal} {Statistics \& Probability Letters}\ }\textbf {\bibinfo
  {volume} {33}},\ \bibinfo {pages} {201 } (\bibinfo {year}
  {1997})}\BibitemShut {NoStop}%
\bibitem [{\citenamefont {Bernard}\ \emph {et~al.}(1992)\citenamefont
  {Bernard}, \citenamefont {Ogilvie}, \citenamefont {DeGrand}, \citenamefont
  {DeTar}, \citenamefont {Gottlieb}, \citenamefont {Krasnitz}, \citenamefont
  {Sugar},\ and\ \citenamefont {Toussaint}}]{Bernard:1991ah}%
  \BibitemOpen
  \bibfield  {author} {\bibinfo {author} {\bibfnamefont {C.~W.}\ \bibnamefont
  {Bernard}}, \bibinfo {author} {\bibfnamefont {M.~C.}\ \bibnamefont
  {Ogilvie}}, \bibinfo {author} {\bibfnamefont {T.~A.}\ \bibnamefont
  {DeGrand}}, \bibinfo {author} {\bibfnamefont {C.~E.}\ \bibnamefont {DeTar}},
  \bibinfo {author} {\bibfnamefont {S.~A.}\ \bibnamefont {Gottlieb}}, \bibinfo
  {author} {\bibfnamefont {A.}~\bibnamefont {Krasnitz}}, \bibinfo {author}
  {\bibfnamefont {R.~L.}\ \bibnamefont {Sugar}}, \ and\ \bibinfo {author}
  {\bibfnamefont {D.}~\bibnamefont {Toussaint}},\ }\href {\doibase
  10.1103/PhysRevLett.68.2125} {\bibfield  {journal} {\bibinfo  {journal}
  {Phys. Rev. Lett.}\ }\textbf {\bibinfo {volume} {68}},\ \bibinfo {pages}
  {2125} (\bibinfo {year} {1992})}\BibitemShut {NoStop}%
\bibitem [{\citenamefont {Bernard}\ \emph {et~al.}(1993)\citenamefont
  {Bernard}, \citenamefont {Blum}, \citenamefont {DeGrand}, \citenamefont
  {Detar}, \citenamefont {Gottlieb}, \citenamefont {Krasnitz}, \citenamefont
  {Sugar},\ and\ \citenamefont {Toussaint}}]{Bernard:1993an}%
  \BibitemOpen
  \bibfield  {author} {\bibinfo {author} {\bibfnamefont {C.~W.}\ \bibnamefont
  {Bernard}}, \bibinfo {author} {\bibfnamefont {T.}~\bibnamefont {Blum}},
  \bibinfo {author} {\bibfnamefont {T.~A.}\ \bibnamefont {DeGrand}}, \bibinfo
  {author} {\bibfnamefont {C.~E.}\ \bibnamefont {Detar}}, \bibinfo {author}
  {\bibfnamefont {S.~A.}\ \bibnamefont {Gottlieb}}, \bibinfo {author}
  {\bibfnamefont {A.}~\bibnamefont {Krasnitz}}, \bibinfo {author}
  {\bibfnamefont {R.~L.}\ \bibnamefont {Sugar}}, \ and\ \bibinfo {author}
  {\bibfnamefont {D.}~\bibnamefont {Toussaint}},\ }\href {\doibase
  10.1103/PhysRevD.48.4419} {\bibfield  {journal} {\bibinfo  {journal} {Phys.
  Rev.}\ }\textbf {\bibinfo {volume} {D48}},\ \bibinfo {pages} {4419} (\bibinfo
  {year} {1993})},\ \Eprint {http://arxiv.org/abs/hep-lat/9305023}
  {arXiv:hep-lat/9305023} \BibitemShut {NoStop}%
\bibitem [{\citenamefont {Bernard}\ \emph {et~al.}(2001)\citenamefont
  {Bernard}, \citenamefont {Burch}, \citenamefont {Orginos}, \citenamefont
  {Toussaint}, \citenamefont {DeGrand}, \citenamefont {Detar}, \citenamefont
  {Datta}, \citenamefont {Gottlieb}, \citenamefont {Heller},\ and\
  \citenamefont {Sugar}}]{Bernard:2001av}%
  \BibitemOpen
  \bibfield  {author} {\bibinfo {author} {\bibfnamefont {C.~W.}\ \bibnamefont
  {Bernard}}, \bibinfo {author} {\bibfnamefont {T.}~\bibnamefont {Burch}},
  \bibinfo {author} {\bibfnamefont {K.}~\bibnamefont {Orginos}}, \bibinfo
  {author} {\bibfnamefont {D.}~\bibnamefont {Toussaint}}, \bibinfo {author}
  {\bibfnamefont {T.~A.}\ \bibnamefont {DeGrand}}, \bibinfo {author}
  {\bibfnamefont {C.~E.}\ \bibnamefont {Detar}}, \bibinfo {author}
  {\bibfnamefont {S.}~\bibnamefont {Datta}}, \bibinfo {author} {\bibfnamefont
  {S.~A.}\ \bibnamefont {Gottlieb}}, \bibinfo {author} {\bibfnamefont {U.~M.}\
  \bibnamefont {Heller}}, \ and\ \bibinfo {author} {\bibfnamefont
  {R.}~\bibnamefont {Sugar}},\ }\href {\doibase 10.1103/PhysRevD.64.054506}
  {\bibfield  {journal} {\bibinfo  {journal} {Phys. Rev.}\ }\textbf {\bibinfo
  {volume} {D64}},\ \bibinfo {pages} {054506} (\bibinfo {year} {2001})},\
  \Eprint {http://arxiv.org/abs/hep-lat/0104002} {arXiv:hep-lat/0104002}
  \BibitemShut {NoStop}%
\bibitem [{\citenamefont {Tanabashi}\ \emph {et~al.}(2018)\citenamefont
  {Tanabashi} \emph {et~al.}}]{Tanabashi:2018oca}%
  \BibitemOpen
  \bibfield  {author} {\bibinfo {author} {\bibfnamefont {M.}~\bibnamefont
  {Tanabashi}} \emph {et~al.} (\bibinfo {collaboration} {Particle Data
  Group}),\ }\href {\doibase 10.1103/PhysRevD.98.030001} {\bibfield  {journal}
  {\bibinfo  {journal} {Phys. Rev.}\ }\textbf {\bibinfo {volume} {D98}},\
  \bibinfo {pages} {030001} (\bibinfo {year} {2018})}\BibitemShut {NoStop}%
\bibitem [{\citenamefont {Prelovsek}(2006)}]{Prelovsek:2005rf}%
  \BibitemOpen
  \bibfield  {author} {\bibinfo {author} {\bibfnamefont {S.}~\bibnamefont
  {Prelovsek}},\ }\href {\doibase 10.1103/PhysRevD.73.014506} {\bibfield
  {journal} {\bibinfo  {journal} {Phys. Rev.}\ }\textbf {\bibinfo {volume}
  {D73}},\ \bibinfo {pages} {014506} (\bibinfo {year} {2006})},\ \Eprint
  {http://arxiv.org/abs/hep-lat/0510080} {arXiv:hep-lat/0510080} \BibitemShut
  {NoStop}%
\bibitem [{\citenamefont {Prelovsek}\ \emph {et~al.}(2004)\citenamefont
  {Prelovsek}, \citenamefont {Dawson}, \citenamefont {Izubuchi}, \citenamefont
  {Orginos},\ and\ \citenamefont {Soni}}]{Prelovsek:2004jp}%
  \BibitemOpen
  \bibfield  {author} {\bibinfo {author} {\bibfnamefont {S.}~\bibnamefont
  {Prelovsek}}, \bibinfo {author} {\bibfnamefont {C.}~\bibnamefont {Dawson}},
  \bibinfo {author} {\bibfnamefont {T.}~\bibnamefont {Izubuchi}}, \bibinfo
  {author} {\bibfnamefont {K.}~\bibnamefont {Orginos}}, \ and\ \bibinfo
  {author} {\bibfnamefont {A.}~\bibnamefont {Soni}},\ }\href {\doibase
  10.1103/PhysRevD.70.094503} {\bibfield  {journal} {\bibinfo  {journal} {Phys.
  Rev.}\ }\textbf {\bibinfo {volume} {D70}},\ \bibinfo {pages} {094503}
  (\bibinfo {year} {2004})},\ \Eprint {http://arxiv.org/abs/hep-lat/0407037}
  {arXiv:hep-lat/0407037} \BibitemShut {NoStop}%
\bibitem [{\citenamefont {Bernard}\ \emph {et~al.}(2007)\citenamefont
  {Bernard}, \citenamefont {DeTar}, \citenamefont {Fu},\ and\ \citenamefont
  {Prelovsek}}]{Bernard:2007qf}%
  \BibitemOpen
  \bibfield  {author} {\bibinfo {author} {\bibfnamefont {C.}~\bibnamefont
  {Bernard}}, \bibinfo {author} {\bibfnamefont {C.~E.}\ \bibnamefont {DeTar}},
  \bibinfo {author} {\bibfnamefont {Z.}~\bibnamefont {Fu}}, \ and\ \bibinfo
  {author} {\bibfnamefont {S.}~\bibnamefont {Prelovsek}},\ }\href {\doibase
  10.1103/PhysRevD.76.094504} {\bibfield  {journal} {\bibinfo  {journal} {Phys.
  Rev.}\ }\textbf {\bibinfo {volume} {D76}},\ \bibinfo {pages} {094504}
  (\bibinfo {year} {2007})},\ \Eprint {http://arxiv.org/abs/0707.2402}
  {arXiv:0707.2402} \BibitemShut {NoStop}%
\bibitem [{\citenamefont {Lee}\ and\ \citenamefont
  {Sharpe}(1999)}]{Lee:1999zxa}%
  \BibitemOpen
  \bibfield  {author} {\bibinfo {author} {\bibfnamefont {W.-J.}\ \bibnamefont
  {Lee}}\ and\ \bibinfo {author} {\bibfnamefont {S.~R.}\ \bibnamefont
  {Sharpe}},\ }\href {\doibase 10.1103/PhysRevD.60.114503} {\bibfield
  {journal} {\bibinfo  {journal} {Phys. Rev.}\ }\textbf {\bibinfo {volume}
  {D60}},\ \bibinfo {pages} {114503} (\bibinfo {year} {1999})},\ \Eprint
  {http://arxiv.org/abs/hep-lat/9905023} {arXiv:hep-lat/9905023} \BibitemShut
  {NoStop}%
\bibitem [{\citenamefont {Aarts}\ \emph
  {et~al.}(2017{\natexlab{b}})\citenamefont {Aarts}, \citenamefont {Allton},
  \citenamefont {de~Boni}, \citenamefont {Hands}, \citenamefont {Jäger},
  \citenamefont {Praki},\ and\ \citenamefont {Skullerud}}]{Aarts:2017iai}%
  \BibitemOpen
  \bibfield  {author} {\bibinfo {author} {\bibfnamefont {G.}~\bibnamefont
  {Aarts}}, \bibinfo {author} {\bibfnamefont {C.}~\bibnamefont {Allton}},
  \bibinfo {author} {\bibfnamefont {D.}~\bibnamefont {de~Boni}}, \bibinfo
  {author} {\bibfnamefont {S.}~\bibnamefont {Hands}}, \bibinfo {author}
  {\bibfnamefont {B.}~\bibnamefont {Jäger}}, \bibinfo {author} {\bibfnamefont
  {C.}~\bibnamefont {Praki}}, \ and\ \bibinfo {author} {\bibfnamefont {J.-I.}\
  \bibnamefont {Skullerud}},\ }\href {\doibase 10.1051/epjconf/201817114005} {\
   (\bibinfo {year} {2017}{\natexlab{b}}),\ 10.1051/epjconf/201817114005},\
  \bibinfo {note} {[EPJ Web Conf.171,14005(2018)]},\ \Eprint
  {http://arxiv.org/abs/1710.00566} {arXiv:1710.00566} \BibitemShut {NoStop}%
\bibitem [{\citenamefont {Datta}\ \emph {et~al.}(2013)\citenamefont {Datta},
  \citenamefont {Gupta}, \citenamefont {Padmanath}, \citenamefont {Maiti},\
  and\ \citenamefont {Mathur}}]{Datta:2012fz}%
  \BibitemOpen
  \bibfield  {author} {\bibinfo {author} {\bibfnamefont {S.}~\bibnamefont
  {Datta}}, \bibinfo {author} {\bibfnamefont {S.}~\bibnamefont {Gupta}},
  \bibinfo {author} {\bibfnamefont {M.}~\bibnamefont {Padmanath}}, \bibinfo
  {author} {\bibfnamefont {J.}~\bibnamefont {Maiti}}, \ and\ \bibinfo {author}
  {\bibfnamefont {N.}~\bibnamefont {Mathur}},\ }\href {\doibase
  10.1007/JHEP02(2013)145} {\bibfield  {journal} {\bibinfo  {journal} {JHEP}\
  }\textbf {\bibinfo {volume} {02}},\ \bibinfo {pages} {145} (\bibinfo {year}
  {2013})},\ \Eprint {http://arxiv.org/abs/1212.2927} {arXiv:1212.2927}
  \BibitemShut {NoStop}%
\bibitem [{\citenamefont {Aoki}\ \emph {et~al.}(2012)\citenamefont {Aoki},
  \citenamefont {Fukaya},\ and\ \citenamefont {Taniguchi}}]{Aoki:2012yj}%
  \BibitemOpen
  \bibfield  {author} {\bibinfo {author} {\bibfnamefont {S.}~\bibnamefont
  {Aoki}}, \bibinfo {author} {\bibfnamefont {H.}~\bibnamefont {Fukaya}}, \ and\
  \bibinfo {author} {\bibfnamefont {Y.}~\bibnamefont {Taniguchi}},\ }\href
  {\doibase 10.1103/PhysRevD.86.114512} {\bibfield  {journal} {\bibinfo
  {journal} {Phys. Rev.}\ }\textbf {\bibinfo {volume} {D86}},\ \bibinfo {pages}
  {114512} (\bibinfo {year} {2012})},\ \Eprint {http://arxiv.org/abs/1209.2061}
  {arXiv:1209.2061} \BibitemShut {NoStop}%
\bibitem [{\citenamefont {Suzuki}\ \emph
  {et~al.}(2018{\natexlab{b}})\citenamefont {Suzuki}, \citenamefont {Aoki},
  \citenamefont {Aoki}, \citenamefont {Cossu}, \citenamefont {Fukaya},\ and\
  \citenamefont {Hashimoto}}]{Suzuki:2017ifu}%
  \BibitemOpen
  \bibfield  {author} {\bibinfo {author} {\bibfnamefont {K.}~\bibnamefont
  {Suzuki}}, \bibinfo {author} {\bibfnamefont {S.}~\bibnamefont {Aoki}},
  \bibinfo {author} {\bibfnamefont {Y.}~\bibnamefont {Aoki}}, \bibinfo {author}
  {\bibfnamefont {G.}~\bibnamefont {Cossu}}, \bibinfo {author} {\bibfnamefont
  {H.}~\bibnamefont {Fukaya}}, \ and\ \bibinfo {author} {\bibfnamefont
  {S.}~\bibnamefont {Hashimoto}} (\bibinfo {collaboration} {JLQCD}),\ }\href
  {\doibase 10.1051/epjconf/201817507025} {\bibfield  {journal} {\bibinfo
  {journal} {EPJ Web Conf.}\ }\textbf {\bibinfo {volume} {175}},\ \bibinfo
  {pages} {07025} (\bibinfo {year} {2018}{\natexlab{b}})},\ \Eprint
  {http://arxiv.org/abs/1711.09239} {arXiv:1711.09239} \BibitemShut {NoStop}%
\bibitem [{\citenamefont {Detar}\ and\ \citenamefont
  {Kogut}(1987{\natexlab{b}})}]{DeTar:1987xb}%
  \BibitemOpen
  \bibfield  {author} {\bibinfo {author} {\bibfnamefont {C.~E.}\ \bibnamefont
  {Detar}}\ and\ \bibinfo {author} {\bibfnamefont {J.~B.}\ \bibnamefont
  {Kogut}},\ }\href {\doibase 10.1103/PhysRevD.36.2828} {\bibfield  {journal}
  {\bibinfo  {journal} {Phys. Rev.}\ }\textbf {\bibinfo {volume} {D36}},\
  \bibinfo {pages} {2828} (\bibinfo {year} {1987}{\natexlab{b}})}\BibitemShut
  {NoStop}%
\bibitem [{\citenamefont {Born}\ \emph {et~al.}(1991)\citenamefont {Born},
  \citenamefont {Gupta}, \citenamefont {Irback}, \citenamefont {Karsch},
  \citenamefont {Laermann}, \citenamefont {Petersson},\ and\ \citenamefont
  {Satz}}]{Born:1991zz}%
  \BibitemOpen
  \bibfield  {author} {\bibinfo {author} {\bibfnamefont {K.~D.}\ \bibnamefont
  {Born}}, \bibinfo {author} {\bibfnamefont {S.}~\bibnamefont {Gupta}},
  \bibinfo {author} {\bibfnamefont {A.}~\bibnamefont {Irback}}, \bibinfo
  {author} {\bibfnamefont {F.}~\bibnamefont {Karsch}}, \bibinfo {author}
  {\bibfnamefont {E.}~\bibnamefont {Laermann}}, \bibinfo {author}
  {\bibfnamefont {B.}~\bibnamefont {Petersson}}, \ and\ \bibinfo {author}
  {\bibfnamefont {H.}~\bibnamefont {Satz}} (\bibinfo {collaboration} {MT(c)}),\
  }\href {\doibase 10.1103/PhysRevLett.67.302} {\bibfield  {journal} {\bibinfo
  {journal} {Phys. Rev. Lett.}\ }\textbf {\bibinfo {volume} {67}},\ \bibinfo
  {pages} {302} (\bibinfo {year} {1991})}\BibitemShut {NoStop}%
\bibitem [{\citenamefont {Banerjee}\ \emph {et~al.}(2011)\citenamefont
  {Banerjee}, \citenamefont {Gavai},\ and\ \citenamefont
  {Gupta}}]{Banerjee:2011yd}%
  \BibitemOpen
  \bibfield  {author} {\bibinfo {author} {\bibfnamefont {D.}~\bibnamefont
  {Banerjee}}, \bibinfo {author} {\bibfnamefont {R.~V.}\ \bibnamefont {Gavai}},
  \ and\ \bibinfo {author} {\bibfnamefont {S.}~\bibnamefont {Gupta}},\ }\href
  {\doibase 10.1103/PhysRevD.83.074510} {\bibfield  {journal} {\bibinfo
  {journal} {Phys. Rev.}\ }\textbf {\bibinfo {volume} {D83}},\ \bibinfo {pages}
  {074510} (\bibinfo {year} {2011})},\ \Eprint {http://arxiv.org/abs/1102.4465}
  {arXiv:1102.4465} \BibitemShut {NoStop}%
\bibitem [{\citenamefont {Laermann}\ and\ \citenamefont
  {Pucci}(2012)}]{Laermann:2012sr}%
  \BibitemOpen
  \bibfield  {author} {\bibinfo {author} {\bibfnamefont {E.}~\bibnamefont
  {Laermann}}\ and\ \bibinfo {author} {\bibfnamefont {F.}~\bibnamefont
  {Pucci}},\ }\href {\doibase 10.1140/epjc/s10052-012-2200-1} {\bibfield
  {journal} {\bibinfo  {journal} {Eur. Phys. J.}\ }\textbf {\bibinfo {volume}
  {C72}},\ \bibinfo {pages} {2200} (\bibinfo {year} {2012})},\ \Eprint
  {http://arxiv.org/abs/1207.6615} {arXiv:1207.6615} \BibitemShut {NoStop}%
\bibitem [{\citenamefont {Gupta}\ and\ \citenamefont
  {Karthik}(2013)}]{Gupta:2013vha}%
  \BibitemOpen
  \bibfield  {author} {\bibinfo {author} {\bibfnamefont {S.}~\bibnamefont
  {Gupta}}\ and\ \bibinfo {author} {\bibfnamefont {N.}~\bibnamefont
  {Karthik}},\ }\href {\doibase 10.1103/PhysRevD.87.094001} {\bibfield
  {journal} {\bibinfo  {journal} {Phys. Rev.}\ }\textbf {\bibinfo {volume}
  {D87}},\ \bibinfo {pages} {094001} (\bibinfo {year} {2013})},\ \Eprint
  {http://arxiv.org/abs/1302.4917} {arXiv:1302.4917} \BibitemShut {NoStop}%
\bibitem [{\citenamefont {Brandt}\ \emph {et~al.}(2014)\citenamefont {Brandt},
  \citenamefont {Francis}, \citenamefont {Laine},\ and\ \citenamefont
  {Meyer}}]{Brandt:2014uda}%
  \BibitemOpen
  \bibfield  {author} {\bibinfo {author} {\bibfnamefont {B.~B.}\ \bibnamefont
  {Brandt}}, \bibinfo {author} {\bibfnamefont {A.}~\bibnamefont {Francis}},
  \bibinfo {author} {\bibfnamefont {M.}~\bibnamefont {Laine}}, \ and\ \bibinfo
  {author} {\bibfnamefont {H.~B.}\ \bibnamefont {Meyer}},\ }\href {\doibase
  10.1007/JHEP05(2014)117} {\bibfield  {journal} {\bibinfo  {journal} {JHEP}\
  }\textbf {\bibinfo {volume} {05}},\ \bibinfo {pages} {117} (\bibinfo {year}
  {2014})},\ \Eprint {http://arxiv.org/abs/1404.2404} {arXiv:1404.2404}
  \BibitemShut {NoStop}%
\bibitem [{\citenamefont {Braaten}\ and\ \citenamefont
  {Nieto}(1996)}]{Braaten:1995jr}%
  \BibitemOpen
  \bibfield  {author} {\bibinfo {author} {\bibfnamefont {E.}~\bibnamefont
  {Braaten}}\ and\ \bibinfo {author} {\bibfnamefont {A.}~\bibnamefont
  {Nieto}},\ }\href {\doibase 10.1103/PhysRevD.53.3421} {\bibfield  {journal}
  {\bibinfo  {journal} {Phys. Rev.}\ }\textbf {\bibinfo {volume} {D53}},\
  \bibinfo {pages} {3421} (\bibinfo {year} {1996})},\ \Eprint
  {http://arxiv.org/abs/hep-ph/9510408} {arXiv:hep-ph/9510408} \BibitemShut
  {NoStop}%
\bibitem [{\citenamefont {Bazavov}\ \emph {et~al.}(2013)\citenamefont
  {Bazavov}, \citenamefont {Ding}, \citenamefont {Hegde}, \citenamefont
  {Karsch}, \citenamefont {Miao}, \citenamefont {Mukherjee}, \citenamefont
  {Petreczky}, \citenamefont {Schmidt},\ and\ \citenamefont
  {Velytsky}}]{Bazavov:2013uja}%
  \BibitemOpen
  \bibfield  {author} {\bibinfo {author} {\bibfnamefont {A.}~\bibnamefont
  {Bazavov}}, \bibinfo {author} {\bibfnamefont {H.~T.}\ \bibnamefont {Ding}},
  \bibinfo {author} {\bibfnamefont {P.}~\bibnamefont {Hegde}}, \bibinfo
  {author} {\bibfnamefont {F.}~\bibnamefont {Karsch}}, \bibinfo {author}
  {\bibfnamefont {C.}~\bibnamefont {Miao}}, \bibinfo {author} {\bibfnamefont
  {S.}~\bibnamefont {Mukherjee}}, \bibinfo {author} {\bibfnamefont
  {P.}~\bibnamefont {Petreczky}}, \bibinfo {author} {\bibfnamefont
  {C.}~\bibnamefont {Schmidt}}, \ and\ \bibinfo {author} {\bibfnamefont
  {A.}~\bibnamefont {Velytsky}},\ }\href {\doibase 10.1103/PhysRevD.88.094021}
  {\bibfield  {journal} {\bibinfo  {journal} {Phys. Rev.}\ }\textbf {\bibinfo
  {volume} {D88}},\ \bibinfo {pages} {094021} (\bibinfo {year} {2013})},\
  \Eprint {http://arxiv.org/abs/1309.2317} {arXiv:1309.2317} \BibitemShut
  {NoStop}%
\bibitem [{\citenamefont {Ding}\ \emph {et~al.}(2015)\citenamefont {Ding},
  \citenamefont {Mukherjee}, \citenamefont {Ohno}, \citenamefont {Petreczky},\
  and\ \citenamefont {Schadler}}]{Ding:2015fca}%
  \BibitemOpen
  \bibfield  {author} {\bibinfo {author} {\bibfnamefont {H.~T.}\ \bibnamefont
  {Ding}}, \bibinfo {author} {\bibfnamefont {S.}~\bibnamefont {Mukherjee}},
  \bibinfo {author} {\bibfnamefont {H.}~\bibnamefont {Ohno}}, \bibinfo {author}
  {\bibfnamefont {P.}~\bibnamefont {Petreczky}}, \ and\ \bibinfo {author}
  {\bibfnamefont {H.~P.}\ \bibnamefont {Schadler}},\ }\href {\doibase
  10.1103/PhysRevD.92.074043} {\bibfield  {journal} {\bibinfo  {journal} {Phys.
  Rev.}\ }\textbf {\bibinfo {volume} {D92}},\ \bibinfo {pages} {074043}
  (\bibinfo {year} {2015})},\ \Eprint {http://arxiv.org/abs/1507.06637}
  {arXiv:1507.06637} \BibitemShut {NoStop}%
\bibitem [{\citenamefont {Koch}\ \emph {et~al.}(1992)\citenamefont {Koch},
  \citenamefont {Shuryak}, \citenamefont {Brown},\ and\ \citenamefont
  {Jackson}}]{Koch:1992nx}%
  \BibitemOpen
  \bibfield  {author} {\bibinfo {author} {\bibfnamefont {V.}~\bibnamefont
  {Koch}}, \bibinfo {author} {\bibfnamefont {E.~V.}\ \bibnamefont {Shuryak}},
  \bibinfo {author} {\bibfnamefont {G.~E.}\ \bibnamefont {Brown}}, \ and\
  \bibinfo {author} {\bibfnamefont {A.~D.}\ \bibnamefont {Jackson}},\ }\href
  {\doibase 10.1103/PhysRevD.46.3169, 10.1103/PhysRevD.47.2157} {\bibfield
  {journal} {\bibinfo  {journal} {Phys. Rev.}\ }\textbf {\bibinfo {volume}
  {D46}},\ \bibinfo {pages} {3169} (\bibinfo {year} {1992})},\ \bibinfo {note}
  {[Erratum: Phys. Rev.D47,2157(1993)]},\ \Eprint
  {http://arxiv.org/abs/hep-ph/9204236} {arXiv:hep-ph/9204236} \BibitemShut
  {NoStop}%
\bibitem [{\citenamefont {Shuryak}(1993)}]{Shuryak:1993kg}%
  \BibitemOpen
  \bibfield  {author} {\bibinfo {author} {\bibfnamefont {E.~V.}\ \bibnamefont
  {Shuryak}},\ }\href {\doibase 10.1103/RevModPhys.65.1} {\bibfield  {journal}
  {\bibinfo  {journal} {Rev. Mod. Phys.}\ }\textbf {\bibinfo {volume} {65}},\
  \bibinfo {pages} {1} (\bibinfo {year} {1993})}\BibitemShut {NoStop}%
\bibitem [{\citenamefont {Laine}\ and\ \citenamefont
  {Vepsalainen}(2004)}]{Laine:2003bd}%
  \BibitemOpen
  \bibfield  {author} {\bibinfo {author} {\bibfnamefont {M.}~\bibnamefont
  {Laine}}\ and\ \bibinfo {author} {\bibfnamefont {M.}~\bibnamefont
  {Vepsalainen}},\ }\href {\doibase 10.1088/1126-6708/2004/02/004} {\bibfield
  {journal} {\bibinfo  {journal} {JHEP}\ }\textbf {\bibinfo {volume} {02}},\
  \bibinfo {pages} {004} (\bibinfo {year} {2004})},\ \Eprint
  {http://arxiv.org/abs/hep-ph/0311268} {arXiv:hep-ph/0311268} \BibitemShut
  {NoStop}%
\bibitem [{\citenamefont {Laine}\ and\ \citenamefont
  {Schroder}(2005)}]{Laine:2005ai}%
  \BibitemOpen
  \bibfield  {author} {\bibinfo {author} {\bibfnamefont {M.}~\bibnamefont
  {Laine}}\ and\ \bibinfo {author} {\bibfnamefont {Y.}~\bibnamefont
  {Schroder}},\ }\href {\doibase 10.1088/1126-6708/2005/03/067} {\bibfield
  {journal} {\bibinfo  {journal} {JHEP}\ }\textbf {\bibinfo {volume} {03}},\
  \bibinfo {pages} {067} (\bibinfo {year} {2005})},\ \Eprint
  {http://arxiv.org/abs/hep-ph/0503061} {arXiv:hep-ph/0503061} \BibitemShut
  {NoStop}%
\bibitem [{\citenamefont {Rohrhofer}\ \emph {et~al.}(2019)\citenamefont
  {Rohrhofer}, \citenamefont {Aoki}, \citenamefont {Cossu}, \citenamefont
  {Fukaya}, \citenamefont {Gattringer}, \citenamefont {Glozman}, \citenamefont
  {Hashimoto}, \citenamefont {Lang},\ and\ \citenamefont
  {Prelovsek}}]{Rohrhofer:2019qwq}%
  \BibitemOpen
  \bibfield  {author} {\bibinfo {author} {\bibfnamefont {C.}~\bibnamefont
  {Rohrhofer}}, \bibinfo {author} {\bibfnamefont {Y.}~\bibnamefont {Aoki}},
  \bibinfo {author} {\bibfnamefont {G.}~\bibnamefont {Cossu}}, \bibinfo
  {author} {\bibfnamefont {H.}~\bibnamefont {Fukaya}}, \bibinfo {author}
  {\bibfnamefont {C.}~\bibnamefont {Gattringer}}, \bibinfo {author}
  {\bibfnamefont {L.~{\relax Ya}.}\ \bibnamefont {Glozman}}, \bibinfo {author}
  {\bibfnamefont {S.}~\bibnamefont {Hashimoto}}, \bibinfo {author}
  {\bibfnamefont {C.~B.}\ \bibnamefont {Lang}}, \ and\ \bibinfo {author}
  {\bibfnamefont {S.}~\bibnamefont {Prelovsek}},\ }\href {\doibase
  10.1103/PhysRevD.100.014502} {\bibfield  {journal} {\bibinfo  {journal}
  {Phys. Rev.}\ }\textbf {\bibinfo {volume} {D100}},\ \bibinfo {pages} {014502}
  (\bibinfo {year} {2019})},\ \Eprint {http://arxiv.org/abs/1902.03191}
  {arXiv:1902.03191} \BibitemShut {NoStop}%
\end{thebibliography}%

\end{document}